\documentclass[preprint,aps,showpacs,floatfix,superscriptaddress,nofootinbib]{revtex4}
\usepackage[dvips]{graphicx}
\usepackage[dvips,usenames]{color}
\usepackage{epsf}
\usepackage{enumerate}
\usepackage{amsmath}
\usepackage{amssymb}
\usepackage{amsfonts}
\usepackage{latexsym}
\usepackage{revsymb}
\usepackage{bm}
\usepackage{multirow}
\begin{document}

\newcommand{\bra}[1]{\left\langle #1 \right|}
\newcommand{\ket}[1]{\left| #1 \right\rangle}
\newcommand{\braket}[2]{\langle #1 | #2 \rangle}
\renewcommand{\tilde}[1]{\overset{\lower1pt\hbox{$\scriptstyle{\sim}$}}{#1}}
\newcommand{\backtilde}[1]{\overset{\lower1pt\hbox{$\scriptstyle{\backsim}$}}{#1}}
\newcommand{\edlit}[1]{\overset{\lower1pt\hbox{$\scriptstyle{\backsim}$}}{#1}}
%%%%%%%%%%%%%%%%%%%%%%%%%%%%%%%%%%%%%%%%%%%%%%%%%%%%%%%%%%%%%%%%%
\preprint{OCHA-PP-252}
\preprint{YITP-05-52}
\preprint{VPI--IPPAP--06--01}

\title{A Simple Parameterization of\\Matter Effects on Neutrino Oscillations}
\author{Minako~Honda}\email{minako@hep.phys.ocha.ac.jp}
\affiliation{Physics Department, Ochanomizu Women's University, Tokyo 112-8610, Japan.}
\author{Yee~Kao}\email{ykao@vt.edu}
\affiliation{Institute for Particle Physics and Astrophysics, Physics Department, Virginia Tech, Blacksburg VA 24061, USA}
\author{Naotoshi~Okamura}\email{okamura@yukawa.kyoto-u.ac.jp}
\affiliation{Yukawa Institute for Theoretical Physics, Kyoto University, Kyoto 606-8502, Japan}
\author{Tatsu~Takeuchi}\email{takeuchi@vt.edu}
\affiliation{Institute for Particle Physics and Astrophysics, Physics Department, Virginia Tech, Blacksburg VA 24061, USA}

\date{February 12, 2006}

\begin{abstract}
\noindent
We present simple analytical approximations to matter-effect corrected
effective neutrino mixing-angles and effective mass-squared-differences.  
The expressions clarify the
dependence of oscillation probabilities in matter to the mixing angles and
mass-squared-differences in vacuum, and are useful for
analyzing long-baseline neutrino oscillation experiments.
\end{abstract}


\maketitle
%%%%%%%%%%%%%%%%%%%%%%%%%%%%%%%%%%%%%%%%%%%%%%%%%%%%%%%%%%%%%%%%%
\section{Introduction}

The discovery of neutrino masses and mixings through solar \cite{solar,KamLAND}, 
atmospheric \cite{atmos}, and reactor \cite{chooz} neutrino oscillation experiments 
has provided new clues to solving the mysteries of the Standard Model.
Flavor mixing in the lepton-sector, together with the well-known
mixing in the quark-sector, may lead to a new understanding of
what `flavor' is, why there are three generations of fermions, and
where the quark- and lepton-mass hierarchies come from.
CP violation in the neutrino-sector could potentially be large enough to 
account for the matter-antimatter asymmetry in the universe \cite{Fukugita:1986hr}.

Due to these possibilities, 
several long-baseline (LBL) neutrino oscillation 
experiments are in progress \cite{K2K,OPERA,MINOS,NOVA,T2K} with
many more being proposed \cite{Diwan,T2B,H2B,TwoBaseline,IKMN,HagiwaraOkamuraSenda,SuperNova}
for the purpose of better determining the neutrino masses 
and mixing parameters.
Since the neutrino beams of LBL experiments necessarily
traverse the Earth, the understanding of matter effects \cite{MSW}
is crucial in extracting the masses and mixing parameters in vacuum
from the measured oscillation probabilities in matter.

The calculation of matter effects requires the diagonalization of the effective
Hamiltonian in matter, which is an energy- and matter-density-dependent $3\times 3$ matrix.
This can be done numerically on a computer with ease, but the dependence of the
oscillation probabilities on the vacuum parameters is not transparent \cite{Lipari:1999wy}.
It is also possible to write down exact analytical expressions for 
the effective parameters in matter in terms of those in vacuum \cite{KTY}, 
but the expressions are too complicated to be illuminating.
Approximate expressions for the oscillation probabilities
have also been worked out, but were limited in their range of
applicability \cite{Arafune:1997hd}, or still too complicated to be of
practical use \cite{Freund:2001pn}.

In this paper, we derive simple approximate expressions for the
effective mass-squared-differences and mixing angles in matter,
in terms of the corresponding parameters in vacuum.
They are simple enough to be used by hand, yet accurate enough in a wide
energy range.

This paper is organized as follows.
In section~II, we review the formalism of neutrino oscillations to
fix our notation, and list the current experimental bounds on the mixing angles
and mass-squared differences.
In section~III, we derive our approximate expressions for the effective
mixing angles and mass-squared differences in matter, for both neutrinos 
and anti-neutrinos.
In section~IV, we provide sample calculations to illustrate the
accuracy of our approximation.
In section~V, we present simple analytical approximations for the
oscillation probabilities and demonstrate their utility in
understanding which parameters are best constrained at what baseline lengths.
Section~VI concludes.

%%%%%%%%%%%%%%%%%%%%%%%%%%%%%%%%%%%%%%%%%%%%%%%%%%%%%%%%%%%%%
\section{Basics}

%%%%%%%%%%%%%%%%%%%%%%%%%%%%%%%%%%%%%%%%%%%%%%%%%%%%%%%%%%%%%
\subsection{The MNS Matrix}

Assuming three-generation neutrino mixing, the flavor eigenstates
$\ket{\nu_\alpha}$ $(\alpha=e,\mu,\tau)$ are related to 
the three mass eigenstates $\ket{\nu_i}$ $(i=1,2,3)$ via the 
Maki-Nakagawa-Sakata (MNS) matrix: \cite{Maki:1962mu}
\begin{equation}
(V_{\mathrm{MNS}})_{\alpha i} \equiv \braket{\nu_\alpha}{\nu_i}\;, \quad
\ket{\nu_i} 
% = \sum_{\alpha=e,\mu,\tau} \ket{\nu_\alpha}\braket{\nu_\alpha}{\nu_i}
= \sum_{\alpha=e,\mu,\tau} (V_{\mathrm{MNS}})_{\alpha i}\ket{\nu_\alpha}\;,\quad
\ket{\nu_\alpha}
% = \sum_{i=1,2,3} \ket{\nu_i}\braket{\nu_i}{\nu_\alpha}
= \sum_{i=1,2,3} (V_{\mathrm{MNS}})^*_{\alpha i}\ket{\nu_i}\;.
\end{equation}
A popular parametrization is given by \cite{PDB} 
\begin{equation}
V_\mathrm{MNS} = U\mathcal{P} \;,
\end{equation}
with
\begin{eqnarray}
U & = & 
\left[ \begin{array}{ccc} 1 & 0 & 0 \\
                          0 &  c_{23} & s_{23} \\
                          0 & -s_{23} & c_{23}
       \end{array}
\right]
\left[ \begin{array}{ccc} c_{13} & 0 & s_{13} e^{-i\delta} \\
                          0 & 1 & 0 \\
                          -s_{13} e^{i\delta} & 0 & c_{13}
       \end{array}
\right]
\left[ \begin{array}{ccc} c_{12} & s_{12} & 0 \\
                         -s_{12} & c_{12} & 0 \\
                          0 & 0 & 1
       \end{array}
\right] \cr
& = &
\left[ \begin{array}{ccc}
c_{12}c_{13} & s_{12}c_{13} & s_{13} e^{-i\delta} \\
-s_{12}c_{23} - c_{12}s_{13}s_{23}e^{i\delta} &
\phantom{-}c_{12}c_{23} - s_{12}s_{13}s_{23}e^{i\delta} & c_{13}s_{23} \\
\phantom{-}s_{12}s_{23} - c_{12}s_{13}c_{23}e^{i\delta} &
-c_{12}s_{23} - s_{12}s_{13}c_{23}e^{i\delta} & c_{13}c_{23}
\end{array} \right] \;,\cr
& & \cr
\mathcal{P}
& = &
%\left[ \begin{array}{ccc} e^{i\alpha_1/2} & 0 & 0 \\
%                          0 & e^{i\alpha_2/2} & 0 \\
%                          0 & 0 & 1
%       \end{array}
%\right] \cr
\mathrm{diag}(e^{i\alpha_1/2},e^{i\alpha_2/2},1) \;,
\label{UPparam}
\end{eqnarray}
where $s_{ij}\equiv\sin\theta_{ij}$, and $c_{ij}\equiv\cos\theta_{ij}$.
Without loss of generality, we can adopt the convention 
$0\le\theta_{ij}\le \pi/2$, $0\le\delta<2\pi$ \cite{HagiwaraOkamura}.
Of the six parameters in this expression and the three masses, 
which add up to a total of nine parameters,
neutrino$\rightarrow$neutrino oscillations are only sensitive to six:
\begin{itemize}
\item the three mixing angles: $\theta_{12}$, $\theta_{23}$, $\theta_{13}$,
\item two mass-squared differences: $\delta m^2_{21}$, $\delta m^2_{31}$,
where $\delta m^2_{ij} = m^2_i - m^2_j$, and
\item the CP-violating phase: $\delta$.
\end{itemize}
The Majorana phases, $\alpha_1$ and $\alpha_2$, only appear
in lepton-number violating processes such as neutrinoless double beta decay, 
and cannot be determined via neutrino$\rightarrow$neutrino oscillations.
The absolute scale of the neutrino masses also remain undetermined since
neutrino oscillation is an interference effect.

%%%%%%%%%%%%%%%%%%%%%%%%%%%%%%%%%%%%%%%%%%%%%%%%%%%%%%%%%%%%%
\subsection{Oscillations}

If a neutrino of flavor $\alpha$ is created at $x=0$ with energy $E$, then
the state of the neutrino at $x=0$ is
\begin{equation}
\ket{\nu_{\alpha,0}(x=0)} 
= \ket{\nu_\alpha}
= \sum_{i=1}^3 (V_{\mathrm{MNS}})^*_{\alpha i} \ket{\nu_i}\;.
\end{equation}
At $x=L$, the same state is
\begin{equation}
\ket{\nu_{\alpha,0}(x=L)}
= \sum_{i=1}^3 e^{i p_i L}\, (V_{\mathrm{MNS}})^*_{\alpha i} \ket{\nu_i} \;.
\end{equation}
Using
\begin{equation}
p_i = \sqrt{E^2-m_i^2} = E - \frac{m_i^2}{2E} + \cdots\;,
\end{equation}
we find
\begin{equation}
\ket{\nu_{\alpha,0}(x=L)}
\;=\; e^{iEL}\sum_{i=1}^3 
\exp\left(-i\dfrac{m_i^2}{2E}L\right)\,
(V_{\mathrm{MNS}})^*_{\alpha i}\ket{\nu_i}\;.
\end{equation}
Therefore, the amplitude of observing the neutrino of flavor $\beta$ at
$x=L$ is given by (dropping the irrelevant overall phase)
\begin{eqnarray}
\lefteqn{\braket{\nu_\beta}{\nu_{\alpha,0}(x=L)}} \cr
& = & \left[\sum_{j=1}^3\bra{\nu_j}(V_{\mathrm{MNS}})_{\beta j} \right]
\left[\sum_{i=1}^3 
\exp\left(-i\dfrac{m_i^2}{2E}L\right)\,
(V_{\mathrm{MNS}})^*_{\alpha i}\ket{\nu_i}\right] \cr
& = & \sum_{j=1}^3
(V_{\mathrm{MNS}})_{\beta j}\exp\left(-i\dfrac{m_j^2}{2E}L\right)\,
(V_{\mathrm{MNS}})^*_{\alpha j}\;,
\end{eqnarray}
and the probability of oscillation from $\ket{\nu_\alpha}$ to
$\ket{\nu_\beta}$ with neutrino energy $E$ and baseline $L$ is given by
\begin{eqnarray}
P(\nu_\alpha\rightarrow\nu_\beta)
& = &
\left| \sum_{j=1}^3 (V_\mathrm{MNS})_{\beta j}
       \exp\left( -i\dfrac{m_j^2}{2E} L\right) (V_\mathrm{MNS})^*_{\alpha j}
\right|^2 \cr
& = & \delta_{\alpha\beta}
      -4\sum_{i>j}
       \Re(U^*_{\alpha i}U_{\beta i}U_{\alpha j}U^*_{\beta j})\sin^2\frac{\Delta_{ij}}{2} 
      +2\sum_{i>j}
       \Im(U^*_{\alpha i}U_{\beta i}U_{\alpha j}U^*_{\beta j})\sin\Delta_{ij} \;, \cr
& & \label{Palphatobeta}
\end{eqnarray}
where
\begin{equation}
\Delta_{ij} 
\;\equiv\; \dfrac{\delta m_{ij}^2}{2E} L
\;=\; 2.534\;\dfrac{\delta m_{ij}^2\,(\mathrm{eV}^2)}{E\,(\mathrm{GeV})}\,L\,(\mathrm{km})
\;,\qquad
\delta m_{ij}^2
\equiv m_i^2-m_j^2\;.
\end{equation}
Since
\begin{equation}
\Delta_{32} = \Delta_{31} - \Delta_{21}\;,
\end{equation}
only two of the three $\Delta_{ij}$'s in Eq.~(\ref{Palphatobeta}) are independent.  
Eliminating $\Delta_{32}$ from Eq.~(\ref{Palphatobeta}), we obtain
\begin{eqnarray}
P(\nu_\alpha\rightarrow\nu_\alpha)
& = & 1 - 4\, |U_{\alpha 2}|^2 \left( 1 - |U_{\alpha 2}|^2 \right)
            \sin^2\frac{\Delta_{21}}{2}
        - 4\, |U_{\alpha 3}|^2 \left( 1 - |U_{\alpha 3}|^2 \right)
            \sin^2\frac{\Delta_{31}}{2} \cr
& &  \phantom{1}
        + 2\, |U_{\alpha 2}|^2 |U_{\alpha 3}|^2
          \left( 4\sin^2\frac{\Delta_{21}}{2}\sin^2\frac{\Delta_{31}}{2}
                + \sin\Delta_{21}\sin\Delta_{31}
          \right) \;,
\label{Palphatoalpha}
\end{eqnarray}
for the $\alpha=\beta$ case, and
\begin{eqnarray}
P(\nu_\alpha \rightarrow \nu_\beta)
& = & 4\, |U_{\alpha 2}|^2 |U_{\beta 2}|^2 \sin^2\frac{\Delta_{21}}{2}
     +4\, |U_{\alpha 3}|^2 |U_{\beta 3}|^2 \sin^2\frac{\Delta_{31}}{2} \cr
&   & +2\,\Re( U^*_{\alpha 3}U_{\beta 3}U_{\alpha 2}U^*_{\beta 2})
      \left(4\sin^2\frac{\Delta_{21}}{2}\sin^2\frac{\Delta_{31}}{2}
           +\sin\Delta_{21}\sin\Delta_{31}
      \right) \cr 
&   & +4\,J_{(\alpha,\beta)}
      \left( \sin^2\frac{\Delta_{21}}{2}\sin\Delta_{31}
            -\sin^2\frac{\Delta_{31}}{2}\sin\Delta_{21}
      \right) \;,    
\label{Palphatonotalpha}       
\end{eqnarray}
for the $\alpha\neq\beta$ case, 
where $J_{(\alpha,\beta)}$ is the Jarskog invariant:
\begin{eqnarray}
J_{(\alpha,\beta)}
& = & +\Im(U^*_{\alpha 1}U_{\beta 1}U_{\alpha 2}U^*_{\beta 2})
\;=\; +\Im(U^*_{\alpha 2}U_{\beta 2}U_{\alpha 3}U^*_{\beta 3})
\;=\; +\Im(U^*_{\alpha 3}U_{\beta 3}U_{\alpha 1}U^*_{\beta 1}) \cr
& = & -\Im(U^*_{\alpha 2}U_{\beta 2}U_{\alpha 1}U^*_{\beta 1})
\;=\; -\Im(U^*_{\alpha 1}U_{\beta 1}U_{\alpha 3}U^*_{\beta 3})
\;=\; -\Im(U^*_{\alpha 3}U_{\beta 3}U_{\alpha 2}U^*_{\beta 2}) \cr
& = & -J_{(\beta,\alpha)}\;.
\end{eqnarray}
In the parametrization given in Eq.~(\ref{UPparam}), we have
\begin{equation}
J_{(\mu,e)} = -J_{(e,\mu)} =
J_{(e,\tau)} = -J_{(\tau,e)} =
J_{(\tau,\mu)} = -J_{(\mu,\tau)} = A\sin\delta\;, 
\end{equation}
with
\begin{equation}
A = s_{12}c_{12}s_{13}c_{13}^2 s_{23}c_{23}\;.
\end{equation}
The oscillation probabilities for the anti-neutrinos are obtained
by replacing $U_{\alpha i}$ with its complex conjugate, which only amounts to
flipping the sign of $\delta$ in the parametrization of Eq.~(\ref{UPparam}).
It is clear from Eq.~(\ref{Palphatoalpha}) that
$P(\bar{\nu}_\alpha\rightarrow\bar{\nu}_\alpha) = P(\nu_\alpha\rightarrow\nu_\alpha)$,
which is to be expected from the CPT theorem.
For flavor changing oscillations, only the Jarskog term in
Eq.~(\ref{Palphatonotalpha}) changes sign.

%%%%%%%%%%%%%%%%%%%%%%%%%%%%%%%%%%%%%%%%%%%%%%%%%%%%%%%%%%%%%
\subsection{Current Experimental Bounds}

When $|\Delta_{21}|\ll|\Delta_{31}|=O(1)$, the expressions given above 
can be expanded in $\Delta_{21}$ to yield
\begin{eqnarray}
\lefteqn{P(\nu_\alpha \rightarrow \nu_\alpha)} \cr
& = & 1 
- 4\;|U_{\alpha 3}|^2 
\left( 1 - |U_{\alpha 3}|^2 \right) \sin^2\dfrac{\Delta_{31}}{2} 
 +\left( 2\;|U_{\alpha 2}|^2 |U_{\alpha 3}|^2 \sin\Delta_{31}\right) \Delta_{21} 
\cr 
& & \phantom{1}
+ |U_{\alpha 2}|^2 \left\{ 2|U_{\alpha 3}|^2 \sin^2\frac{\Delta_{31}}{2}
                          -\left( 1 - |U_{\alpha 2}|^2 \right)
                   \right\} \Delta_{21}^2
- \left(\frac{1}{3}\,|U_{\alpha 2}|^2 |U_{\alpha 3}|^2 \sin\Delta_{31}\right) \Delta_{21}^3
+ O(\Delta_{21}^4) \cr
& = & 1
-4\;|U_{\alpha 3}|^2 \left( 1 - |U_{\alpha 3}|^2 \right)
\sin^2\left( \frac{ \Delta_{31} - \kappa_{\alpha\alpha}\Delta_{21} }{ 2 }
      \right)
\cr
& & \phantom{1}
-|U_{\alpha 1}|^2 |U_{\alpha 2}|^2 
 \left( 1 + \dfrac{|U_{\alpha 3}|^2}{1-|U_{\alpha 3}|^2}\cos\Delta_{31} \right) \Delta_{21}^2
\cr
& & \phantom{1}
-|U_{\alpha 1}|^2 |U_{\alpha 2}|^2 |U_{\alpha 3}|^2
     \left\{\dfrac{ 1 + |U_{\alpha 2}|^2 - |U_{\alpha 3}|^2 }{ 3(1-|U_{\alpha 3}|^2)^2 }
            \sin\Delta_{31}
     \right\} \Delta_{21}^3    
+ O(\Delta_{21}^4) \;, \\
\label{PalphatoalphaExpand}
%%%%%%%%%
& & \cr
\lefteqn{P(\nu_\alpha \rightarrow \nu_\beta)} \cr
& = &
4\;|U_{\alpha 3}|^2 |U_{\beta 3}|^2 \sin^2\dfrac{\Delta_{31}}{2} \cr
& &
+\left\{ 2\,\Re\left( U^*_{\alpha 3}U_{\beta 3}U_{\alpha 2}U^*_{\beta 2}\right)\sin\Delta_{31}
        -4\,J_{(\alpha,\beta)}\sin^2\dfrac{\Delta_{31}}{2}
 \right\}\Delta_{21} 
\cr
& & 
+\left\{ |U_{\alpha 2}|^2 |U_{\beta 2}|^2
        +2\Re\left( U^*_{\alpha 3}U_{\beta 3}U_{\alpha 2}U^*_{\beta 2}\right)\sin^2\dfrac{\Delta_{31}}{2} 
        +J_{(\alpha,\beta)}\sin\Delta_{31}
 \right\} \Delta_{21}^2
\cr
& &
+ \frac{1}{3}\left\{ 2 J_{(\alpha,\beta)}\sin^2\dfrac{\Delta_{31}}{2}
                   - \Re\left( U^*_{\alpha 3}U_{\beta 3}U_{\alpha 2}U^*_{\beta 2}\right)\sin\Delta_{31}
             \right\} \Delta_{21}^3
+ O(\Delta_{21}^4) \cr
& = &
4\left( |U_{\alpha 3}|^2 |U_{\beta 3}|^2 - J_{(\alpha,\beta)}\Delta_{21} \right)
\sin^2
\left(\frac{\Delta_{31} - \kappa_{\alpha\beta}\Delta_{21}}{2}
\right)
\cr
& &
+ \left\{ \dfrac{J_{(\alpha,\beta)}^2}{|U_{\alpha 3}|^2 |U_{\beta 3}|^2}
        - 2|U_{\alpha 3}|^2 |U_{\beta 3}|^2 \kappa_{\alpha\beta}
          \left( 1 - \kappa_{\alpha\beta} \right) \sin^2\frac{\Delta_{31}}{2}
        - J_{(\alpha,\beta)}
          \left( 1 - 2\kappa_{\alpha\beta} \right) \sin\Delta_{31}
  \right\} \Delta_{21}^2
\cr
& & 
- \frac{1}{3}
  \left\{ 3 J_{(\alpha,\beta)}\kappa_{\alpha\beta}^2 
        + 2 J_{(\alpha,\beta)}(1-3\kappa_{\alpha\beta}^2)\sin^2\frac{\Delta_{31}}{2}
        - |U_{\alpha 3}|^2 |U_{\beta 3}|^2 \kappa_{\alpha\beta} (1-\kappa_{\alpha\beta}^2)\sin\Delta_{31}
  \right\} \Delta_{21}^3 \cr
& & +O(\Delta_{21}^4) \;,
\label{PalphatobetaExpand}
\end{eqnarray} 
where
\begin{equation}
\kappa_{\alpha\alpha} \equiv \dfrac{ |U_{\alpha 2}|^2 }{ 1 - |U_{\alpha 3}|^2 } 
\;,\qquad
\kappa_{\alpha\beta} \equiv
       - \dfrac{\Re\left( U^*_{\alpha 3}U_{\beta 3}U_{\alpha 2}U^*_{\beta 2}\right)}
               {|U_{\alpha 3}|^2 |U_{\beta 3}|^2}
\;.
\end{equation}
Neglecting terms of order $O(\Delta_{21}^2)$ and higher, we obtain
the following simplified expressions for a few specific processes:
\begin{eqnarray}
P(\nu_e \rightarrow \nu_e)
& = & 1 - 4 s_{13}^2 (1-s_{13}^2)
      \sin^2\left(\frac{\Delta_{31} - s_{12}^2\Delta_{21}}{2}
            \right) \cr
& = & 1 - \sin^2(2\theta_\mathrm{rct})
          \sin^2\left(\frac{\Delta_{31} - s_{12}^2\Delta_{21}}{2}
                \right) \;,\cr       
P(\nu_\mu\rightarrow \nu_\mu)
& = & 1 - 4c_{13}^2 s_{23}^2(1-c_{13}^2 s_{23}^2)
      \sin^2\left(\frac{\Delta_{31} - \kappa_{\mu\mu}\Delta_{21}}{2}
            \right) \cr
& = & 1 - \sin^2(2\theta_\mathrm{atm})
      \sin^2\left(\frac{\Delta_{31} - \kappa_{\mu\mu}\Delta_{21}}{2}
            \right) \;,\cr
P(\nu_\mu\rightarrow \nu_e)
& = & 4 (c_{13}^2 s_{13}^2 s_{23}^2 - J_{(\mu,e)}\Delta_{21})
      \sin^2\left(\frac{\Delta_{31} - \kappa_{\mu e}\Delta_{21}}{2}
            \right) \cr
& = & 4 (\sin^2\theta_\mathrm{rct}\,\sin^2\theta_\mathrm{atm}
         - A \sin\delta \,\Delta_{21})
      \sin^2\left(\frac{\Delta_{31} - \kappa_{\mu e}\Delta_{21}}{2}
            \right) \;,
\label{Probabilities}
\end{eqnarray}
where 
\begin{eqnarray}
A 
%& = & s_{12}c_{12}s_{13}c_{13}^2 s_{23}c_{23} \cr
& = & \frac{1}{8}\sin(2\theta_{12})
\sin(2\theta_\mathrm{rct})
\sin(2\theta_\mathrm{atm})
\sqrt{1-\tan^2\theta_\mathrm{rct}\tan^2\theta_\mathrm{atm}}
\;, \cr
\kappa_{\mu\mu}
%& = & \frac{(c_{12}^2 c_{23}^2 + s_{12}^2 s_{13}^2 s_{23}^2)
%           - 2c_{12}c_{23}s_{12}s_{13}s_{23}\cos\delta}
%           {1-c_{13}^2 s_{23}^2} \cr
& = & c_{12}^2 - (c_{12}^2-s_{12}^2)
      \tan^2\theta_\mathrm{rct}\tan^2\theta_\mathrm{atm}
     -\left(\frac{2A}{\cos^2\theta_{\mathrm{rct}}\cos^2\theta_{\mathrm{atm}}}\right)
      \cos\delta \;,\cr
\kappa_{\mu e}
& = & s_{12}^2 
- \left(\frac{A}{\sin^2\theta_\mathrm{rct}\sin^2\theta_{atm}}\right)\cos\delta \;.
%J_{(\mu,e)} & = & A\sin\delta\;, \cr
\label{VacuumParams}
\end{eqnarray}
%
%For anti-neutrinos, 
%$P(\bar{\nu}_\alpha\rightarrow\bar{\nu}_\alpha) = 
%P(\nu_\alpha\rightarrow\nu_\alpha)$ from the CPT theorem.
%$P(\bar{\nu}_\alpha\rightarrow\bar{\nu}_\beta)$ $(\alpha\neq\beta)$
%can be obtained from $P(\nu_\alpha\rightarrow\nu_\beta)$ by
%flipping the sign of $\delta$.
%
%In Eq.~(\ref{Probabilities}),
We have made the identifications
\begin{eqnarray}
\sin\theta_\mathrm{atm} & = & s_{23}c_{13} \;=\; \sin\theta_{23}\cos\theta_{13}\;,\cr
\sin\theta_\mathrm{rct} & = & s_{13} \;=\; \sin\theta_{13}\;,
\label{Identification}
\end{eqnarray}
where $\theta_\mathrm{atm}$ and $\theta_\mathrm{rct}$ are the
mixing angles extracted from atmospheric \cite{atmos} and reactor \cite{chooz} 
neutrino oscillation experiments, respectively, based on two-flavor oscillation analyses.
The current experimental bounds from atmospheric neutrinos
at the 90\% confidence level are
\begin{eqnarray}
|\delta m^2_{31}|
& = & (1.5\sim 3.4)\times 10^{-3}\;\mathrm{eV}^2\;, \cr
\sin^2(2\theta_\mathrm{atm}) & > & 0.92\;.
\label{atmdata}
\end{eqnarray}
Only the absolute value of $\delta m^2_{31}$ is known since 
$P(\nu_\mu\rightarrow\nu_\mu) = 1 -\sin^2(2\theta_\mathrm{atm})\sin^2(\Delta_{31}/2)$
at leading order in $\Delta_{21}$.
The 90\% confidence limit on $\theta_\mathrm{rct} = \theta_{13}$ 
from the CHOOZ experiment, which measured 
$P(\bar{\nu}_e\rightarrow\bar{\nu}_e) = 1 -\sin^2(2\theta_\mathrm{rct})\sin^2(\Delta_{31}/2)$, 
depends on the not-yet-well-known value of $|\delta m^2_{31}|$:
\begin{eqnarray}
\sin^2(2\theta_{rct}) & < 0.20 \;\qquad\mbox{for $|\delta m^2_{31}| = 2.0\times 10^{-3}\;\mathrm{eV}^2$} \;, \cr
\sin^2(2\theta_{rct}) & < 0.16 \;\qquad\mbox{for $|\delta m^2_{31}| = 2.5\times 10^{-3}\;\mathrm{eV}^2$} \;, \cr
\sin^2(2\theta_{rct}) & < 0.14 \;\qquad\mbox{for $|\delta m^2_{31}| = 3.0\times 10^{-3}\;\mathrm{eV}^2$} \;.
\label{reactordata} 
\end{eqnarray}
When $|\Delta_{31}|\gg|\Delta_{21}|=O(1)$, $P(\nu_e\rightarrow\nu_e)$ 
simplifies to
\begin{eqnarray}
P(\nu_e\rightarrow \nu_e)
%& = & 1
%- 4\left( |U_{e 2}|^2 |U_{e 1}|^2
%        + |U_{e 2}|^2 |U_{e 3}|^2
%   \right)\sin^2\dfrac{\Delta_{21}}{2} \cr
%& & 
%- 2\left( |U_{e 3}|^2 |U_{e 1}|^2
%        + |U_{e 3}|^2 |U_{e 2}|^2
%   \right) 
%+ 4\; |U_{e 2}|^2 |U_{e 3}|^2 \sin^2\dfrac{\Delta_{21}}{2} \cr
& = & 1 - 2|U_{e3}|^2\left(1-|U_{e3}|^2\right)
- 4 |U_{e1}|^2 |U_{e2}|^2 \sin^2\frac{\Delta_{21}}{2} \cr
& = & 1 - 2 s_{13}^2 (1-s_{13}^2)
- 4c_{12}^2s_{12}^2c_{13}^4\sin^2\frac{\Delta_{21}}{2} \cr
%& = & 1 - \frac{1}{2}\sin^2(2\theta_{13})
%- c_{13}^4 \sin^2(2\theta_{12}) \sin^2\frac{\Delta_{21}}{2} \;. 
& = & 1 - \sin^2(2\theta_{12}) \sin^2\frac{\Delta_{21}}{2} + O(s_{13}^2) \;.
%& = & 1 - \sin^2(2\theta_\mathrm{sol})\sin^2\frac{\Delta_{21}}{2} \;.
\end{eqnarray}
%
%Solar neutrino experiments fit their data to the 2 flavor oscillation formula 
%$P(\nu_e\rightarrow \nu_e) = 1 - \sin^2(2\theta_\mathrm{sol})\sin^2(\Delta_{21}/{2})$,
Since $s_{13}$ is known to be small from CHOOZ,
we can identify $\theta_{12}$ with the 2-flavor solar mixing angle 
$\theta_\mathrm{sol}$. 
The current experimental bounds at the 90\% confidence level from
solar neutrinos are \cite{solar,KamLAND}
\begin{eqnarray}
\delta m^2_{21} & = & 8.2^{+0.6}_{-0.5}\times 10^{-5}\,\mathrm{eV}^2\;,\cr
\tan^2\theta_\mathrm{sol} & = & 0.40^{+0.09}_{-0.07}\;.
\label{solardata}
\end{eqnarray}
The sign of $\delta m^2_{21} = m_2^2-m_1^2$ is known to be positive since
$m_2^2 > m_1^2$ is required for the MSW effect \cite{MSW} to work.

%%%%%%%%%%%%%%%%%%%%%%%%%%%%%%%%%%%%%%%%%%%%%%%%%%%%%%%%%%%%%
\subsection{The Sizes of $\theta_{13}$, $\theta_{12}$, and $\theta_{23}$}

\begin{table}
\begin{tabular}{|c||c|c|}
\hline
\;\;$|\delta m^2_{31}|$ ($\mathrm{eV}^2$)\;\;
& \;\;$\varepsilon=\sqrt{\dfrac{\delta m^2_{21}}{|\delta m^2_{31}|}}$\;\; 
& \;\;Upper bound on $\theta_{13}$\;\; \\
\hline\hline
$2.0\times 10^{-3}$ & \;\;$0.20\sim 0.21$\;\;  & $0.23$ \\
$2.5\times 10^{-3}$ & \;\;$0.18\sim 0.19$\;\;  & $0.21$ \\
$3.0\times 10^{-3}$ & \;\;$0.16\sim 0.17$\;\;  & $0.19$ \\
\hline
\end{tabular}
\caption{Comparison of the size of the ratio $\delta m^2_{21}/|\delta m^2_{31}|$
and the 90\% confidence limit on $\theta_{13}$.
}
\end{table}

If we allow both $\delta m^2_{21}$ and $|\delta m^2_{31}|$ to
move within their respective 90\% confidence limits given in Eqs.~(\ref{atmdata})
and (\ref{solardata}), the ratio of the two is in the range
\begin{equation}
\dfrac{\delta m^2_{21}}{|\delta m^2_{31}|} = 0.023\sim 0.059\;.
\end{equation}
Thus, the approximation
$\Delta_{21}\ll|\Delta_{31}|$ that was used above is justified.
The square-root of this ratio, which we will call $\varepsilon$, 
is in the range
\begin{equation}
\varepsilon \equiv \sqrt{\dfrac{\delta m^2_{21}}{|\delta m^2_{31}|}} = 0.15\sim 0.24\;.
\end{equation}
Note that $\varepsilon$ is roughly in the same range as that of the
90\% upper bound on $\theta_{13}$ which
can be obtained from Eq.~(\ref{reactordata}).
In fact, the two numbers 
are positively correlated as shown in TABLE~I. 

While we currently have no experimental lower bound on $\theta_{13}$,
many analyses of LBL experiments only consider cases where
$\sin^2(2\theta_{13}) > 0.04$, which corresponds to $\theta_{13}>0.1$,
since smaller values of $\theta_{13}$ would put
$P(\nu_\mu\rightarrow \nu_e)$ below detectable range. 
Therefore, we will assume that $\theta_{13}$ is of order $\varepsilon$
in the following.  
However, the formulae we derive below can be applied 
as is to cases in which $\theta_{13}$ is smaller, since we only 
use the size of $\theta_{13}$ to decide when terms containing 
$s_{13}$ can be neglected.  A smaller $\theta_{13}$ will simply
make those terms even more negligible.

%%%%%%%%%%%%%%%%%%%%%%%%%%%%
%\subsection{The Size of $\theta_{12}$}

The 90\% confidence limits on the solar mixing angle given in Eq.~(\ref{solardata})
translates into
\begin{equation}
\theta_{12} = 0.56^{+0.05}_{-0.04} = (0.18\pm 0.01)\pi \;,
\end{equation}
and
\begin{eqnarray}
s_{12} & = & 0.50 \sim 0.57\;, \cr
c_{12} & = & 0.82 \sim 0.87\;, \cr
\sin(2\theta_{12}) & = & 0.86 \sim 0.94\;, \cr
\cos(2\theta_{12}) & = & 0.34 \sim 0.50\;.
\end{eqnarray}
Being sines and cosines, the upper end of these ranges are always smaller than one. 
However, $s_{12}$, $c_{12}$, and $\sin(2\theta_{12})$ are still much larger than 
$\varepsilon$ so we will treat them, and also $\theta_{12}$, as numbers of order 1.
On the other hand, $\cos(2\theta_{12})$ is only slightly larger than $\varepsilon$.
Its range is roughly equal to that of $2\varepsilon$.
Therefore, we can consider $\cos(2\theta_{12})/2$ as a number of order $\varepsilon$.
%Therefore, we will treat it as a quantity of order $\varepsilon$ with a
%caveat that it could be slightly larger.

%%%%%%%%%%%%%%%%%%%%%%%%%%%%
%\subsection{The Size of $\theta_{23}$}

The 90\% confidence limits on the atmospheric mixing angle given in
Eq.~(\ref{atmdata}) translates to
\begin{equation}
\theta_\mathrm{atm} 
= 0.64 \sim 0.93
= (0.25\pm 0.05)\pi \;.
\end{equation}
Though we made the identification $\sin\theta_\mathrm{atm} = s_{23}c_{13}$ 
in Eq.~(\ref{Identification}), the limits on $\theta_{23}$ are virtually
identical to those of $\theta_\mathrm{atm}$ due to the smallness of 
$\theta_{13}$.  We can therefore assume
\begin{eqnarray}
s_{23} & = & 0.60\sim 0.80\;,\cr
c_{23} & = & 0.60\sim 0.80\;,\cr
\sin(2\theta_{23}) & > & 0.96\;,\cr 
|\cos(2\theta_{23})| & < & 0.28\;.  
\end{eqnarray}
As in the case of $\theta_{12}$, we can assume
$s_{23}$, $c_{23}$, $\sin(2\theta_{23})$, and $\theta_{23}$ to be
numbers of order 1, while $\cos(2\theta_{23})$ is of order $\varepsilon$ or
smaller.

%%%%%%%%%%%%%%%%%%%%%%%%%%%%%%%%%%%%%%%%%%%%%%%%%%%%%%%%%%%%%
\section{Matter Effects}

%%%%%%%%%%%%%%%%%%%%%%%%%%%%%%%%%%%%%%%%%%%%%%%%%%%%%%%%%%%%%
\subsection{Diagonalization of the Effecive Neutrino Hamiltonian}

If the matter density along the baseline is constant, matter effects 
on neutrino oscillations can be taken into account by simply 
replacing the MNS matrix elements and mass-squared 
differences with their ``effective'' values in matter:
\begin{equation}
\Delta_{ij}\rightarrow\tilde{\Delta}_{ij}\;,\quad
U_{\alpha i}\rightarrow \tilde{U}_{\alpha i}\;,
\end{equation}
where $\tilde{U}$ is the unitary matrix that diagonalizes the modified
Hamiltonian,
\begin{equation}
H = 
\tilde{U}
\left[ \begin{array}{ccc} \lambda_1 & 0 & 0 \\
                          0 & \lambda_2 & 0 \\
                          0 & 0 & \lambda_3
       \end{array}
\right]
\tilde{U}^\dagger
= U
\left[ \begin{array}{ccc} 0 & 0 & 0 \\
                          0 & \delta m^2_{21} & 0 \\
                          0 & 0 & \delta m^2_{31}
       \end{array}
\right]
U^\dagger +
\left[ \begin{array}{ccc} a & 0 & 0 \\
                          0 & 0 & 0 \\
                          0 & 0 & 0 
       \end{array}
\right] \;,
\end{equation}
and 
\begin{equation}
\tilde{\Delta}_{ij}
= \dfrac{\delta\lambda_{ij}}{2E}\,L\;,\qquad
\delta\lambda_{ij} = \lambda_i - \lambda_j\;.
\end{equation}
The factor $a$ is due to the interaction of the $\ket{\nu_e}$ component
of the neutrinos with the electrons in matter via $W$-exchange:
\begin{equation}
a = 2\sqrt{2}\,G_F N_e E
= 7.63\times 10^{-5}(\mathrm{eV})^2
\left(\dfrac{\rho}{\mathrm{g/cm^3}}\right)
\left(\dfrac{E}{\mathrm{GeV}}\right)\;.
\end{equation}
%
%For anti-neutrino beams, $a$ reverses sign.
Note that $a$ is $E$-dependent, which means that both $\tilde{U}$ and
$\tilde{\Delta}_{ij}$ are also $E$-dependent.
It is also assumed that $E\ll M_W$ since the $W$-exchange interaction is
approximated by a point-like four-fermion interaction in deriving this expression.

To see the effect of $a$, we introduce the matrix
\begin{equation}
\mathcal{Q} = \mathrm{diag}(1,1,e^{i\delta})\;,
\end{equation}
and start with the partially diagonalized Hamiltonian:
\begin{eqnarray}
H' & = & \mathcal{Q}^\dagger U^\dagger H U\mathcal{Q} \cr
& = & \mathcal{Q}^\dagger \left\{\;
\left[ \begin{array}{ccc} 0 & 0 & 0 \\
                          0 & \delta m^2_{21} & 0 \\
                          0 & 0 & \delta m^2_{31}
       \end{array}
\right] +
U^\dagger
\left[ \begin{array}{ccc} a & 0 & 0 \\
                          0 & 0 & 0 \\
                          0 & 0 & 0 
       \end{array}
\right] 
U \;\right\} \mathcal{Q} \cr
& = & \mathcal{Q}^\dagger
\left[ \begin{array}{ccc} 0 & 0 & 0 \\
                          0 & \delta m^2_{21} & 0 \\
                          0 & 0 & \delta m^2_{31}
       \end{array}
\right] \mathcal{Q}
+ a \,\mathcal{Q}^\dagger
\left[ \begin{array}{ccc} U^*_{e1}U_{e1} & U^*_{e1}U_{e2} & U^*_{e1}U_{e3} \\
                          U^*_{e2}U_{e1} & U^*_{e2}U_{e2} & U^*_{e2}U_{e3} \\
                          U^*_{e3}U_{e1} & U^*_{e3}U_{e2} & U^*_{e3}U_{e3} 
       \end{array}
\right] \mathcal{Q} \cr
& = &
\left[ \begin{array}{ccc} 0 & 0 & 0 \\
                          0 & \delta m^2_{21} & 0 \\
                          0 & 0 & \delta m^2_{31}
       \end{array}
\right] + a 
\left[ \begin{array}{ccc} 
        c_{12}^2 c_{13}^2    & c_{12}s_{12}c_{13}^2 & c_{12}c_{13}s_{13}  \\
        c_{12}s_{12}c_{13}^2 & s_{12}^2c_{13}^2     & s_{12}c_{13}s_{13}  \\
        c_{12}c_{13}s_{13}   & s_{12}c_{13}s_{13}   & s_{13}^2 
       \end{array}
\right] \cr
& = & 
\left[ 
\begin{array}{ccc} 
a c_{12}^2 c_{13}^2    & a c_{12}s_{12}c_{13}^2 & a c_{12}c_{13}s_{13} \\
a c_{12}s_{12}c_{13}^2 & a s_{12}^2c_{13}^2 + \delta m^2_{21} & a s_{12}c_{13}s_{13} \\
a c_{12}c_{13}s_{13}   & a s_{12}c_{13}s_{13}   & a s_{13}^2 + \delta m^2_{31} 
\end{array}
\right] \;.
\label{Hprime}
\end{eqnarray}
The matrix $\mathcal{Q}$ serves to rid $H'$ of any reference to the
CP violating phase $\delta$.
Our strategy is to approximately diagonalize $H'$ through the Jacobi method
using $\varepsilon=\sqrt{\delta m^2_{21}/|\delta m^2_{31}|}$ as the expansion parameter.
We assume $\theta_{13}= O(\varepsilon)$ as discussed above.
Corrections to the eigenvalues of $H'$ of order $\varepsilon^3 |\delta m^2_{31}|$ and higher,
and those to the elements of $\tilde{U}$ of order $\varepsilon^3$ and higher will be 
neglected. 
For $\varepsilon=0.15\sim 0.24$, we have $\varepsilon^3 = 0.0034 \sim 0.014$.

%%%%%%%%%%%%%%%%%%%%%%%%%%%%%%%%%%%%%%%%%%%%%
Recall that for $2\times 2$ real symmetric matrices, such as
\begin{equation}
M = \left[ \begin{array}{cc} \alpha & \beta \\ \beta & \gamma \end{array}\right]\;,\qquad
\alpha, \beta, \gamma \in\mathbb{R}\;,
\end{equation}
diagonalization is trivial. Just define
\begin{equation}
R = \left[ \begin{array}{rr} c_\omega & s_\omega \\
                            -s_\omega & c_\omega 
           \end{array}
    \right]\;,\quad
\mbox{where}\qquad
c_\omega = \cos\omega\;,\qquad
s_\omega = \sin\omega\;,\qquad
\tan 2\omega \equiv \dfrac{2\beta}{\gamma-\alpha}\;,
\end{equation}
and we obtain
\begin{equation}
R^\dagger M R
= \left[ \begin{array}{cc} \Lambda_1 & 0 \\ 0 & \Lambda_2 \end{array}\right]\;,
\end{equation} 
with
\begin{equation}
\Lambda_1 = \dfrac{\alpha c^2_\omega - \gamma s^2_\omega}{c^2_\omega-s^2_\omega}\;,\qquad
\Lambda_2 = \dfrac{\gamma c^2_\omega - \alpha s^2_\omega}{c^2_\omega-s^2_\omega}\;.
\end{equation}
The Jacobi method entails iteratively diagonalizing $2\times 2$ submatrices
of a larger matrix in the order that requires the largest rotation angle at each step.
In the case of $H'$, as we will show below, two iterations at most are sufficient to
achieve approximate diagonalization to the order required,
regardless of the size of $a$.

In the following, we will consider the five cases
$a/|\delta m^2_{31}| = O(\varepsilon^3)$, $O(\varepsilon^2)$, $O(\varepsilon)$,
$O(1)$, and $O(\varepsilon^{-1})$ separately.  
Values of $a/|\delta m^2_{31}|$ outside this range can be treated in a similar manner as
the $O(\varepsilon^3)$ and $O(\varepsilon^{-1})$ cases. 
(For conditions on the Earth, however,
$a/|\delta m^2_{31}| \ge O(\varepsilon^{-2})$ implies $E > M_W$, invalidating
our approximation.)

%%%%%%%%%%%%%%%%%%%%%%%%%%%%%%%%%%%%%
%%%%%%%%%%%%%%%%%%%%%%%%%%%%%%%%%%%%%
\renewcommand{\thesubsubsection}{\thesubsection\arabic{subsubsection}}
\subsubsection{$a/|\delta m^2_{31}|=O(\varepsilon^3)$ case}

We begin by considering the case $a/|\delta m^2_{31}| = O(\varepsilon^3)$.
Treating both $c_{12}$ and $s_{12}$ as numbers of order $1$, and
$s_{13}$ as a number of order $\varepsilon$,
the relative sizes of the elements of $H'$ are given by
\begin{equation}
H' = 
\left[ 
\begin{array}{ccc} 
a c_{12}^2 c_{13}^2    & a c_{12}s_{12}c_{13}^2 & a c_{12}c_{13}s_{13} \\
a c_{12}s_{12}c_{13}^2 & a s_{12}^2c_{13}^2 + \delta m^2_{21} & a s_{12}c_{13}s_{13} \\
a c_{12}c_{13}s_{13}   & a s_{12}c_{13}s_{13}   & a s_{13}^2 + \delta m^2_{31} 
\end{array}
\right] =
|\delta m^2_{31}|
\left[ 
\begin{array}{ccc}
O(\varepsilon^{3}) & O(\varepsilon^{3}) & O(\varepsilon^{4}) \\
O(\varepsilon^{3}) & O(\varepsilon^{2}) & O(\varepsilon^{4}) \\
O(\varepsilon^{4}) & O(\varepsilon^{4}) & O(1) \\
\end{array}
\right]
\;.
\end{equation}
Of the three $2\times 2$ submatrices of $H'$,
the one which is farthest from diagonal is the $(1,2)$ submatrix since
it requires an $O(\varepsilon)$ rotation to diagonalize, 
while the other two submatrices only require $O(\varepsilon^4)$ rotations.
To diagonalize the $(1,2)$ submatrix of $H'$, we define
\begin{equation}
V = 
\left[ \begin{array}{ccc} c_{\varphi} &  s_{\varphi} & 0 \\
	                     -s_{\varphi} &  c_{\varphi} & 0 \\
	                      0 & 0 & 1
	   \end{array}
\right]\;,
\label{Vdef}
\end{equation}
where
\begin{equation}
c_{\varphi} =\cos\varphi\;,\quad
s_{\varphi} =\sin\varphi\;,\quad
\tan 2\varphi \equiv 
\dfrac{a c_{13}^2\sin2\theta_{12}}{\delta m^2_{21}-a c_{13}^2\cos2\theta_{12}}\;,\quad
\left(0\le\varphi<\frac{\pi}{2}\right)\;.
\label{phi1def}
\end{equation}
Using $V$, we find
\begin{equation}
H'' 
= V^\dagger H' V 
=
\left[ \begin{array}{ccc}
       \lambda'_1 & 0 & a {c}_{12}' c_{13}s_{13} \\
       0 & \lambda'_2 & a {s}_{12}' c_{13}s_{13} \\
       a {c}_{12}' c_{13}s_{13} & a {s}_{12}' c_{13}s_{13} &
       a s_{13}^2 + \delta m^2_{31}
       \end{array}
\right] \;,
\label{Hdoubleprimedef}
\end{equation}
where
\begin{equation}
{c}_{12}' = \cos\theta_{12}' \;,\quad
{s}_{12}' = \sin\theta_{12}' \;,\quad
{\theta}_{12}' = \theta_{12} + \varphi \;,
\label{theta12primedef}
\end{equation}
and
\begin{eqnarray}
\lambda'_1 
& = & \dfrac{(a c_{12}^2 c_{13}^2) c^2_\varphi
           - (a s_{12}^2 c_{13}^2 + \delta m^2_{21}) s^2_\varphi}
            {c^2_\varphi - s^2_\varphi} 
\;=\; \lambda'_{-} \;,\cr
\lambda'_2 
& = & \dfrac{(a s_{12}^2 c_{13}^2 + \delta m^2_{21}) c^2_\varphi
           - (a c_{12}^2 c_{13}^2) s^2_\varphi}
            {c^2_\varphi - s^2_\varphi}
\;=\; \lambda'_{+} \;,
\label{lambdaprimesdef}
\end{eqnarray}
with
\begin{equation}
\lambda'_{\pm}
= \dfrac{ (a c_{13}^2+\delta m^2_{21})
          \pm\sqrt{ (a c_{13}^2-\delta m^2_{21})^2 + 4 a c_{13}^2 s_{12}^2 \delta m^2_{21} }
        }
        { 2 }\;.
\label{lambdaprimeplusminusdef}
\end{equation}
Note that
\begin{equation}
\tan 2\theta_{12}' 
= \dfrac{\tan 2\theta_{12}+\tan 2\varphi}
        {1 - \tan 2\theta_{12} \tan 2\varphi}
= \dfrac{\delta m^2_{21}\sin 2\theta_{12}}
        {\delta m^2_{21}\cos 2\theta_{12} - a c_{13}^2}\;,
\label{tan2theta12prime}
\end{equation}
which can be used to calculate $\theta'_{12}$ without calculating $\varphi$ first.

Since $a/|\delta m^2_{31}| = O(\varepsilon^3)$, which means
that $a/\delta m^2_{21} = O(\varepsilon)$, we can expand 
$\lambda_1'$, $\lambda_2'$, and $\varphi$
in powers of $a/\delta m^2_{21}$ and $s_{13} = O(\varepsilon)$.
We find
\begin{eqnarray}
\lambda'_1 & = & a c_{12}^2 + O(\varepsilon^4 |\delta m^2_{31}|) 
\;=\; O(\varepsilon^3 |\delta m^2_{31}|) \;, \cr
\lambda'_2 & = & \delta m^2_{21} + a s_{12}^2 + O(\varepsilon^4 |\delta m^2_{31}|)
\;=\; \delta m^2_{21} + O(\varepsilon^3 |\delta m^2_{31}|)
\;=\; O(\varepsilon^2 |\delta m^2_{31}|) \;,
\end{eqnarray}
and,
\begin{eqnarray}
\varphi      
& = & \frac{1}{2}\left(\frac{a c_{13}^2}{\delta m^2_{21}}\right)\sin(2\theta_{12})
+ \frac{1}{2}\left(\frac{a c_{13}^2}{\delta m^2_{21}}\right)^2\sin(2\theta_{12})\cos(2\theta_{12})
+ \cdots  \cr
& = & \frac{a}{2\,\delta m^2_{21}}\sin(2\theta_{12}) +  O(\varepsilon^3) 
\;=\; O(\varepsilon) \;,\cr
& & \cr
\theta_{12}' 
& = & \theta_{12} + \varphi 
\;=\; \theta_{12} + \frac{a}{2\,\delta m^2_{21}}\sin(2\theta_{12}) + O(\varepsilon^3) 
\;=\; O(1) \;.
\end{eqnarray}
Note that the second term in the expansion of $\varphi$ can be
considered to be of order $\varepsilon^3$ since $\cos(2\theta_{12})/2$ is of
order $\varepsilon$ as discussed in the previous section.
Therefore, the sizes of the elements of $H''$ are 
\begin{equation}
H'' = 
\left[ \begin{array}{ccc}
       \lambda'_1 & 0 & a {c}_{12}' c_{13}s_{13} \\
       0 & \lambda'_2 & a {s}_{12}' c_{13}s_{13} \\
       a {c}_{12}' c_{13}s_{13} & a {s}_{12}' c_{13}s_{13} &
       a s_{13}^2 + \delta m^2_{31}
       \end{array}
\right] =
|\delta m^2_{31}|
\left[ \begin{array}{ccc}
       O(\varepsilon^3) & 0                & O(\varepsilon^4) \\
       0                & O(\varepsilon^2) & O(\varepsilon^4) \\
       O(\varepsilon^4) & O(\varepsilon^4) & O(1)
       \end{array}
\right]\;.
\end{equation}
The rotation angles required to diagonalize the $(1,3)$ or $(2,3)$ 
submatrices are of order $\varepsilon^4$, which we will neglect.
So in this case, the Hamiltonian is approximately diagonalized by
one $(1,2)$ rotation of angle $O(\varepsilon)$, and the
eigenvalues are
\begin{eqnarray}
\lambda_1 & \approx & \lambda'_1 \;\approx\; a c_{12}^2\;,\cr
\lambda_2 & \approx & \lambda'_2 \;\approx\; \delta m^2_{21} + a s_{12}^2\;,\cr
\lambda_3 & \approx & \delta m^2_{31} \;.
\end{eqnarray}
We have kept terms up to $O(\varepsilon^3|\delta m^2_{31}|)$ here
to show how the eigenvalues are shifted away from their vacuum values.

%%%%%%%%%%%%%%%%%%%%%%%%%%%%%%%%%%%%%%%%%%%%%%%%%%%%%%%%%%%%%%%%%
\subsubsection{$a/|\delta m^2_{31}|=O(\varepsilon^2)$ case}

Next, we consider the case $a/|\delta m^2_{31}| = O(\varepsilon^2)$.
The relative sizes of the elements of $H'$ in this case are
\begin{equation}
H' = 
\left[ 
\begin{array}{ccc} 
a c_{12}^2 c_{13}^2    & a c_{12}s_{12}c_{13}^2 & a c_{12}c_{13}s_{13} \\
a c_{12}s_{12}c_{13}^2 & a s_{12}^2c_{13}^2 + \delta m^2_{21} & a s_{12}c_{13}s_{13} \\
a c_{12}c_{13}s_{13}   & a s_{12}c_{13}s_{13}   & a s_{13}^2 + \delta m^2_{31} 
\end{array}
\right] =
|\delta m^2_{31}|
\left[ 
\begin{array}{ccc}
O(\varepsilon^{2}) & O(\varepsilon^{2}) & O(\varepsilon^{3}) \\
O(\varepsilon^{2}) & O(\varepsilon^{2}) & O(\varepsilon^{3}) \\
O(\varepsilon^{3}) & O(\varepsilon^{3}) & O(1) \\
\end{array}
\right]
\;,
\end{equation}
and again we find that we must diagonalize the $(1,2)$ submatrix first.
We define $V$ and $\varphi$ as in Eqs.~(\ref{Vdef}) and (\ref{phi1def}), 
and the matrix will be partially diagonalized to
$H''$ in Eq.~(\ref{Hdoubleprimedef}), with $\lambda'_1$, $\lambda'_2$,
and $\theta'_{12}$ defined as in Eqs.~(\ref{theta12primedef}) and 
(\ref{lambdaprimesdef}).
Since $a$ and $\delta m^2_{21}$ are of the same order in this case,
both $\tan(2\varphi)$ of Eq.~(\ref{phi1def}) and $\tan(2\theta_{12}')$ of
Eq.~(\ref{tan2theta12prime}) can be expected to be large.
Therefore, both $s_{12}'$ and $c_{12}'$ can be considered to be numbers of order $1$ 
in this case also, and the relative sizes of the elements of $H''$ are
\begin{equation}
H'' = 
\left[ \begin{array}{ccc}
       \lambda'_1 & 0 & a {c}_{12}' c_{13}s_{13} \\
       0 & \lambda'_2 & a {s}_{12}' c_{13}s_{13} \\
       a {c}_{12}' c_{13}s_{13} & a {s}_{12}' c_{13}s_{13} &
       a s_{13}^2 + \delta m^2_{31}
       \end{array}
\right] =
|\delta m^2_{31}|
\left[ \begin{array}{ccc}
       O(\varepsilon^2) & 0                & O(\varepsilon^3) \\
       0                & O(\varepsilon^2) & O(\varepsilon^3) \\
       O(\varepsilon^3) & O(\varepsilon^3) & O(1)
       \end{array}
\right]\;.
\end{equation}
Further diagonalization requires rotations by angles of order $\varepsilon^3$,
which we will neglect. 

In this case,
we cannot expand $\varphi$, $\lambda'_1$, and $\lambda'_2$ in powers of 
$a/\delta m^2_{21}$ or its inverse.
However, we can still expand in $s_{13} = O(\varepsilon)$ and find
\begin{equation}
\lambda'_{\pm} =
\dfrac{ (a+\delta m^2_{21})\pm
 \sqrt{ (a-\delta m^2_{21})^2+4a\,\delta m^2_{21} s_{12}^2 } }
      { 2 }
+ O(\varepsilon^4 |\delta m^2_{31}|)\;,
\end{equation}
and
\begin{equation}
\varphi 
= \frac{1}{2}\tan^{-1}\left(\dfrac{a\sin 2\theta_{12}}{\delta m^2_{21} - a\cos 2\theta_{12}}\right)
- \dfrac{a\,\delta m^2_{21} s_{12} c_{12} s_{13}^2 }
        {(a-\delta m^2_{21})^2 + 4 a\,\delta m^2_{21} s_{12}^2} + O(\varepsilon^4)\;.
\end{equation}
The coefficient of $s_{13}^2$ in the second term is bounded from above by
\begin{equation}
\dfrac{a\,\delta m^2_{21} s_{12} c_{12}}
      {(a-\delta m^2_{21})^2 + 4 a\,\delta m^2_{21} s_{12}^2}
\le \dfrac{c_{12}}{4 s_{12}} 
= 0.38 \sim 0.42 \approx 2\varepsilon\;.
\end{equation}
Therefore, though this factor is formally of $O(1)$, we can consider it to
be a number of order $\varepsilon$ and approximate
\begin{equation}
\varphi \approx \frac{1}{2}\tan^{-1}
\left(\dfrac{a\sin 2\theta_{12}}{\delta m^2_{21} - a\cos 2\theta_{12}}\right)\;,\qquad
\theta'_{12} \approx \frac{1}{2}\tan^{-1}
\left(\dfrac{\delta m^2_{21}\sin 2\theta_{12}}{\delta m^2_{21}\cos 2\theta_{12} - a}\right)\;.
\end{equation}
The eigenvalues in this case are
\begin{eqnarray}
\lambda_1 & \approx & \lambda'_{-} \;\approx\; 
\dfrac{ (a+\delta m^2_{21})-
 \sqrt{ (a-\delta m^2_{21})^2+4a\,\delta m^2_{21} s_{12}^2 } }
      { 2 }\;,\cr
\lambda_2 & \approx & \lambda'_{+} \;\approx\;
\dfrac{ (a+\delta m^2_{21})+
 \sqrt{ (a-\delta m^2_{21})^2+4a\,\delta m^2_{21} s_{12}^2 } }
      { 2 }\;,\cr
\lambda_3 & \approx & \delta m^2_{31}\;.
\end{eqnarray}
%

%%%%%%%%%%%%%%%%%%%%%%%%%%%%%%%%%%%%%%%%%%%%%%%%%%%%%%%%%%%
\subsubsection{$a/|\delta m^2_{31}|=O(\varepsilon)$ case}

The relative sizes of the elements of $H'$ in this case are
\begin{equation}
H' = 
\left[ 
\begin{array}{ccc} 
a c_{12}^2 c_{13}^2    & a c_{12}s_{12}c_{13}^2 & a c_{12}c_{13}s_{13} \\
a c_{12}s_{12}c_{13}^2 & a s_{12}^2c_{13}^2 + \delta m^2_{21} & a s_{12}c_{13}s_{13} \\
a c_{12}c_{13}s_{13}   & a s_{12}c_{13}s_{13}   & a s_{13}^2 + \delta m^2_{31} 
\end{array}
\right] =
|\delta m^2_{31}|
\left[ 
\begin{array}{ccc}
O(\varepsilon    ) & O(\varepsilon    ) & O(\varepsilon^{2}) \\
O(\varepsilon    ) & O(\varepsilon    ) & O(\varepsilon^{2}) \\
O(\varepsilon^{2}) & O(\varepsilon^{2}) & O(1) \\
\end{array}
\right]
\;.
\end{equation}
As in the previous two cases, the $(1,2)$ submatrix is diagonalized by
$V$ defined as in Eq.~(\ref{Vdef}) with $\varphi$ defined as in (\ref{phi1def}). 
The partially diagonalized form is
$H''$ in Eq.~(\ref{Hdoubleprimedef}), with $\lambda'_1$, $\lambda'_2$,
and $\theta'_{12}$ defined as in Eqs.~(\ref{theta12primedef}) and 
(\ref{lambdaprimesdef}).

Since $\delta m^2_{21}/a = O(\varepsilon)$ in this case, we can expand 
$\lambda'_1$, $\lambda'_2$, and $\varphi$ in powers of
$\delta m^2_{21}/a$ and find
\begin{eqnarray}
\lambda'_1 
& = &
\delta m^2_{21} c^2_{12} + O(\varepsilon^3 \delta m^2_{31}) 
\;=\; O(\varepsilon^2 |\delta m^2_{31}|) \;, \cr
\lambda'_2 
& = &
a + \delta m^2_{21} s_{12}^2 + O(\varepsilon^3 \delta m^2_{31})
\;=\; O(\varepsilon\,|\delta m^2_{31}|) \;,
\label{lambdaprimeexpand1} 
\end{eqnarray}
and
\begin{eqnarray}
\varphi 
& = & \left(\frac{\pi}{2}-\theta_{12}\right)
    - \frac{1}{2}\left(\frac{\delta m^2_{21}}{a c_{13}^2}\right)\sin(2\theta_{12})
    - \frac{1}{2}\left(\frac{\delta m^2_{21}}{a c_{13}^2}\right)^2
                 \sin(2\theta_{12})\cos(2\theta_{12})
    + \cdots \;, \cr
& = & \left(\frac{\pi}{2}-\theta_{12}\right)
    - \frac{\delta m^2_{21}}{2a}\sin(2\theta_{12}) + O(\varepsilon^3) \;, \cr
& & \cr
\theta'_{12} 
& = & \theta_{12} + \varphi
\;=\; \frac{\pi}{2}
    - \frac{\delta m^2_{21}}{2a}\sin(2\theta_{12}) + O(\varepsilon^3) \;,
\label{phithetaexpand1}
\end{eqnarray}
which shows that
\begin{eqnarray}
c'_{12}
& \approx & \cos\left(\frac{\pi}{2} - \frac{\delta m^2_{21}}{2a}\sin(2\theta_{12})\right)
\;=\; \sin\left(\frac{\delta m^2_{21}}{2a}\sin(2\theta_{12})\right)
\;=\; O(\varepsilon) \;,\cr
s'_{12}
& \approx & \sin\left(\frac{\pi}{2} - \frac{\delta m^2_{21}}{2a}\sin(2\theta_{12})\right)
\;=\; \cos\left(\frac{\delta m^2_{21}}{2a}\sin(2\theta_{12})\right)
\;=\; O(1) \;.
\label{orderofs12prime}
\end{eqnarray}
Note that
\begin{equation}
a c'_{12}
\approx a\left(\frac{\delta m^2_{21}}{2a}\sin(2\theta_{12})\right)
= |\delta m^2_{31}|\left(\frac{\delta m^2_{21}}{2|\delta m^2_{31}|}\sin(2\theta_{12})\right)
= |\delta m^2_{31}| \,O(\varepsilon^2) \;,
\label{orderofac12prime}
\end{equation}
for all $a\gg\delta m^2_{21}$.
We will use this relation repeatedly in the following.

Thus, we find the sizes of the elements of $H''$ in this case to be
\begin{equation}
H'' = 
\left[ \begin{array}{ccc}
       \lambda'_1 & 0 & a {c}_{12}' c_{13}s_{13} \\
       0 & \lambda'_2 & a {s}_{12}' c_{13}s_{13} \\
       a {c}_{12}' c_{13}s_{13} & a {s}_{12}' c_{13}s_{13} &
       a s_{13}^2 + \delta m^2_{31}
       \end{array}
\right] =
|\delta m^2_{31}|
\left[ \begin{array}{ccc}
O(\varepsilon^2) & 0                & O(\varepsilon^3) \\
0                & O(\varepsilon)   & O(\varepsilon^2) \\
O(\varepsilon^3) & O(\varepsilon^2) & O(1)
\end{array}\right]\;.
\end{equation}
This time, the $(2,3)$ submatrix requires an $O(\varepsilon^2)$ rotation to be
diagonalized.
We define
\begin{equation}
W = 
\left[ \begin{array}{ccc} 1 & 0 & 0 \\
                          0 &  c_{\phi} &  s_{\phi} \\
	                      0 & -s_{\phi} &  c_{\phi} \\
	   \end{array}
\right]\;,
\label{Wdef}
\end{equation}
where
\begin{equation}
c_{\phi} = \cos\phi \;,\quad
s_{\phi} = \sin\phi \;,\quad
\tan 2\phi \equiv 
\dfrac{a s'_{12}\sin2\theta_{13}}
      {\delta m^2_{31}+a s_{13}^2 - \lambda'_2}
%\;\approx\;
%\dfrac{a\sin 2\theta_{13}}{\delta m^2_{31}}
\;.
\label{phi2def}
\end{equation}
The angle $\phi$ is in the first quadrant when $\delta m^2_{31} > 0$,
and in the fourth quadrant when $\delta m^2_{31} < 0$.
Then,
\begin{equation}
H'''
= W^\dagger H'' W 
=
\left[ \begin{array}{ccc}
\lambda'_1 & -a c'_{12}c_{13}s_{13}s_{\phi} & a c'_{12}c_{13}s_{13}c_{\phi} \\
-a c'_{12}c_{13}s_{13}s_{\phi} & \lambda''_2 & 0  \\
 a c'_{12}c_{13}s_{13}c_{\phi} & 0 & \lambda''_3
\end{array} \right] \;,
\label{Htripleprime}
\end{equation}
where
\begin{eqnarray}
\lambda''_2 & = &  
\dfrac{\lambda'_2 c^2_\phi - (a s^2_{13}+\delta m^2_{31})s^2_\phi}{c^2_\phi-s^2_\phi} \;,\cr 
\lambda''_3 & = &  
\dfrac{(a s^2_{13}+\delta m^2_{31})c^2_\phi - \lambda'_2 s^2_\phi}{c^2_\phi-s^2_\phi} \;.
\label{lambdadoubleprimes}
\end{eqnarray}
If we define
\begin{equation}
\lambda''_{\pm} \equiv
   \dfrac{ [ \lambda'_{2} + (a s_{13}^2+\delta m^2_{31}) ]
\pm \sqrt{ [ \lambda'_{2} - (a s_{13}^2+\delta m^2_{31}) ]^2 
         + 4 a^2 {s'_{12}}^2 c_{13}^2 s_{13}^2 }
         }
         { 2 } \;,
\label{lambdadoubleprimeplusminusdef}
\end{equation}
then 
\begin{eqnarray}
\lambda''_2 = \lambda''_{-}\;,\quad
\lambda''_3 = \lambda''_{+}\;,\qquad \mbox{if $\delta m^2_{31} > 0$}\;, \cr
\lambda''_2 = \lambda''_{+}\;,\quad
\lambda''_3 = \lambda''_{-}\;,\qquad \mbox{if $\delta m^2_{31} < 0$}\;.
\end{eqnarray}
Expanding $\lambda''_2$, $\lambda''_3$, and $\phi$ in powers of 
$a/\delta m^2_{31} = O(\varepsilon)$, we find
\begin{eqnarray}
\lambda''_2 & = & 
\lambda'_2 + O(\varepsilon^4 |\delta m^2_{31}|)
\;=\; a + \delta m^2_{21} s^2_{12} + O(\varepsilon^3 |\delta m^2_{31}|)
\;=\; O(\varepsilon\,|\delta m^2_{31}|) 
\;,\cr
\lambda''_3 & = & 
(\delta m^2_{31} + a s_{13}^2) + O(\varepsilon^4 |\delta m^2_{31}|)
\;=\; \delta m^2_{31} + O(\varepsilon^3 |\delta m^2_{31}|)
\;=\; O(|\delta m^2_{31}|) \;,
\end{eqnarray}
and
\begin{equation}
\phi 
\;=\; \frac{a}{2\,\delta m^2_{31}}\sin(2\theta_{13}) + \cdots
\;=\; \frac{a}{\delta m^2_{31}}\theta_{13} + O(\varepsilon^3)
\;=\; O(\varepsilon^2)\;,
\label{alphaminus1phi}
\end{equation}
which means that $s_\phi = O(\varepsilon^2)$, $c_\phi = O(1)$, and
\begin{equation}
H''' = 
\left[ \begin{array}{ccc}
\lambda'_1 & -a c'_{12}c_{13}s_{13}s_{\phi} & a c'_{12}c_{13}s_{13}c_{\phi} \\
-a c'_{12}c_{13}s_{13}s_{\phi} & \lambda''_2 & 0  \\
 a c'_{12}c_{13}s_{13}c_{\phi} & 0 & \lambda''_3
\end{array} \right] =
|\delta m^2_{31}|
\left[ \begin{array}{ccc}
O(\varepsilon^2) & O(\varepsilon^5) & O(\varepsilon^3) \\
O(\varepsilon^5) & O(\varepsilon)   & 0 \\
O(\varepsilon^3) & 0                & O(1)
\end{array} \right]
\;.
\end{equation}
Further diagonalization require rotations of order 
$O(\varepsilon^3)$ and higher, which we neglect.
So in this case, the Hamiltonian is approximately diagonalized by
a $(1,2)$ rotation of angle $O(1)$ followed by
a $(2,3)$ rotation of angle $O(\varepsilon^2)$, and the
eigenvalues are
\begin{eqnarray}
\lambda_1 & \approx & \lambda'_1 \;\approx\; \delta m^2_{21} c_{12}^2 \;,\cr
\lambda_2 & \approx & \lambda''_2 \;\approx\; a + \delta m^2_{21} s_{12}^2 \;,\cr
\lambda_3 & \approx & \lambda''_3 \;\approx\; \delta m^2_{31}\;.
\label{alphaminus1lambda}
\end{eqnarray}

%%%%%%%%%%%%%%%%%%%%%%%%%%%%%%%%%%%%%%%%%%%%%%%%%%%%
\subsubsection{$a/|\delta m^2_{31}| = O(1)$ case}\label{A4}

The relative sizes of the elements of $H'$ in this case are
\begin{equation}
H' = 
\left[ 
\begin{array}{ccc} 
a c_{12}^2 c_{13}^2    & a c_{12}s_{12}c_{13}^2 & a c_{12}c_{13}s_{13} \\
a c_{12}s_{12}c_{13}^2 & a s_{12}^2c_{13}^2 + \delta m^2_{21} & a s_{12}c_{13}s_{13} \\
a c_{12}c_{13}s_{13}   & a s_{12}c_{13}s_{13}   & a s_{13}^2 + \delta m^2_{31} 
\end{array}
\right] =
|\delta m^2_{31}|
\left[ 
\begin{array}{ccc}
O(1) & O(1) & O(\varepsilon) \\
O(1) & O(1) & O(\varepsilon) \\
O(\varepsilon) & O(\varepsilon) & O(1) \\
\end{array}
\right]
\;.
\end{equation}
As in the three previous cases, the $(1,2)$ submatrix is diagonalized by
$V$ defined as in Eq.~(\ref{Vdef}) with $\varphi$ defined as in (\ref{phi1def}). 
The partially diagonalized form is
$H''$ in Eq.~(\ref{Hdoubleprimedef}), with $\lambda'_1$, $\lambda'_2$,
and $\theta'_{12}$ defined as in Eqs.~(\ref{theta12primedef}) and 
(\ref{lambdaprimesdef}).
The expansions of $\lambda'_1$ and $\lambda'_2$ in powers of
$\delta m^2_{21}/a$ yield
\begin{eqnarray}
\lambda'_1 
& = &
\delta m^2_{21} c^2_{12} + O(\varepsilon^4 |\delta m^2_{31}|) 
\;=\; O(\varepsilon^2 |\delta m^2_{31}|) \;, \cr
\lambda'_2 
& = &
a c_{13}^2 + \delta m^2_{21} s_{12}^2 + O(\varepsilon^4 |\delta m^2_{31}|)
\;=\; O(|\delta m^2_{31}|) \;.
\label{lambdaprimeexpand0} 
\end{eqnarray}
Note that we cannot replace $c_{13}$ in this expression
with $1$ without introducing an error of order $\varepsilon^2 |\delta m^2_{31}|$
which is the same order as the second term.
The expansions of $\varphi$ and $\theta'_{12}$ are
\begin{eqnarray}
\varphi 
& = & \left(\frac{\pi}{2}-\theta_{12}\right)
    - \frac{\delta m^2_{21}}{2a}\sin(2\theta_{12}) + O(\varepsilon^4) \;, \cr
\theta'_{12} 
& = & \theta_{12} + \varphi
\;=\; \frac{\pi}{2}
    - \frac{\delta m^2_{21}}{2a}\sin(2\theta_{12}) + O(\varepsilon^4) \;.
\label{phithetaexpand0}
\end{eqnarray}
From Eqs.~(\ref{orderofs12prime}) and (\ref{orderofac12prime}), we can tell
that $s'_{12} = O(1)$, and $a c'_{12} = O(\varepsilon^2 |\delta m^2_{31}|)$. 
Therefore,
\begin{equation}
H'' = 
\left[ \begin{array}{ccc}
       \lambda'_1 & 0 & a {c}_{12}' c_{13}s_{13} \\
       0 & \lambda'_2 & a {s}_{12}' c_{13}s_{13} \\
       a {c}_{12}' c_{13}s_{13} & a {s}_{12}' c_{13}s_{13} &
       a s_{13}^2 + \delta m^2_{31}
       \end{array}
\right] =
|\delta m^2_{31}|
\left[ \begin{array}{ccc}
O(\varepsilon^2) & 0                & O(\varepsilon^3) \\
0                & O(1)             & O(\varepsilon)   \\
O(\varepsilon^3) & O(\varepsilon)   & O(1)
\end{array}\right]\;.
\end{equation}
Again, we need to diagonalize the $(2,3)$ submatrix with the matrix $W$ 
defined in Eq.~(\ref{Wdef}) with the angle $\phi$ defined in Eq.(\ref{phi2def}).
The resulting matrix is $H'''$ given in Eq.~(\ref{Htripleprime}).
From this point on, 
we must treat the $\delta m^2_{31}>0$ and $\delta m^2_{31}<0$
cases separately since level crossing between $a$ and $\delta m^2_{31}$ occurs
for the $\delta m^2_{31}>0$ case but not for the $\delta m^2_{31}<0$ case.

When $\delta m^2_{31}>0$, we can use
$1-s_{12}'=O(\varepsilon^4)$, and 
$\lambda'_2 = ac_{13}^2 + O(\varepsilon^2 |\delta m^2_{31}|)$
to approximate
\begin{equation}
\tan 2\phi 
= \dfrac{a \sin 2\theta_{13}}{\delta m^2_{31} - a \cos 2\theta_{13}}
+ O(\varepsilon^3)\;.
\end{equation}
In this case, we expect $\phi = O(1)$, $s_\phi = O(1)$, and $c_\phi = O(1)$.
The $\lambda''$s can also be expanded in $\delta m^2_{21}/a = O(\varepsilon^2)$ and
we find
\begin{eqnarray}
\lambda''_2 & = & \lambda''_{-} \cr
& \approx &
\dfrac{ (a+\delta m^2_{31})
         - \sqrt{ (a-\delta m^2_{31})^2 
                 + 4 a\,\delta m^2_{31} s_{13}^2 }
       }
       { 2 } 
+ \Theta(\delta m^2_{31}-a) \delta m^2_{21} s_{12}^2
+ O(\varepsilon^4 |\delta m^2_{31}|) \;,\cr
\lambda''_3 & = & \lambda''_{+} \cr
& \approx &
\dfrac{ (a+\delta m^2_{31})
         + \sqrt{ (a-\delta m^2_{31})^2 
                 + 4 a\,\delta m^2_{31} s_{13}^2 }
       }
       { 2 } 
+ \Theta(a-\delta m^2_{31}) \delta m^2_{21} s_{12}^2
+ O(\varepsilon^4 |\delta m^2_{31}|) \;,\cr
& &
\end{eqnarray}
where $\Theta$ is the Heaviside step function.

For the $\delta m^2_{31} <0$ case, the denominator in the definition
of $\tan 2\phi$ in Eq.~(\ref{phi2def})
is always negative and never crosses zero for any value of $a$.  
Therefore, the smallness of $\sin 2\theta_{13}$ in the numerator
is never cancelled by an equally small denominator.
This allows us to expand $\phi$ in $\theta_{13}$ and we find
\begin{equation}
\phi 
= \left(\dfrac{a}{\delta m^2_{31}-a}\right)\theta_{13} + O(\varepsilon^3)
= O(\varepsilon)\;,
\label{phiInverted}
\end{equation}
which is actually valid for all values of $a$ when $\delta m^2_{31}<0$.
Therefore, $s_\phi = O(\varepsilon)$ and $c_\phi = O(1)$.
The $\lambda''$s are expanded as
\begin{eqnarray}
\lambda''_2 & = & \lambda''_{+} \;\approx\;
a c_{13}^2 + \delta m^2_{21} s_{12}^2 + O(\varepsilon^4 |\delta m^2_{31}|) \;,\cr
\lambda''_3 & = & \lambda''_{-} \;\approx\;
\delta m^2_{31} + a s_{13}^2 + O(\varepsilon^6 |\delta m^2_{31}|)\;.
\end{eqnarray}

For both the $\delta m^2_{31}>0$ and $\delta m^2_{31}<0$ cases,
both $\lambda''_2$ and $\lambda''_3$ are of the same order as $|\delta m^3_{31}|$.
Therefore,
\begin{equation}
H''' = 
\left[ \begin{array}{ccc}
\lambda'_1 & -a c'_{12}c_{13}s_{13}s_{\phi} & a c'_{12}c_{13}s_{13}c_{\phi} \\
-a c'_{12}c_{13}s_{13}s_{\phi} & \lambda''_2 & 0  \\
 a c'_{12}c_{13}s_{13}c_{\phi} & 0 & \lambda''_3
\end{array} \right] =
|\delta m^2_{31}|
\left[ \begin{array}{ccc}
O(\varepsilon^2) & O(\varepsilon^{3,4}) & O(\varepsilon^3) \\
O(\varepsilon^{3,4}) & O(1)   & 0 \\
O(\varepsilon^3) & 0                & O(1)
\end{array} \right]
\;,
\end{equation}
where the sizes of the $(1,2)$ and $(2,1)$ elements depend on 
whether $\phi = O(\varepsilon)$ ($\delta m^2_{31}<0$) or 
$\phi = O(1)$ ($\delta m^2_{31}>0$).
In either case, further diagonalization is not necessary. 

If we relax our accuracy requirement and allow for errors of 
$O(\varepsilon^2 |\delta m^2_{31}|)$, then
the $\lambda''$s for both the $\delta m^2_{31}>0$ and
$\delta m^2_{31}<0$ cases can be approximated by
\begin{equation}
\lambda''_{\pm} \approx
   \dfrac{ (a+\delta m^2_{31})
\pm \sqrt{ (a-\delta m^2_{31})^2 
         + 4 a\,\delta m^2_{31} s_{13}^2 }
         }
         { 2 } \;.
\end{equation}
We will argue that this approximation is sufficient later.
The eigenvalues are then:
\begin{eqnarray}
\lambda_1 & \approx & \lambda'_1 \;\approx\; \delta m^2_{21} c_{12}^2 \;,\cr 
\lambda_2 
& \approx & \lambda''_\mp 
\;\approx\; \dfrac{ (a+\delta m^2_{31})
                    \mp \sqrt{ (a-\delta m^2_{31})^2 
                               + 4 a\, \delta m^2_{31} s_{13}^2 }
                  }
                  { 2 } \;,\cr
\lambda_3
& \approx & \lambda''_\pm 
\;\approx\; \dfrac{ (a+\delta m^2_{31})
                    \pm \sqrt{ (a-\delta m^2_{31})^2 
                               + 4 a\, \delta m^2_{31} s_{13}^2 }
                  }
                  { 2 } \;,
\end{eqnarray}
where the upper sign corresponds to the $\delta m^2_{31}>0$ case, and the
lower sign corresponds to the $\delta m^2_{31}<0$ case.

%%%%%%%%%%%%%%%%%%%%%%%%%%%
\subsubsection{$a/|\delta m^2_{31}| = O(\varepsilon^{-1})$}

The relative sizes of the elements of $H'$ in this case are
\begin{equation}
H' = 
\left[ 
\begin{array}{ccc} 
a c_{12}^2 c_{13}^2    & a c_{12}s_{12}c_{13}^2 & a c_{12}c_{13}s_{13} \\
a c_{12}s_{12}c_{13}^2 & a s_{12}^2c_{13}^2 + \delta m^2_{21} & a s_{12}c_{13}s_{13} \\
a c_{12}c_{13}s_{13}   & a s_{12}c_{13}s_{13}   & a s_{13}^2 + \delta m^2_{31} 
\end{array}
\right] =
|\delta m^2_{31}|
\left[ 
\begin{array}{ccc}
O(\varepsilon^{-1}) & O(\varepsilon^{-1}) & O(1) \\
O(\varepsilon^{-1}) & O(\varepsilon^{-1}) & O(1) \\
O(1) & O(1) & O(1) \\
\end{array}
\right]
\;.
\end{equation}
As in all the previous cases, the $(1,2)$ submatrix is diagonalized by
$V$ defined as in Eq.~(\ref{Vdef}) with $\varphi$ defined as in (\ref{phi1def}). 
The partially diagonalized form is
$H''$ in Eq.~(\ref{Hdoubleprimedef}), with $\lambda'_1$, $\lambda'_2$,
and $\theta'_{12}$ defined as in Eqs.~(\ref{theta12primedef}) and 
(\ref{lambdaprimesdef}).
The expansions of these quantities are:
\begin{eqnarray}
\lambda'_1 
& = &
\delta m^2_{21} c^2_{12} + O(\varepsilon^5 |\delta m^2_{31}|) 
\;=\; O(\varepsilon^2 |\delta m^2_{31}|) \;, \cr
\lambda'_2 
& = &
a c_{13}^2 + \delta m^2_{21} s_{12}^2 + O(\varepsilon^5 |\delta m^2_{31}|)
\;=\; O(\varepsilon^{-1}|\delta m^2_{31}|) \;,
\label{lambdaprimeexpandminus1} 
\end{eqnarray}
and
\begin{eqnarray}
\varphi 
& = & \left(\frac{\pi}{2}-\theta_{12}\right)
    - \frac{\delta m^2_{21}}{2a}\sin(2\theta_{12}) + O(\varepsilon^5) \;, \cr
\theta'_{12} 
& = & \theta_{12} + \varphi
\;=\; \frac{\pi}{2}
    - \frac{\delta m^2_{21}}{2a}\sin(2\theta_{12}) + O(\varepsilon^5) \;.
\label{phithetaexpandminus1}
\end{eqnarray}
Since $s'_{12} = O(1)$,
and $a c'_{12} = O(\varepsilon^2 |\delta m^2_{31}|)$, we find
\begin{equation}
H'' = 
\left[ \begin{array}{ccc}
       \lambda'_1 & 0 & a {c}_{12}' c_{13}s_{13} \\
       0 & \lambda'_2 & a {s}_{12}' c_{13}s_{13} \\
       a {c}_{12}' c_{13}s_{13} & a {s}_{12}' c_{13}s_{13} &
       a s_{13}^2 + \delta m^2_{31}
       \end{array}
\right] =
|\delta m^2_{31}|
\left[ \begin{array}{ccc}
O(\varepsilon^2) & 0                   & O(\varepsilon^3) \\
0                & O(\varepsilon^{-1}) & O(1)             \\
O(\varepsilon^3) & O(1)                & O(1)
\end{array}\right]\;.
\end{equation}
We diagonalize the $(2,3)$ submatrix with the matrix $W$ 
defined in Eq.~(\ref{Wdef}) with the angle $\phi$ defined in Eq.(\ref{phi2def}).
The resulting matrix is $H'''$ given in Eq.~(\ref{Htripleprime}).

Recall that $2\phi$ is in the second quadrant if $\delta m^2_{31} > 0$ 
(level crossing occurs),
and in the fourth quadrant if $\delta m^2_{31} <0$ (no level crossing).
The expansions of $\lambda''_2$, $\lambda''_3$, and $\phi$ in powers of
$\delta m^2_{31}/a = O(\varepsilon)$ differ accordingly.
For the $\delta m^2_{31} > 0$ case, we find
\begin{eqnarray}
\lambda''_2 
& = & \lambda''_{-}
\;=\; \delta m^2_{31} c_{13}^2 + O(\varepsilon^3 |\delta m^2_{31}|) 
\;=\; O(|\delta m^2_{31}|) \;,\cr
\lambda''_3 
& = & \lambda''_{+}
\;=\; a + \delta m^2_{31} s_{13}^2 + \delta m^2_{21} s_{12}^2 + O(\varepsilon^3 |\delta m^2_{31}|) 
\;=\; O(\varepsilon^{-1}|\delta m^2_{31}|) \;,
\end{eqnarray}
and
\begin{equation}
\phi 
%& = & \left( \frac{\pi}{2} - \theta_{13} \right)
%- \dfrac{\delta m^2_{31}}{2a}\sin(2\theta_{13})
%+ O(\varepsilon^3) \cr
\;=\; \left( \frac{\pi}{2} - \theta_{13} \right)
- \dfrac{\delta m^2_{31}}{a}\,\theta_{13} + O(\varepsilon^3)
\;=\; O(1) \;,
\end{equation}
in which case both $s_\phi$ and $c_\phi$ are of order 1 and 
\begin{equation}
H''' = 
\left[ \begin{array}{ccc}
\lambda'_1 & -a c'_{12}c_{13}s_{13}s_{\phi} & a c'_{12}c_{13}s_{13}c_{\phi} \\
-a c'_{12}c_{13}s_{13}s_{\phi} & \lambda''_2 & 0  \\
 a c'_{12}c_{13}s_{13}c_{\phi} & 0 & \lambda''_3
\end{array} \right] =
|\delta m^2_{31}|
\left[ \begin{array}{ccc}
O(\varepsilon^2) & O(\varepsilon^3) & O(\varepsilon^3) \\
O(\varepsilon^3) & O(1)   & 0 \\
O(\varepsilon^3) & 0                & O(\varepsilon^{-1})
\end{array} \right]
\;.
\end{equation}
For the $\delta m^2_{31} < 0$ case, we have
\begin{eqnarray}
\lambda''_2
& = & \lambda''_{+}
\;=\; a + \delta m^2_{31} s_{13}^2 + \delta m^2_{21} s_{12}^2 + O(\varepsilon^3 |\delta m^2_{31}|) 
\;=\; O(\varepsilon^{-1}|\delta m^2_{31}|) \;,\cr
\lambda''_3 
& = & \lambda''_{-}
\;=\; \delta m^2_{31} c_{13}^2 + O(\varepsilon^3 |\delta m^2_{31}|) 
\;=\; O(|\delta m^2_{31}|) \;,
\end{eqnarray}
and
\begin{equation}
\phi = - \theta_{13}
- \dfrac{\delta m^2_{31}}{a}\,\theta_{13}
+ O(\varepsilon^3) 
= O(\varepsilon) \;, 
\end{equation}
which is just Eq.~(\ref{phiInverted}) expanded in powers of $\delta m^2_{31}/a$.
In this case, $s_\phi=O(\varepsilon)$ and $c_\phi=O(1)$.
Therefore, 
\begin{equation}
H''' = 
\left[ \begin{array}{ccc}
\lambda'_1 & -a c'_{12}c_{13}s_{13}s_{\phi} & a c'_{12}c_{13}s_{13}c_{\phi} \\
-a c'_{12}c_{13}s_{13}s_{\phi} & \lambda''_2 & 0  \\
 a c'_{12}c_{13}s_{13}c_{\phi} & 0 & \lambda''_3
\end{array} \right] =
|\delta m^2_{31}|
\left[ \begin{array}{ccc}
O(\varepsilon^2) & O(\varepsilon^4)    & O(\varepsilon^3) \\
O(\varepsilon^4) & O(\varepsilon^{-1}) & 0 \\
O(\varepsilon^3) & 0                   & O(1)
\end{array} \right]
\;.
\end{equation}
In either case, the Hamiltonian has been approximately diagonalized.
The eigenvalues for the $\delta m^2_{31}>0$ case are
\begin{eqnarray}
\lambda_1 
& \approx & \lambda'_1
\;\approx\; \delta m^2_{21} c_{12}^2 \;,\cr
\lambda_2
& \approx & \lambda''_{-}
\;\approx\; \delta m^2_{31} c_{13}^2 \;,\cr
\lambda_3
& \approx & \lambda''_{+}
\;\approx\; a + \delta m^2_{31} s_{13}^2 + \delta m^2_{21} s_{12}^2\;,
\end{eqnarray}
while for the $\delta m^2_{31}<0$ case, they are
\begin{eqnarray}
\lambda_1 
& \approx & \lambda'_1
\;\approx\; \delta m^2_{21} c_{12}^2 \;,\cr
\lambda_2
& \approx & \lambda''_{+}
\;\approx\; a + \delta m^2_{31} s_{13}^2 + \delta m^2_{21} s_{12}^2\;,\cr
\lambda_3
& \approx & \lambda''_{-}
\;\approx\; \delta m^2_{31} c_{13}^2 \;.
\end{eqnarray}
%

%%%%%%%%%%%%%%%%%%%%%%%%%%%%%%%%%%%%%%%%%%%%%%%%%%%%%%%%%%%%%%%%%%%%%%%%
\subsection{Effective Mixing Angles}

To summarize what we have learned above, when $a/|\delta m^2_{31}| = O(\varepsilon^2)$ or
smaller, $H'$ can be approximately diagonalized by a single $(1,2)$ rotation using the 
matrix
\begin{equation}
V = 
\left[ \begin{array}{ccc} c_{\varphi} &  s_{\varphi} & 0 \\
	                     -s_{\varphi} &  c_{\varphi} & 0 \\
	                      0 & 0 & 1
	   \end{array}
\right]\;,
\end{equation}
where
\begin{equation}
c_{\varphi} =\cos\varphi\;,\quad
s_{\varphi} =\sin\varphi\;,\quad
\tan 2\varphi \equiv 
\dfrac{a c_{13}^2\sin 2\theta_{12}}{\delta m^2_{21}-a c_{13}^2\cos 2\theta_{12}}
\approx \dfrac{a\sin 2\theta_{12}}{\delta m^2_{21} - a \cos 2\theta_{12}} 
\;,
\end{equation}
with $0\le\varphi<\frac{\pi}{2}$.
When $a/|\delta m^2_{31}|=O(\varepsilon)$ or larger, this must be followed by a $(2,3)$
rotation using the matrix
\begin{equation}
W = 
\left[ \begin{array}{ccc} 1 & 0 & 0 \\
                          0 &  c_{\phi} &  s_{\phi} \\
	                      0 & -s_{\phi} &  c_{\phi} \\
	   \end{array}
\right]\;,
\end{equation}
where
\begin{equation}
c_{\phi} = \cos\phi \;,\quad
s_{\phi} = \sin\phi \;,\quad
\tan 2\phi \equiv
\dfrac{a s_{12}' \sin 2\theta_{13}}{\delta m^2_{31} + a s_{13}^2 - \lambda'_2} 
\approx
\dfrac{a \sin 2\theta_{13}}
      {\delta m^2_{31}-a \cos 2\theta_{13}}
%\;\approx\;
%\dfrac{a\sin 2\theta_{13}}{\delta m^2_{31}}
\;,
\end{equation}
with $s'_{12}$ and $\lambda'_2$ defined in Eqs.~(\ref{theta12primedef}) and 
(\ref{lambdaprimesdef}).
If $\delta m^2_{31} >0$, then $0<\phi<\frac{\pi}{2}$.
If $\delta m^2_{31} <0$, then $-\frac{\pi}{4}<\phi<0$.

Note that
the first case is encompassed in the second, since in the first case the second
rotation angle $\phi$ becomes negligibly small.
Therefore, the matrix which approximately diagonalizes the effective Hamiltonian is
\begin{eqnarray}
\lefteqn{U'} \cr 
& = & U\mathcal{Q}VW \cr
& = & 
\left[ \begin{array}{ccc} 1 & 0 & 0 \\
                          0 &  c_{23} & s_{23} \\
                          0 & -s_{23} & c_{23}
       \end{array}
\right]
\left[ \begin{array}{ccc} c_{13} & 0 & s_{13} e^{-i\delta} \\
                          0 & 1 & 0 \\
                          -s_{13} e^{i\delta} & 0 & c_{13}
       \end{array}
\right]
\left[ \begin{array}{ccc} c_{12} & s_{12} & 0 \\
                         -s_{12} & c_{12} & 0 \\
                          0 & 0 & 1
       \end{array}
\right] 
%\cr & & \times
\left[ \begin{array}{ccc} 1 & 0 & 0 \\
                          0 & 1 & 0 \\
                          0 & 0 & e^{i\delta}
       \end{array}
\right]
\left[ \begin{array}{ccc} c_{\varphi} &  s_{\varphi} & 0 \\
	                     -s_{\varphi} &  c_{\varphi} & 0 \\
	                      0 & 0 & 1
	   \end{array}
\right]
W \cr
& = & 
\left[ \begin{array}{ccc} 1 & 0 & 0 \\
                          0 &  c_{23} & s_{23} \\
                          0 & -s_{23} & c_{23}
       \end{array}
\right]
\left[ \begin{array}{ccc} c_{13} & 0 & s_{13} e^{-i\delta} \\
                          0 & 1 & 0 \\
                          -s_{13} e^{i\delta} & 0 & c_{13}
       \end{array}
\right]
\left[ \begin{array}{ccc} c_{12} & s_{12} & 0 \\
                         -s_{12} & c_{12} & 0 \\
                          0 & 0 & 1
       \end{array}
\right] 
%\cr & & \times
\left[ \begin{array}{ccc} c_{\varphi} &  s_{\varphi} & 0 \\
	                     -s_{\varphi} &  c_{\varphi} & 0 \\
	                      0 & 0 & 1
	   \end{array}
\right]
\left[ \begin{array}{ccc} 1 & 0 & 0 \\
                          0 & 1 & 0 \\
                          0 & 0 & e^{i\delta}
       \end{array}
\right]
W \cr
& = & 
\left[ \begin{array}{ccc} 1 & 0 & 0 \\
                          0 &  c_{23} & s_{23} \\
                          0 & -s_{23} & c_{23}
       \end{array}
\right]
\left[ \begin{array}{ccc} c_{13} & 0 & s_{13} e^{-i\delta} \\
                          0 & 1 & 0 \\
                          -s_{13} e^{i\delta} & 0 & c_{13}
       \end{array}
\right]
\left[ \begin{array}{ccc} c'_{12} & s'_{12} & 0 \\
                         -s'_{12} & c'_{12} & 0 \\
                          0 & 0 & 1
       \end{array}
\right] 
\left[ \begin{array}{ccc} 1 & 0 & 0 \\
                          0 & 1 & 0 \\
                          0 & 0 & e^{i\delta}
       \end{array}
\right]
\left[ \begin{array}{ccc} 1 & 0 & 0 \\
                          0 &  c_{\phi} &  s_{\phi} \\
	                      0 & -s_{\phi} &  c_{\phi} \\
	   \end{array}
\right] \cr
& = &
\left[ \begin{array}{ccc}
       c_{13}c'_{12} & 
       c_{13}s'_{12}c_\phi - s_{13}s_\phi &
       c_{13}s'_{12}s_\phi+s_{13}c_\phi \\
      -c_{23}s'_{12}-s_{23}s_{13}c'_{12} e^{i\delta} &
       c_{23}c'_{12}c_\phi-s_{23}(s_{13}s'_{12}c_\phi+c_{13}s_\phi)e^{i\delta} &
       c_{23}c'_{12}s_\phi-s_{23}(s_{13}s'_{12}s_\phi-c_{13}c_\phi)e^{i\delta} \\
       s_{23}s'_{12}-c_{23}s_{13}c'_{12} e^{i\delta} &
      -s_{23}c'_{12}c_\phi-c_{23}(s_{13}s'_{12}c_\phi+c_{13}s_\phi)e^{i\delta} &
      -s_{23}c'_{12}s_\phi-c_{23}(s_{13}s'_{12}s_\phi-c_{13}c_\phi)e^{i\delta}
      \end{array}
\right]
\cr & &
\label{UQVW}
\end{eqnarray}
We would like to identify this matrix with
\begin{eqnarray}
\tilde{U} & = &
\left[ \begin{array}{ccc} 1 & 0 & 0 \\
                          0 &  \tilde{c}_{23} & \tilde{s}_{23} \\
                          0 & -\tilde{s}_{23} & \tilde{c}_{23}
       \end{array}
\right]
\left[ \begin{array}{ccc} \tilde{c}_{13} & 0 & \tilde{s}_{13} e^{-i\tilde{\delta}} \\
                          0 & 1 & 0 \\
                          -\tilde{s}_{13} e^{i\tilde{\delta}} & 0 & \tilde{c}_{13}
       \end{array}
\right]
\left[ \begin{array}{ccc} \tilde{c}_{12} & \tilde{s}_{12} & 0 \\
                         -\tilde{s}_{12} & \tilde{c}_{12} & 0 \\
                          0 & 0 & 1
       \end{array}
\right]  \cr
& = &
\left[ \begin{array}{ccc}
       \tilde{c}_{12}\tilde{c}_{13} & 
       \tilde{s}_{12}\tilde{c}_{13} & 
       \tilde{s}_{13} e^{-i\tilde{\delta}} \\
      -\tilde{s}_{12}\tilde{c}_{23} - \tilde{c}_{12}\tilde{s}_{13}\tilde{s}_{23}e^{i\tilde{\delta}} &
       \phantom{-}\tilde{c}_{12}\tilde{c}_{23} - \tilde{s}_{12}\tilde{s}_{13}\tilde{s}_{23}e^{i\tilde{\delta}} &
       \tilde{c}_{13}\tilde{s}_{23} \\
       \phantom{-}\tilde{s}_{12}\tilde{s}_{23} - \tilde{c}_{12}\tilde{s}_{13}\tilde{c}_{23}e^{i\tilde{\delta}} &
      -\tilde{c}_{12}\tilde{s}_{23} - \tilde{s}_{12}\tilde{s}_{13}\tilde{c}_{23}e^{i\tilde{\delta}} &
       \tilde{c}_{13}\tilde{c}_{23}
       \end{array} 
\right] \;,
\end{eqnarray}
up to phases that can be absorbed into the Majorana phases of the neutrinos 
and redefinitions of the charged lepton fields.
Comparing the $(1,3)$ elements of the matrices, we can make the identification 
\begin{eqnarray}
\tilde{s}_{13}
& = & c_{13}s'_{12}s_\phi + s_{13}c_\phi \cr
& = & (c_{13}s_\phi+s_{13}c_\phi) - (1-s'_{12})s_\phi c_{13} \cr
& = & s'_{13} + c'_{13} \left\{ -\frac{ (1-s'_{12})s_\phi c_{13} }{ c'_{13} } \right\} \;,
\end{eqnarray}
where we have defined
\begin{equation}
s'_{13} = \sin\theta'_{13}\;,\qquad
c'_{13} = \cos\theta'_{13}\;,\qquad
\theta'_{13} = \theta_{13} + \phi\;.
\end{equation}
Note that the factor $(1-s'_{12})s_\phi/c'_{13}$ is of order $\varepsilon^3$ or smaller
regardless of the value of $a$, as shown in Table~\ref{OrderEstimate}.
Therefore, 
\begin{equation}
\sin\tilde{\theta}_{13} 
=  \sin\left[ \theta'_{13} + O(\varepsilon^3) \right] \;,
\end{equation}
which implies
\begin{equation}
\tilde{\theta}_{13} = \theta'_{13} + O(\varepsilon^3)\;.
\label{tildetheta13def}
\end{equation}
%
%
%%%%%%%%%%%%%%%%%%%%%%%%%%%%%%%%%%%%%
\begingroup
\squeezetable
\begin{center}
\begin{table}
\begin{tabular}{|c|c||c|c|c|c||c|c|c|c|c|}
\hline
$\dfrac{a}{|\delta m^2_{31}|}$ & $\delta m^2_{31}$ &
$c'_{12}$ & $(1-s'_{12})$ & $s_\phi$ & $c'_{13}$ & 
$\dfrac{(1-s'_{12})s_\phi}{c'_{13}}$ & 
$\dfrac{s_{13}c'_{12}(1-s'_{12})s_\phi}{c'_{13}}$ &
$c'_{12}\left(1-\dfrac{c'_{13}}{c_{13}}\right)$ & 
$c'_{12}s_\phi$ & $c'_{12}s_\phi/c'_{13}$ \\
\hline\hline
$O(\varepsilon^3)$ & $\pm$ & $O(1)$ & $O(1)$ & $O(\varepsilon^4)$ & $O(1)$ & $O(\varepsilon^4)$ & $O(\varepsilon^5)$ & $O(\varepsilon^5)$ & $O(\varepsilon^4)$ & $O(\varepsilon^4)$ \\ 
\hline
$O(\varepsilon^2)$ & $\pm$ & $O(1)$ & $O(1)$ & $O(\varepsilon^3)$ & $O(1)$ & $O(\varepsilon^3)$ & $O(\varepsilon^4)$ & $O(\varepsilon^4)$ & $O(\varepsilon^3)$ & $O(\varepsilon^3)$ \\
\hline
$O(\varepsilon)  $ & $\pm$ & $O(\varepsilon)$ & $O(\varepsilon^2)$ & $O(\varepsilon^2)$ & $O(1)$ & $O(\varepsilon^4)$ & $O(\varepsilon^6)$ & $O(\varepsilon^4)$ & $O(\varepsilon^3)$  & $O(\varepsilon^3)$ \\
\hline
$O(1)$ & $-$ & $O(\varepsilon^2)$ & $O(\varepsilon^4)$ & $O(\varepsilon)$ & $O(1)$ & $O(\varepsilon^{5})$ & $O(\varepsilon^{8})$ & $O(\varepsilon^4)$ & $O(\varepsilon^{3})$ & $O(\varepsilon^{3})$ \\
& $+$ & $O(\varepsilon^2)$ & $O(\varepsilon^4)$ & $O(1)$ & $O(1)$ & $O(\varepsilon^{4})$ & $O(\varepsilon^{7})$ & --- & $O(\varepsilon^{2})$ & $O(\varepsilon^{2})$ \\
\hline
$O(\varepsilon^{-1})$ & $-$ & $O(\varepsilon^3)$ & $O(\varepsilon^6)$ & $O(\varepsilon)$ & $O(1)$ & $O(\varepsilon^{7})$ & $O(\varepsilon^{11})$ & $O(\varepsilon^5)$ & $O(\varepsilon^{4})$ & $O(\varepsilon^{4})$ \\
& $+$ & $O(\varepsilon^3)$ & $O(\varepsilon^6)$ & $O(1)$ & $O(\varepsilon^2)$ & $O(\varepsilon^{4})$ & $O(\varepsilon^{8})$ & --- & $O(\varepsilon^{3})$ & $O(\varepsilon)$ \\
\hline
\end{tabular}
\caption{The sizes of the factors $(1-s'_{12})s_\phi/c'_{13}$, $s_{13}c'_{12}(1-s'_{12})s_\phi/c'_{13}$, $c'_{12}(1-c'_{13}/c_{13})$,
$c'_{12}s_\phi$, and $c'_{12}s_{\phi}/c'_{13}$.}
\label{OrderEstimate}
\end{table}
\end{center}
\endgroup
%%%%%%%%%%%%%%%%%%%%%%%%%%%%%%%%%%%%%
%
Next, looking at the $(1,1)$ and $(1,2)$ elements, we find
\begin{eqnarray}
\tan{\tilde{\theta}_{12}}
& = & \dfrac{\tilde{s}_{12}\tilde{c}_{13}}{\tilde{c}_{12}\tilde{c}_{13}} \cr
& = & \dfrac{c_{13}s'_{12}c_\phi - s_{13}s_\phi}{c_{13}c'_{12}} \cr
& = & \dfrac{(c_{13}c_\phi - s_{13}s_\phi)s'_{12} - s_{13}(1-s'_{12})s_\phi}{c_{13}c'_{12}} \cr
& = & \left(\dfrac{c'_{13}}{c_{13}}\right)
\left[ \tan\theta'_{12}
     + \frac{1}{\cos^2\theta'_{12}}
       \left\{-\frac{s_{13}c'_{12}(1-s'_{12})s_\phi}{c'_{13}}\right\}
\right] \;.
\label{tildetheta12def}
\end{eqnarray}
From Table~\ref{OrderEstimate}, we find that
the factor $s_{13}c'_{12}(1-s'_{12})s_\phi/c'_{13}$ is of order 
$\varepsilon^4$ or smaller for all $a$. Therefore,
\begin{equation}
\tan{\tilde{\theta}_{12}}
= \left(\frac{c'_{13}}{c_{13}}\right)
  \tan\left[\theta'_{12} + O(\varepsilon^4) \right] \;.
\label{tantildetheta12}
\end{equation}
Since $\theta'_{13}=\theta_{13} + \phi$, we can expect the ratio $c'_{13}/c_{13}$ to
be roughly equal to one when $\phi$ is small, and consequently, 
$\tilde{\theta}_{12} \approx \theta'_{12}$.
Indeed, if $\delta m^2_{31}>0$ with $a/|\delta m^2_{31}|\le O(\varepsilon)$, or
$\delta m^2_{31}<0$ with any $a$, then $s_\phi \le O(\varepsilon)$, 
and we find
\begin{equation}
1-\frac{c'_{13}}{c_{13}}
= (1-c_\phi) + s_\phi \tan\theta_{13}
\le O(\varepsilon^2)\;.
\end{equation}
In these cases, we can treat $(1-c'_{13}/c_{13})$ as a small quantity and expand
\begin{equation}
\tilde{\theta}_{12}
\;=\; \theta'_{12} + s'_{12}c'_{12}\left(1-\frac{c'_{13}}{c_{13}}\right) + \cdots
\;=\; \theta'_{12} + O(\varepsilon^4) \;.
\end{equation}
The $\delta m^2_{31}>0$ case with $a/|\delta m^2_{31}|= O(1)$ or
$a/|\delta m^2_{31}|=O(\varepsilon^{-1})$ 
must be considered separately. First, taking the reciprocal of both sides of
Eq.~(\ref{tantildetheta12}), we obtain
\begin{equation}
\cot\tilde{\theta}_{12} 
= \frac{c_{13}}{c'_{13}}\cot\theta'_{12}\;,
\label{cottildetheta12}
\end{equation}
where we have dropped the shift in $\theta'_{12}$ on the right hand side which is of 
order $\varepsilon^{7,8}$ in these particular cases. (cf. Table~\ref{OrderEstimate}.)
Recall that when $a/|\delta m^2_{31}| = O(\varepsilon^{0,-1})$, we have
\begin{equation}
\theta'_{12} = \frac{\pi}{2} - \frac{\delta m^2_{21}}{2a}\sin(2\theta_{12}) 
+ O(\varepsilon^{4,5}) \;,
\end{equation}
while
\begin{equation}
\dfrac{c_{13}}{c'_{13}} 
= O(\varepsilon^{0,-2})\;.
\end{equation}
Therefore, from Eq.~(\ref{cottildetheta12}) we find
\begin{eqnarray}
\tan\left(\frac{\pi}{2}-\tilde{\theta}_{12}\right)
& = & \frac{c_{13}}{c'_{13}}\tan\left(\frac{\pi}{2}-\theta'_{12}\right) \cr
& = & \frac{c_{13}}{c'_{13}}\tan\left(\frac{\delta m^2_{21}}{2a}\sin(2\theta_{12}) + O(\varepsilon^{4,5}) \right) \cr
& = & \frac{c_{13}}{c'_{13}}\left(\frac{\delta m^2_{21}}{2a}\sin(2\theta_{12})\right) + O(\varepsilon^{4,3}) \cr 
& = & \tan\left(\frac{c_{13}}{c'_{13}}\,\frac{\delta m^2_{21}}{2a}\sin(2\theta_{12}) + O(\varepsilon^{4,3})\right)\;,
\end{eqnarray}
from which we can conclude
\begin{equation}
\tilde{\theta}_{12} = \dfrac{\pi}{2}
-\frac{c_{13}}{c'_{13}}\left(\frac{\delta m^2_{21}}{2a}\right)\sin(2\theta_{12})
+ O(\varepsilon^{4,3})\;.
\end{equation}

Next, using the relation
\begin{equation}
s_{13}s'_{12}s_\phi - c_{13}c_\phi
\;=\; -(c_{13}c_\phi-s_{13}s_\phi)-s_{13}(1-s'_{12})s_\phi 
\;=\; -c'_{13} + O(\varepsilon^4)\;,
\end{equation}
we simplify the $(2,3)$ and $(3,3)$ elements of $U'$ as
\begin{eqnarray}
c_{23}c'_{12}s_\phi - s_{23}(s_{13}s'_{12}s_\phi-c_{13}c_\phi)e^{i\delta}
& = & c_{23}c'_{12}s_\phi + s_{23}c'_{13} e^{i\delta} + O(\varepsilon^4) \;,\cr
-s_{23}c'_{12}s_\phi - c_{23}(s_{13}s'_{12}s_\phi-c_{13}c_\phi)e^{i\delta}
& = & -s_{23}c'_{12}s_\phi + c_{23}c'_{13} e^{i\delta} + O(\varepsilon^4) \;.
\end{eqnarray}
Then, using the fact that $c'_{12}s_\phi = O(\varepsilon^2)$ or smaller, we find
\begin{eqnarray}
\lefteqn{ |c_{23}c'_{12}s_\phi - s_{23}(s_{13}s'_{12}s_\phi-c_{13}c_\phi)e^{i\delta}| } \cr
& = & \sqrt{ s_{23}^2 {c'_{13}}^2 + 2 s_{23}c_{23}c'_{13}c'_{12}s_\phi\cos\delta + c_{23}^2{c'_{12}}^2 s_\phi^2 +O(\varepsilon^4) } \cr
& = & \sqrt{ s_{23}^2 {c'_{13}}^2 + 2 s_{23}c_{23}c'_{13}c'_{12}s_\phi\cos\delta +O(\varepsilon^4) } \cr
& = & s_{23} {c'_{13}} + c_{23}c'_{12}s_\phi\cos\delta +O(\varepsilon^4)  \cr
\lefteqn{ |-s_{23}c'_{12}s_\phi - c_{23}(s_{13}s'_{12}s_\phi-c_{13}c_\phi)e^{i\delta}| } \cr
& = & \sqrt{ c_{23}^2 {c'_{13}}^2 - 2 s_{23}c_{23}c'_{13}c'_{12}s_\phi\cos\delta + s_{23}^2{c'_{12}}^2 s_\phi^2 +O(\varepsilon^4) } \cr
& = & \sqrt{ c_{23}^2 {c'_{13}}^2 - 2 s_{23}c_{23}c'_{13}c'_{12}s_\phi\cos\delta +O(\varepsilon^4) } \cr
& = & c_{23} {c'_{13}} - s_{23}c'_{12}s_\phi\cos\delta +O(\varepsilon^4) 
\end{eqnarray}
Therefore, we can make the identification
\begin{eqnarray}
\tan\tilde{\theta}_{23}
& = & \frac{\tilde{c}_{13}\tilde{s}_{23}}{\tilde{c}_{13}\tilde{c}_{23}} \cr
& = & \frac{ s_{23} c'_{13} + c_{23} c'_{12} s_\phi \cos\delta }
           { c_{23} c'_{13} - s_{23} c'_{12} s_\phi \cos\delta } + O(\varepsilon^4) \cr
& = & \dfrac{ t_{23} + \left(\dfrac{c'_{12}s_\phi}{c'_{13}}\right)\cos\delta }
            { 1 - t_{23} \left(\dfrac{c'_{12}s_\phi}{c'_{13}}\right)\cos\delta } + O(\varepsilon^4) \cr
& = & \tan\left[\theta_{23} + \left(\dfrac{c'_{12}s_\phi}{c'_{13}}\right)\cos\delta\right] + O(\varepsilon^4) \;,
\end{eqnarray}
and we obtain
\begin{equation}
\tilde{\theta}_{23} = \theta_{23} + \left(\dfrac{c'_{12}s_\phi}{c'_{13}}\right)\cos\delta + O(\varepsilon^4)\;.
\label{tildetheta23def}
\end{equation}
The factor $c'_{12}s_\phi/c'_{13}$ is of order $\varepsilon^3$ or smaller if
$\delta m^2_{31}>0$ with $a/|\delta m^2_{31}|\le O(\varepsilon)$, or 
$\delta m^2_{31}<0$ with any $a$.
In those cases, we have
\begin{equation}
\tilde{\theta}_{23} = \theta_{23} + O(\varepsilon^3)\;.
\end{equation}
For the case of $\delta m^2_{31} > 0$ with $a/|\delta m^2_{31}|\ge O(1)$,
we can expand $c'_{12}$ and approximate
\begin{equation}
\tilde{\theta}_{23} = \theta_{23} 
+ \frac{s_\phi}{c'_{13}}\left(\frac{\delta m^2_{21}}{2a}\right)\sin(2\theta_{12})\cos\delta
+ O(\varepsilon^4)\;.
\label{tildetheta23approx}
\end{equation}
Finally, we calculate the CP violating phase. The Jarskog invariant of $U'$ is
\begin{eqnarray}
J' & = & 
(c_{13}s'_{12}s_\phi+s_{13}c_\phi)(c_{13}s'_{12}c_\phi-s_{13}s_\phi)(c_{13}c'_{12})s_{23}c_{23}\sin\delta \cr
& = & \tilde{s}_{13}(\tilde{c}_{13}\tilde{s}_{12})(\tilde{c}_{13}\tilde{c}_{12})s_{23}c_{23}\sin\delta \cr
& = & (\tilde{s}_{13}\tilde{c}_{13}^2\tilde{s}_{12}\tilde{c}_{12})s_{23}c_{23}\sin\delta \;.
\end{eqnarray}
On the other hand, the Jarskog invariant of $\tilde{U}$ is
\begin{equation}
\tilde{J} = \tilde{s}_{13}\tilde{c}_{13}^2\tilde{s}_{12}\tilde{c}_{12}\tilde{s}_{23}\tilde{c}_{23}\sin\tilde{\delta}\;.
\end{equation}
Comparison with $J'$ shows that
\begin{equation}
\sin(2\tilde{\theta}_{23})\sin\tilde{\delta} = \sin(2\theta_{23})\sin\delta\;,
\end{equation}
which is actually an exact relation as discussed in Ref.~\cite{Freund:2001pn}.
Since $\tilde{\theta}_{23}=\theta_{23}+O(\varepsilon^3)$ 
when $\delta m^2_{31} > 0$ with $a/|\delta m^2_{31}|\le O(\varepsilon)$, or 
$\delta m^2_{31} < 0$ with any $a$, for these cases we have
\begin{equation}
\tilde{\delta} = \delta + O(\varepsilon^3)\;.
\end{equation}
For the case $\delta m^2_{31} > 0$ with $a/|\delta m^2_{31}|\ge O(1)$,
we can use Eq.~(\ref{tildetheta23approx}) to obtain
\begin{equation}
\tilde{\delta} = \delta - \frac{s_\phi}{c_{13}'}
\left(\frac{\delta m^2_{21}}{a}\right) 
\frac{\sin(2\theta_{12})}{\tan(2\theta_{23})}\,\sin\delta 
+ O(\varepsilon^6) \;.
\end{equation}
%

%%%%%%%%%%%%%%%%%%%%%%%%%%%%%%%%%%%%%%%%%%%%%%%%%%%%%%%%%%%%%%%%%%%%%%%%%%
\subsection{Summary of Neutrino Results and Sample Calculation}

Let us summarize the results of the two previous subsections.

\begin{figure}[p]
\begin{center}
\includegraphics[scale=0.75]{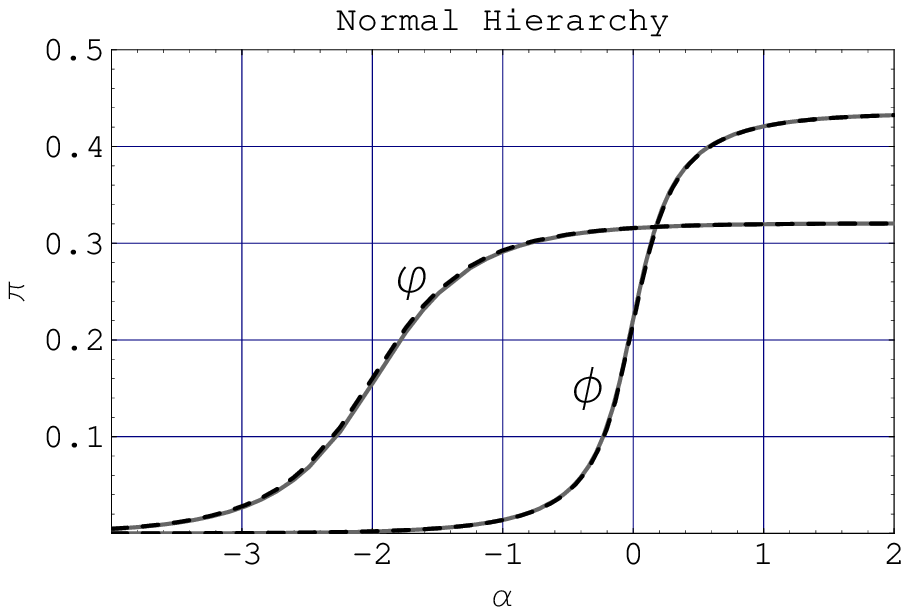}
\includegraphics[scale=0.75]{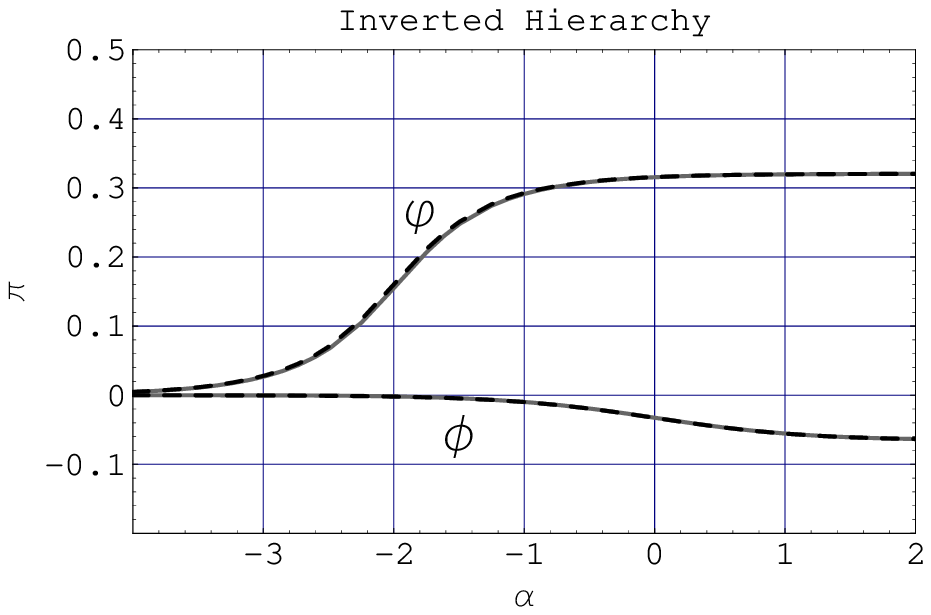}
\caption{The exact (gray solid line) and approximate (black dashed line) values of 
$\varphi$ and $\phi$ plotted against 
$\alpha = \log_{1/\varepsilon}(a/|\delta m^2_{31}|)$.
The parameter choice was $\tan^2\theta_{12} = 0.4$, $\sin^2(2\theta_{13}) = 0.16$,
$\delta m^2_{21}=8.2\times 10^{-5}\mathrm{eV}^2$ and
$|\delta m^2_{31}|=2.5\times 10^{-3}\mathrm{eV}^2$.}
\label{phifigure}
\end{center}
\end{figure}

Approximate values of the effective mixing angles in matter 
can be obtained from the relations
\begin{eqnarray}
\tilde{\theta}_{13} & \approx & \theta'_{13}\;, \cr
\tan\tilde{\theta}_{12} 
& \approx & \frac{c'_{13}}{c_{13}}\tan\theta'_{12} \;,\cr
\tilde{\theta}_{23} 
& \approx & \theta_{23} + \left(\frac{c'_{12}s_\phi}{c'_{13}}\right)\cos\delta\;,\cr
\sin(2\tilde{\theta}_{23})\sin\tilde{\delta} 
& = & \sin(2\theta_{23})\sin\delta \;,
\label{tildetheta0}
\end{eqnarray}
where
\begin{eqnarray}
\theta'_{12} & = & \theta_{12} + \varphi \;, \cr
\theta'_{13} & = & \theta_{13} + \phi \;,
\end{eqnarray}
and the angles $\varphi$ and $\phi$ were defined in 
Eqs.~(\ref{phi1def}) and (\ref{phi2def}), respectively.
The $a$-dependence of
$\varphi$ and $\phi$ for the sample case of 
$\tan^2\theta_{12}=0.4$, $\sin^2(2\theta_{13}) = 0.16$,
$\delta m^2_{21}=8.2\times 10^{-5}\mathrm{eV}^2$, and
$|\delta m^2_{31}|=2.5\times 10^{-3}\mathrm{eV}^2$ is shown in
Fig~\ref{phifigure} with gray solid lines.
In the figure,
the angles are in units of $\pi$, and they are plotted 
against the variable $\alpha$ defined as:
\begin{equation}
\alpha \equiv \log_{1/\varepsilon} \frac{a}{|\delta m^2_{31}|}\;,\qquad
\frac{a}{|\delta m^2_{31}|} = \varepsilon^{-\alpha}\;.
\end{equation}
$\alpha=0$ corresponds to $a=|\delta m^2_{31}|$, and 
$\alpha=-2$ corresponds to $a=\delta m^2_{21}$.

%%%%%%%%%%%%%%%%%%%
%%%%%%%%%%%%%%%%%%%
%%%%%%%%%%%%%%%%%%%
\begin{figure}[p]
\begin{center}
\includegraphics[scale=0.75]{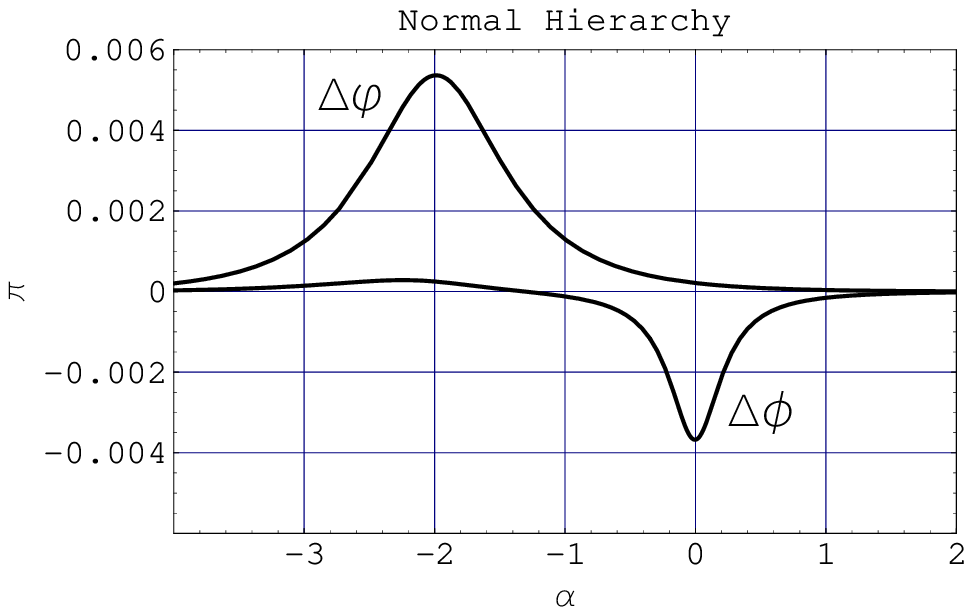}
\includegraphics[scale=0.75]{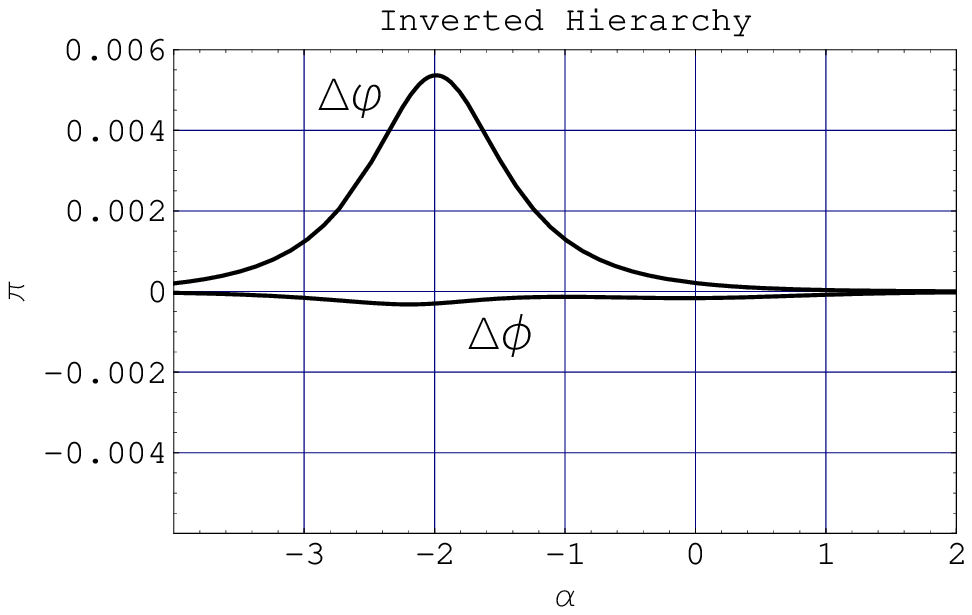}
\caption{$\Delta\varphi=\varphi_\mathrm{approx}-\varphi_\mathrm{exact}$ and 
$\Delta\phi=\phi_\mathrm{approx}-\phi_\mathrm{exact}$ plotted against 
$\alpha = \log_{1/\varepsilon}(a/|\delta m^2_{31}|)$ for the same parameter
choice as Fig.~\protect{\ref{phifigure}}.}
\label{Deltaphifigure}
\end{center}
\end{figure}
%%%%%%%%%%%%%%%%%%%

When either $\delta m^2_{31}>0$ (normal hierarchy) with $\alpha \alt -1$, which corresponds to $a/|\delta m^2_{31}|\le O(\varepsilon)$,
or $\delta m^2_{31} <0$ (inverted hierarchy) with any $\alpha$, the angle $\phi$ is small,
and Eq.~(\ref{tildetheta0}) reduces to
\begin{eqnarray}
\tilde{\theta}_{13} & \approx & \theta'_{13} \;=\; \theta_{13}+\phi \;, \cr
\tilde{\theta}_{12} & \approx & \theta'_{12} \;=\; \theta_{12}+\varphi \;, \cr
\tilde{\theta}_{23} & \approx & \theta_{23} \;, \cr
\tilde{\delta} & \approx & \delta\;.
\label{tildetheta1}
\end{eqnarray}
For the $\delta m^2_{31}>0$ case (normal hierarchy) 
with $\alpha \agt 0$, which corresponds to
$a/|\delta m^2_{31}|\ge O(1)$, the angles can be approximated as
\begin{eqnarray}
\tilde{\theta}_{13} & \approx & \theta'_{13} \;,\cr
\tilde{\theta}_{12} & \approx & \dfrac{\pi}{2}
-\frac{c_{13}}{c'_{13}}\left(\frac{\delta m^2_{21}}{2a}\right)\sin(2\theta_{12})\;,\cr
\tilde{\theta}_{23} & \approx & \theta_{23}
+ \frac{s_\phi}{c'_{13}}\left(\frac{\delta m^2_{21}}{2a}\right)\sin(2\theta_{12})\cos\delta  \;,\cr
\tilde{\delta} & \approx & \delta - \frac{s_\phi}{c_{13}'}
\left(\frac{\delta m^2_{21}}{a}\right) \frac{\sin(2\theta_{12})}{\tan(2\theta_{23})}\,\sin\delta \;.
\label{tildetheta2}
\end{eqnarray}
To make use of these expressions, we must first calculate $\varphi$ and $\phi$, and
then $\theta'_{12}=\theta_{12}+\varphi$ and $\theta'_{13}=\theta_{13}+\phi$.
Simple approximations to $\varphi$ and $\phi$ are provided by
\begin{equation}
\tan 2\varphi 
\;\approx\; \dfrac{ a \sin 2\theta_{12} }{ \delta m^2_{21} - a \cos 2\theta_{12} } \;,\qquad
\tan 2\phi
\;\approx\; \dfrac{ a \sin 2\theta_{13} }{ \delta m^2_{31} - a \cos 2\theta_{13} } \;.
\label{phiapprox}
\end{equation}
The approximate values of $\varphi$ and $\phi$ obtained from these
expressions are also shown in Fig.~\ref{phifigure} with black dashed lines.
As is clear from the figure, the graphs
of the exact and approximate values are virtually indistinguishable at this scale.
In Fig.~\ref{Deltaphifigure} we plot the differences between the approximate and exact values
of $\varphi$ and $\phi$:
\begin{equation}
\Delta\varphi \equiv \varphi_\mathrm{approx} - \varphi_\mathrm{exact}\;,\qquad
\Delta\phi    \equiv \phi_\mathrm{approx}    - \phi_\mathrm{exact}\;.
\end{equation}
The maximum deviation from the exact values occur at
the level-crossing points $\alpha=-2$ and $\alpha=0$ (there is no level-crossing
at $\alpha=0$ when $\delta m^2_{31}<0$) but even then, it is a mere fraction of
a percent of $\pi$.
Using Eq.~(\ref{phiapprox}), 
we can also obtain simple approximate formulae for
$\theta'_{12}=\theta_{12}+\varphi$ and $\theta'_{13}=\theta_{13}+\phi$:
\begin{equation}
\tan 2\theta'_{12} \;\approx\; 
\dfrac{ \delta m^2_{21} \sin 2\theta_{12} }{ \delta m^2_{21} \cos 2\theta_{12} - a } \;,\qquad
\tan 2\theta'_{13} \;\approx\; 
\dfrac{ \delta m^2_{31} \sin 2\theta_{13} }{ \delta m^2_{31} \cos 2\theta_{13} - a } \;.
\label{thetaprimeapprox}
\end{equation}
These allow us to calculate $\theta'_{12}$ and $\theta'_{13}$ directly without
going through $\varphi$ and $\phi$.
Eqs.~(\ref{phiapprox}) and (\ref{thetaprimeapprox}) provide a
quick and easy way to obtain the input angles necessary to utilize
Eqs.~(\ref{tildetheta0}), (\ref{tildetheta1}), and (\ref{tildetheta2}).

To demonstrate the accuracy of these approximations, we present a sample calculation 
using the following parameter choice:
\begin{eqnarray}
\delta m^2_{21}      & = & 8.2\times 10^{-5}\,\mathrm{eV}^2\;,\cr
|\delta m^2_{31}|    & = & 2.5\times 10^{-3}\,\mathrm{eV}^2\;,\cr
\tan^2\theta_{12}    & = & 0.4\;,\cr
\sin^2(2\theta_{13}) & = & 0.16\;,\cr
\theta_{23}          & = & 0.2\,\pi\;,\cr
\delta               & = & 0.25\,\pi\;.
\label{exampleparameterset}
\end{eqnarray}
The values of $\delta m^2_{21}$, $|\delta m^2_{31}|$, and $\tan^2\theta_{12}$
are the experimental central values.
The value of $\sin^2(2\theta_{13})$ is taken to be the 90\% upper limit corresponding 
to our choice of $|\delta m^2_{31}|$ so that the $O(\theta_{13})$ terms that we neglect
are maximized.  The value of $\theta_{23}$ is also chosen to be one of the
90\% confidence limits since if we set $\theta_{23}$ to the experimentally
preferred central value of $0.25\pi$, then the shift of $\tilde{\delta}$ away from 
$\delta$ would be suppressed.
Similarly, we chose $\delta = 0.25\pi$ so that both $\tilde{\theta}_{23}$ and
$\tilde{\delta}$ will be shifted from their vacuum values.

%

%%%%%%%%%%%%%%%%%%%
\begin{figure}[p]
\begin{center}
\includegraphics[scale=0.75]{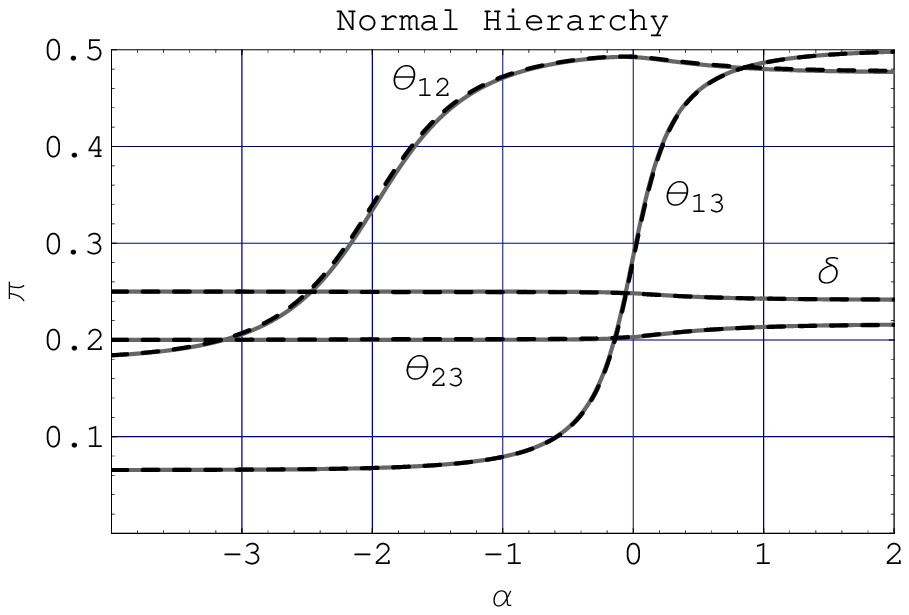}
\includegraphics[scale=0.75]{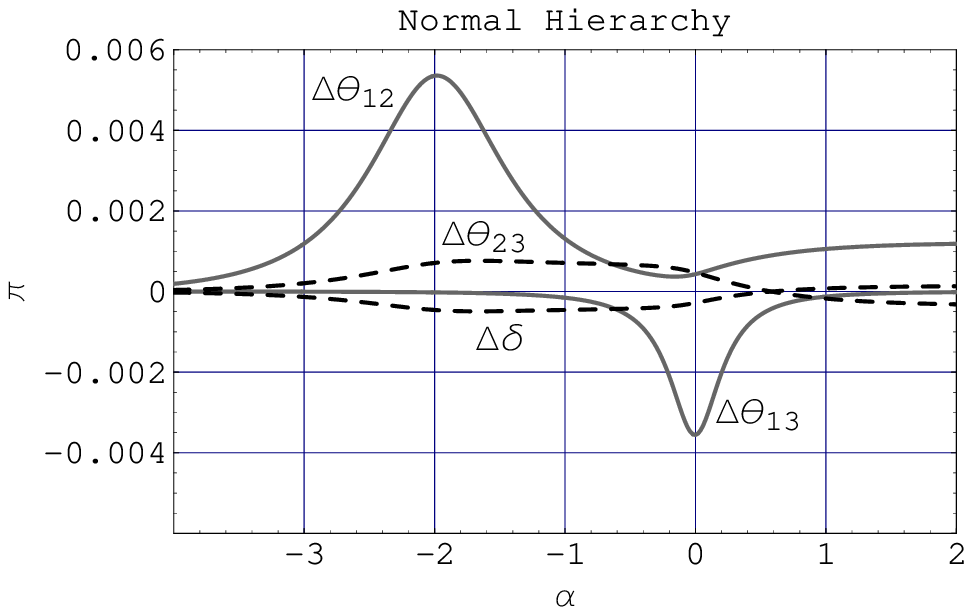}
\caption{(a) The exact values of $\tilde{\theta}_{12}$, $\tilde{\theta}_{13}$, $\tilde{\theta}_{23}$, and $\tilde{\delta}$ (solid gray lines) plotted against their approximate values (black dashed lines) obtained using Eq.~(\protect{\ref{tildetheta0}}), with
Eqs.~(\protect{\ref{phiapprox}}) and (\protect{\ref{thetaprimeapprox}}), as functions of
$\alpha=\log_{1/\varepsilon}(a/|\delta m^2_{31}|)$.
(b) The differences 
$\Delta\theta_{12} = \tilde{\theta}_{12,\mathrm{approx}}-\tilde{\theta}_{12,\mathrm{exact}}$
and $\Delta\theta_{13} = \tilde{\theta}_{13,\mathrm{approx}}-\tilde{\theta}_{13,\mathrm{exact}}$ (solid gray lines), and the differences
$\Delta\theta_{23} = \tilde{\theta}_{23,\mathrm{approx}}-\tilde{\theta}_{23,\mathrm{exact}}$
and $\Delta\delta = \tilde{\delta}_{\mathrm{approx}}-\tilde{\delta}_{\mathrm{exact}}$ (black dashed lines) of this approximation plotted against $\alpha$.}
\label{thetaTildeNormal1}
\end{center}
\end{figure}
%%%%%%%%%%%%%%%%%%%
%%%%%%%%%%%%%%%%%%%
\begin{figure}[p]
\begin{center}
\includegraphics[scale=0.75]{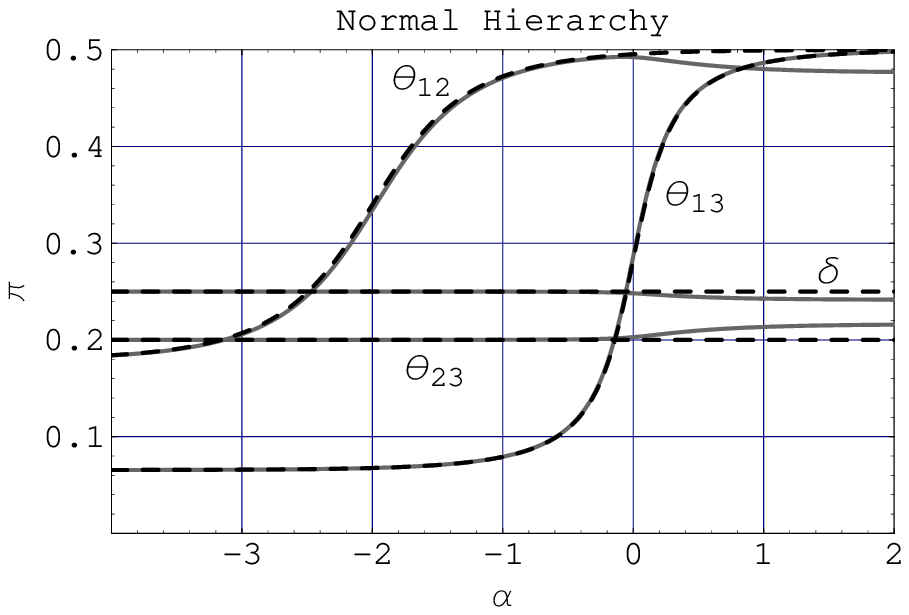}
\includegraphics[scale=0.75]{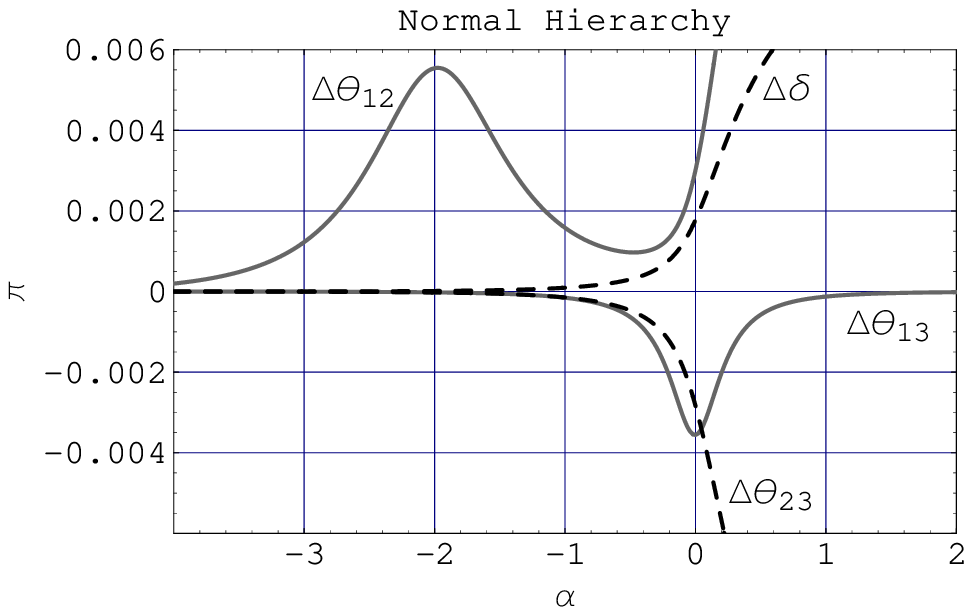}
\caption{(a) The exact values of $\tilde{\theta}_{12}$, $\tilde{\theta}_{13}$, $\tilde{\theta}_{23}$, and $\tilde{\delta}$ (solid gray lines) plotted against their approximate values (black dashed lines) obtained using Eq.~(\protect{\ref{tildetheta1}}) 
as functions of $\alpha=\log_{1/\varepsilon}(a/|\delta m^2_{31}|)$.
(b) $\Delta\theta_{12}$ and $\Delta\theta_{13}$ (solid gray lines), and
$\Delta\theta_{23}$ and $\Delta\delta$ (black dashed lines) of this approximation plotted as functions of $\alpha$. This approximation is applicable when $\alpha\alt -1$.}
\label{thetaTildeNormal2}
\end{center}
\end{figure}
%%%%%%%%%%%%%%%%%%%
%%%%%%%%%%%%%%%%%%%
\begin{figure}[p]
\begin{center}
\includegraphics[scale=0.75]{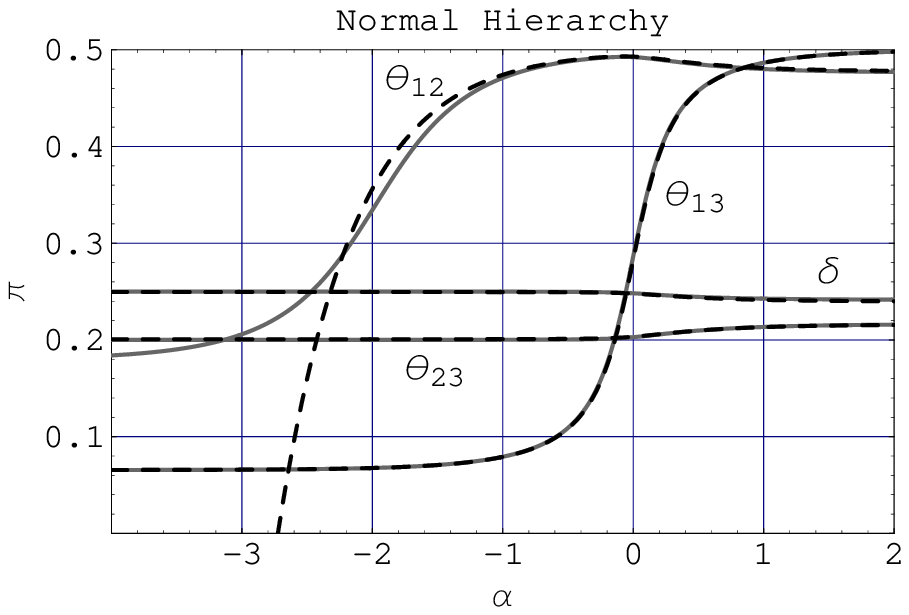}
\includegraphics[scale=0.75]{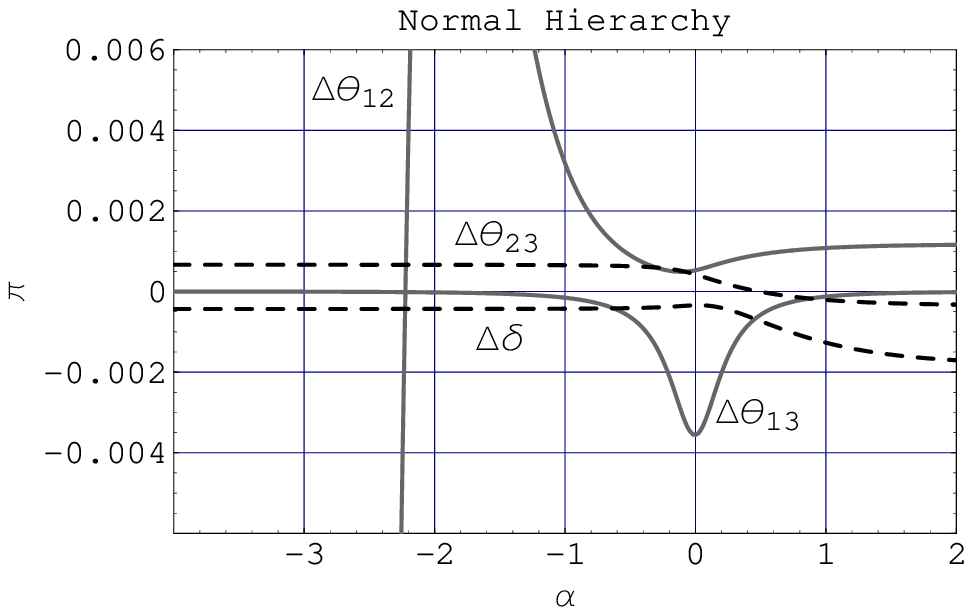}
\caption{(a) The exact values of $\tilde{\theta}_{12}$, $\tilde{\theta}_{13}$, $\tilde{\theta}_{23}$, and $\tilde{\delta}$ (solid gray lines) plotted against their approximate values (black dashed lines) obtained using Eq.~(\protect{\ref{tildetheta2}}) 
as functions of $\alpha=\log_{1/\varepsilon}(a/|\delta m^2_{31}|)$.
(b) $\Delta\theta_{12}$ and $\Delta\theta_{13}$ (solid gray lines), and
$\Delta\theta_{23}$ and $\Delta\delta$ (black dashed lines) of this approximation plotted as functions of $\alpha$. This approximation is applicable when $\alpha\agt 0$.}
\label{thetaTildeNormal3}
\end{center}
\end{figure}
%%%%%%%%%%%%%%%%%%%
%%%%%%%%%%%%%%%%%%%
\begin{figure}[p]
\begin{center}
\includegraphics[scale=0.75]{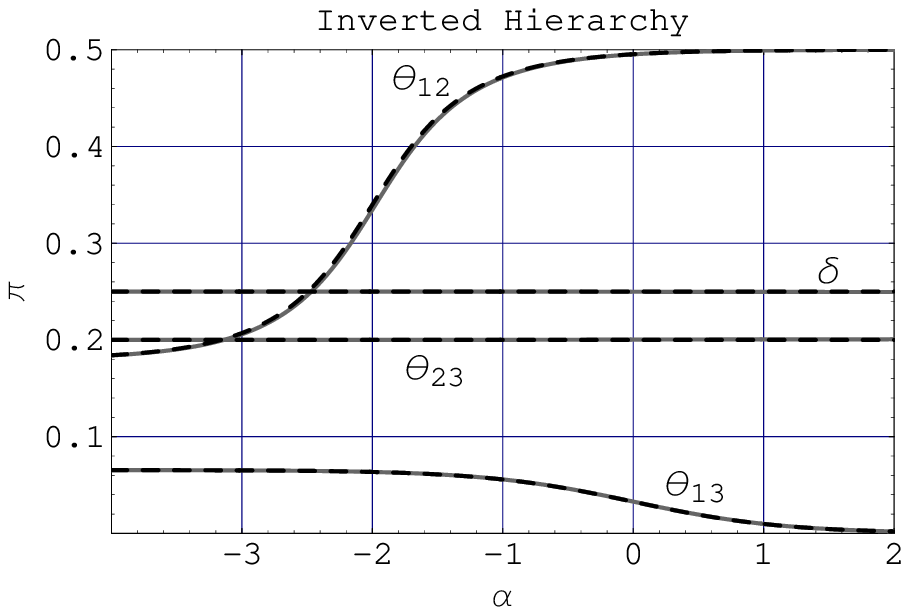}
\includegraphics[scale=0.75]{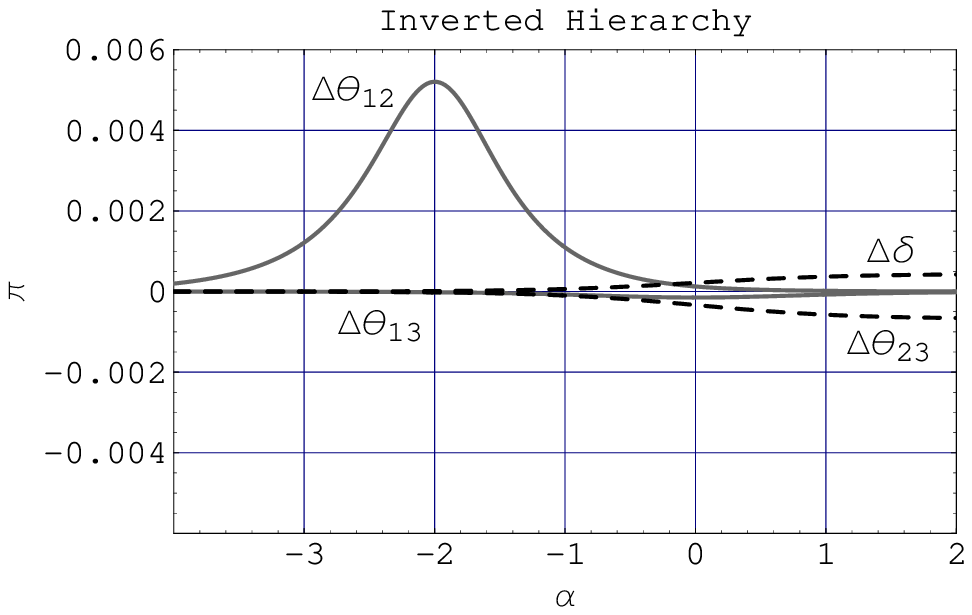}
\caption{(a) The exact values of $\tilde{\theta}_{12}$, $\tilde{\theta}_{13}$, $\tilde{\theta}_{23}$, and $\tilde{\delta}$ (solid gray lines) plotted against their approximate values (black dashed lines) obtained using Eq.~(\protect{\ref{tildetheta1}}) 
as functions of $\alpha=\log_{1/\varepsilon}(a/|\delta m^2_{31}|)$ for the inverted
hierarchy case ($\delta m^2_{31}<0$).
(b) $\Delta\theta_{12}$ and $\Delta\theta_{13}$ (solid gray lines), and
$\Delta\theta_{23}$ and $\Delta\delta$ (black dashed lines) of this approximation plotted as functions of $\alpha$.}
\label{thetaTildeInverted}
\end{center}
\end{figure}
%%%%%%%%%%%%%%%%%%%

We first consider the normal hierarchy case ($\delta m^2_{31}>0$).
In Fig.~\ref{thetaTildeNormal1}a, we plot the exact values of $\tilde{\theta}_{12}$,
$\tilde{\theta}_{13}$, $\tilde{\theta}_{23}$, and $\tilde{\delta}$ calculated 
numerically with gray solid lines, 
together with the approximate values obtained from Eq.~(\ref{tildetheta0}),
using Eqs.~(\ref{phiapprox}) and (\ref{thetaprimeapprox}) to calculate the input
angles, with dashed black lines.
Fig.~\ref{thetaTildeNormal1}b shows the errors:
\begin{equation}
\Delta\theta_{ij} \equiv \tilde{\theta}_{ij,\mathrm{approx}}-\tilde{\theta}_{ij,\mathrm{exact}}\;,\qquad
\Delta\delta \equiv \tilde{\delta}_\mathrm{approx}-\tilde{\delta}_\mathrm{exact}\;.
\end{equation}
$\Delta\theta_{12}$ and $\Delta\theta_{13}$ are indicated with solid gray lines,
while $\Delta\theta_{23}$ and $\Delta\delta$ are indicated with dashed black lines.
The errors are never larger than a fraction of a percent of $\pi$, and 
comparison with Fig.~\ref{Deltaphifigure} makes it apparent that the majority of it
was inherited from having calculated $\varphi$ and $\phi$ using Eq.~(\ref{phiapprox}).
In Figs.~\ref{thetaTildeNormal2}a and \ref{thetaTildeNormal3}a, we plot the
exact values against the approximate values obtained using 
Eqs.~(\ref{tildetheta1}) and (\ref{tildetheta2}), respectively,
together with the errors in Figs.~\ref{thetaTildeNormal2}b and \ref{thetaTildeNormal3}b.
Clearly, the approximations using Eqs.~(\ref{tildetheta1}) and (\ref{tildetheta2}) 
are good in their respective ranges of applicability.

For the inverted hierarchy case ($\delta m^2_{31}<0$), we only need to
consider Eq.~(\ref{tildetheta1}).
In Fig.~\ref{thetaTildeInverted}a, we show the comparison between the
numerically calculated exact values and the approximate values obtained from
Eq.~(\ref{tildetheta1}).  The errors are shown in 
Fig.~\ref{thetaTildeInverted}b.

%%%%%%%%%%%%%%%%%%%
\begin{figure}[p]
\begin{center}
\includegraphics[scale=0.75]{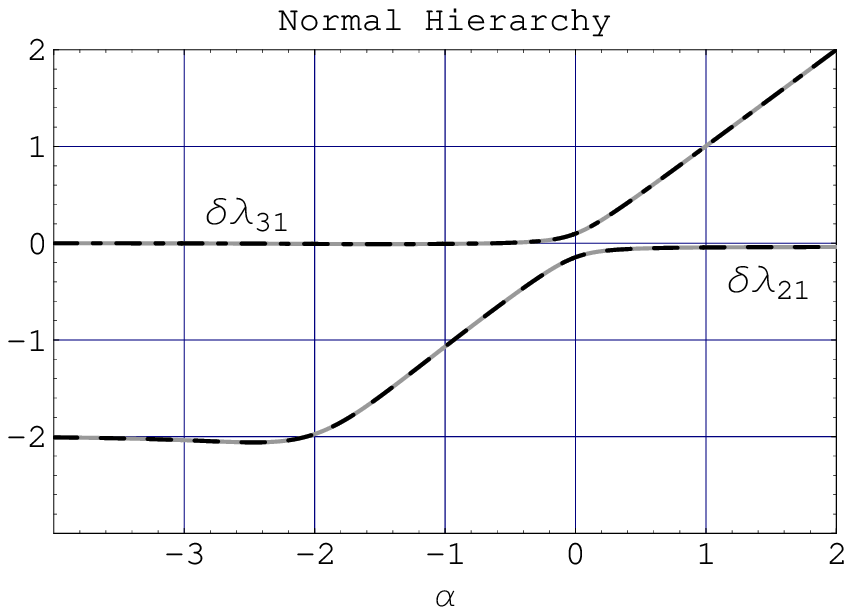}
\includegraphics[scale=0.75]{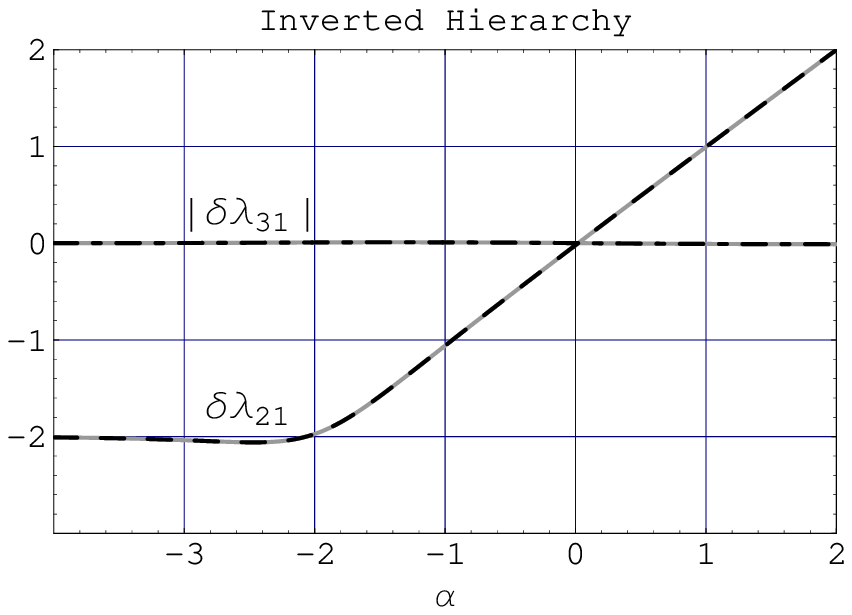}
\caption{The exact and approximate values of 
$\log_{1/\varepsilon}(\delta\lambda_{21}/|\delta m^2_{31}|)$ and
$\log_{1/\varepsilon}(|\delta\lambda_{31}|/|\delta m^2_{31}|)$ 
for the parameter set of Eq.~(\protect{\ref{exampleparameterset}})
plotted against $\alpha=\log_{1/\varepsilon}(a/|\delta m^2_{31}|)$.
The exact values are in gray solid lines, whereas the approximate values are
in black dashed ($\delta\lambda_{21}$) and black dot-dashed ($|\delta\lambda_{31}|$) lines.
}
\label{deltalambdaXY1}
\end{center}
\end{figure}
%%%%%%%%%%%%%%%%%%%
%%%%%%%%%%%%%%%%%%%
\begin{figure}[p]
\begin{center}
\includegraphics[scale=0.75]{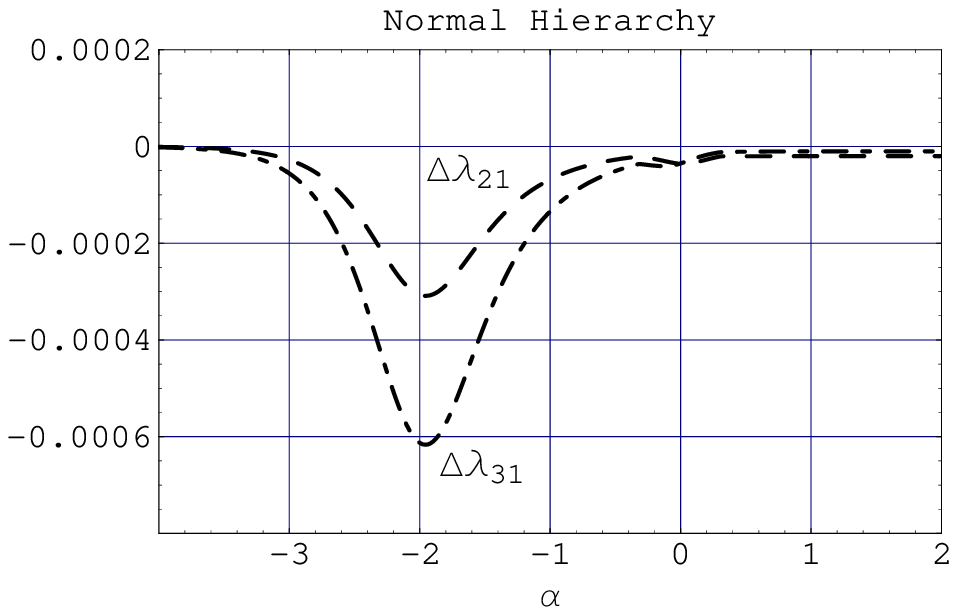}
\includegraphics[scale=0.75]{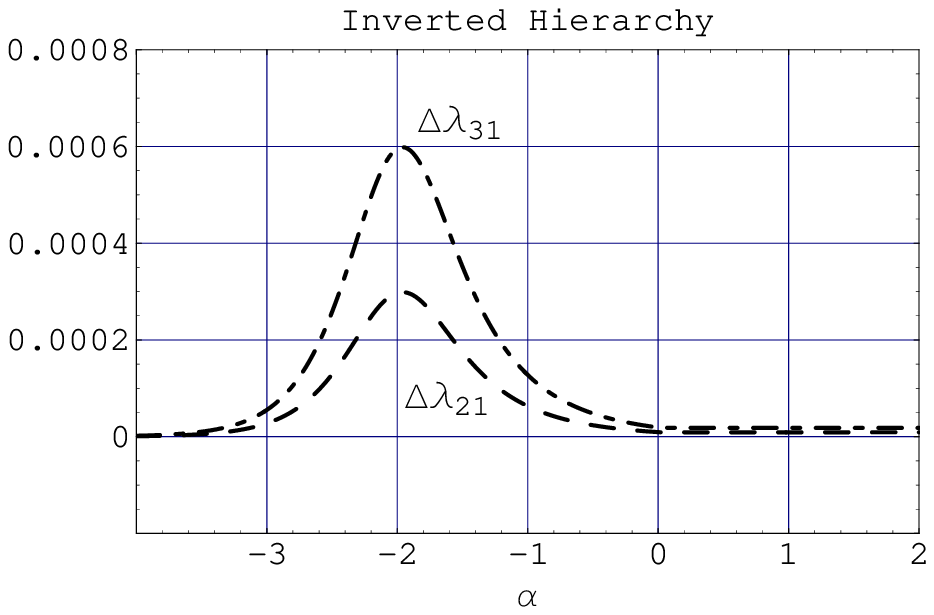}
\caption{The rescaled errors $\Delta\lambda_{21}$ (dashed) and 
$\Delta\lambda_{31}$ (dot-dashed), as 
defined in Eq.~(\protect{\ref{DeltalambdaXYdef}}), for the approximation
of Fig.~\protect{\ref{deltalambdaXY1}}.}
\label{errorlambdaXY1}
\end{center}
\end{figure}
%%%%%%%%%%%%%%%%%%%
%%%%%%%%%%%%%%%%%%%
\begin{figure}[p]
\begin{center}
\includegraphics[scale=0.75]{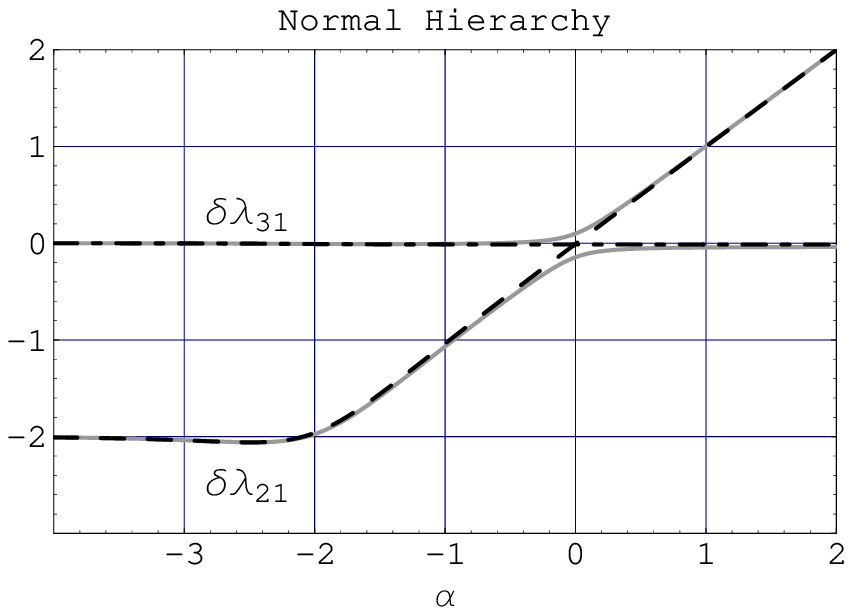}
\includegraphics[scale=0.75]{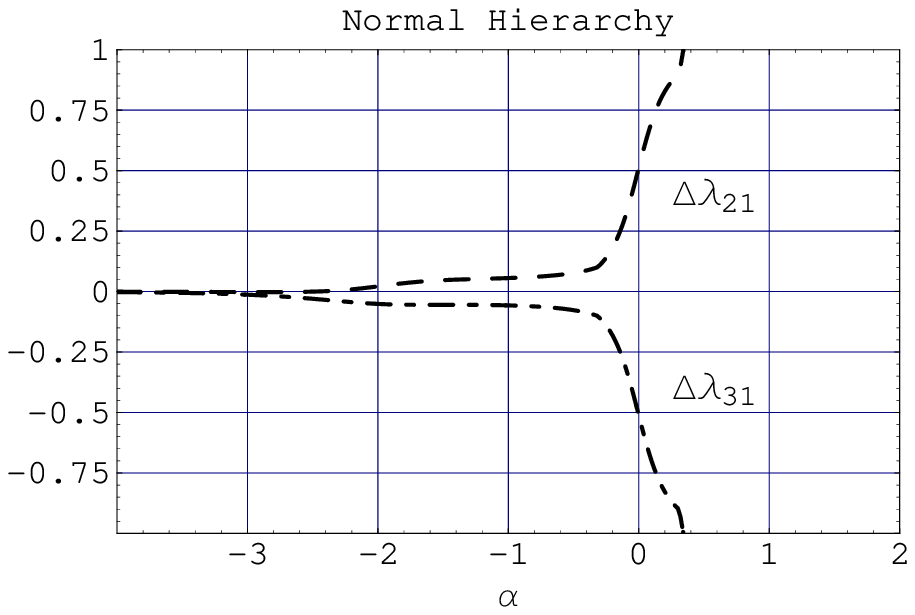}
\caption{Comparison of exact and approximate values using 
Eq.~(\protect{\ref{lambdaapprox2}}) for the normal hierarchy case.
The approximation is applicable when $\alpha\alt -1$.}
\label{deltalambdaXYnormal2}
\end{center}
\end{figure}
%%%%%%%%%%%%%%%%%%%
%%%%%%%%%%%%%%%%%%%
\begin{figure}[p]
\begin{center}
\includegraphics[scale=0.75]{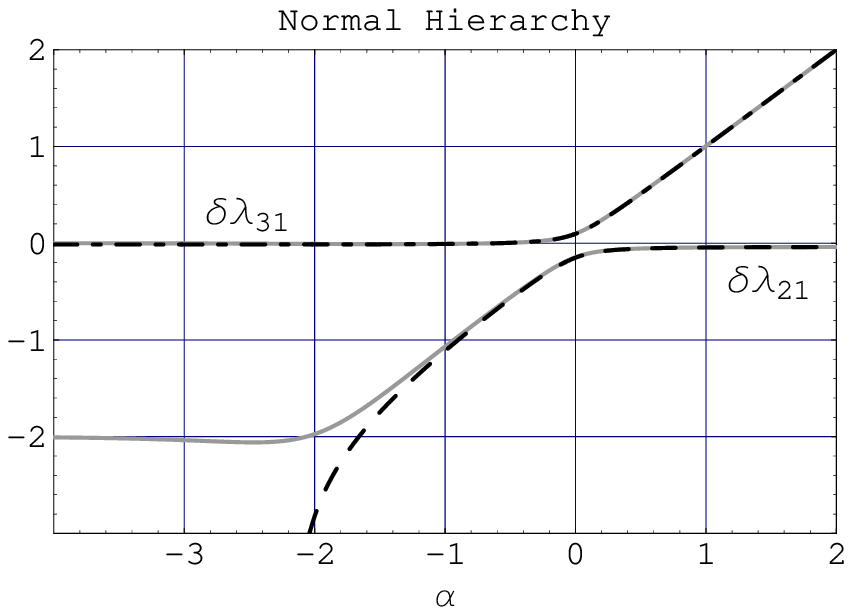}
\includegraphics[scale=0.75]{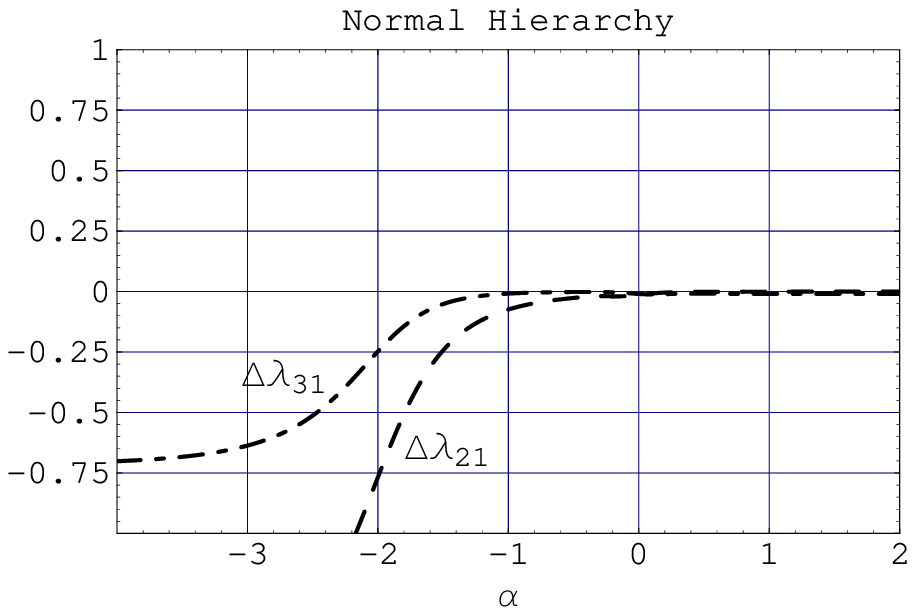}
\caption{Comparison of exact and approximate values using 
Eq.~(\protect{\ref{lambdaapprox3}}) for the normal hierarchy case.
The approximation is applicable when $\alpha\agt -1$.}
\label{deltalambdaXYnormal3}
\end{center}
\end{figure}
%%%%%%%%%%%%%%%%%%%
%%%%%%%%%%%%%%%%%%%
\begin{figure}[p]
\begin{center}
\includegraphics[scale=0.75]{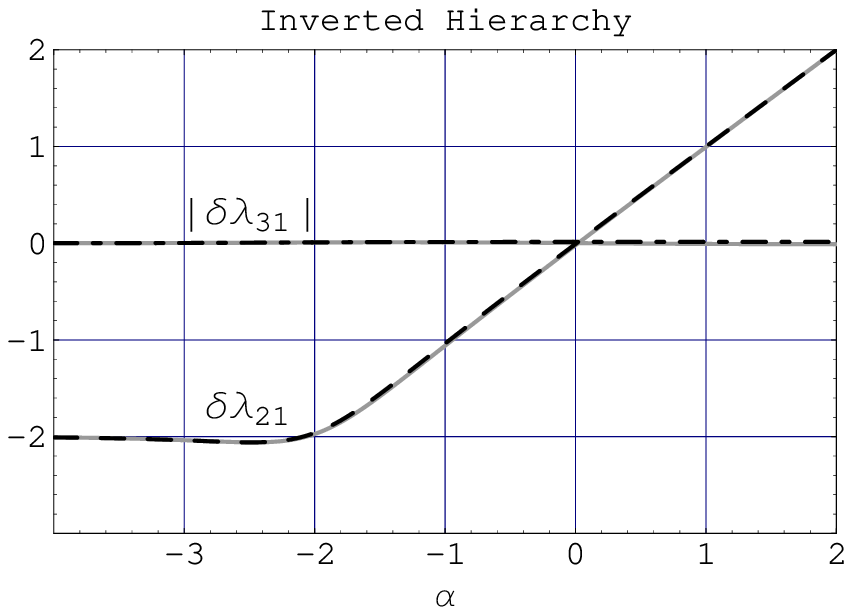}
\includegraphics[scale=0.75]{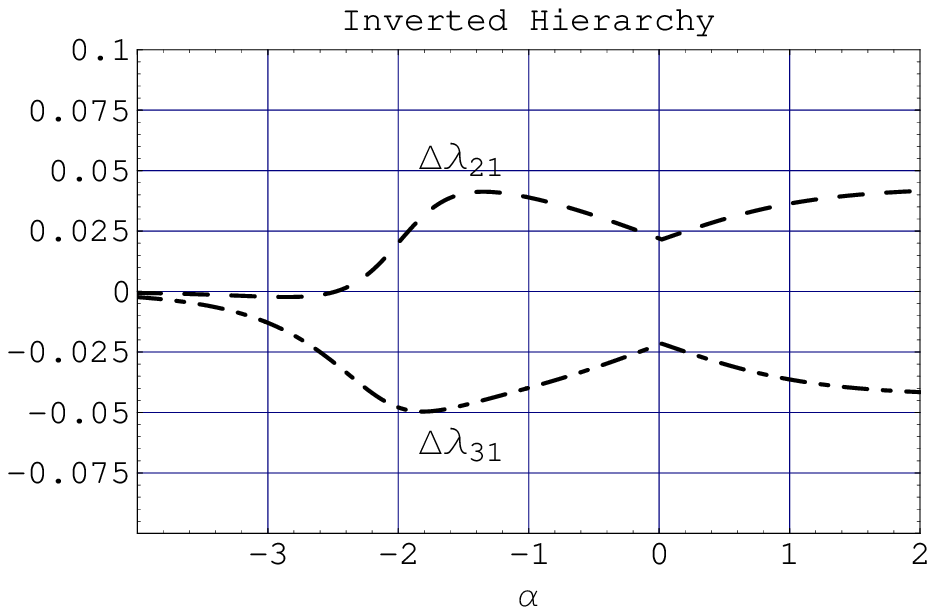}
\caption{Comparison of exact and approximate values using 
Eq.~(\protect{\ref{lambdaapprox2}}) for the inverted hierarchy case.}
\label{deltalambdaXYinverted2}
\end{center}
\end{figure}
%%%%%%%%%%%%%%%%%%%

The approximate values for the mass-squared eigenvalues are given by
\begin{eqnarray}
\lambda_1 & \approx & \lambda'_{-}\;,\cr
\lambda_2 & \approx & \lambda''_{-}\;,\cr
\lambda_3 & \approx & \lambda''_{+}\;,
\label{lambdaapprox1normal}
\end{eqnarray}
for $\delta m^2_{31}>0$ (normal hierarchy), and by
\begin{eqnarray}
\lambda_1 & \approx & \lambda'_{-}\;,\cr
\lambda_2 & \approx & \lambda''_{+}\;,\cr
\lambda_3 & \approx & \lambda''_{-}\;,
\label{lambdaapprox1inverted}
\end{eqnarray}
for $\delta m^2_{31}<0$ (inverted hierarchy), where $\lambda'_{\pm}$ and
$\lambda''_{\pm}$ are defined in Eqs,~(\ref{lambdaprimeplusminusdef}) and
(\ref{lambdadoubleprimeplusminusdef}), respectively.
The accuracy of this approximation is illustrated in 
Fig~\ref{deltalambdaXY1} using the parameter values of 
Eq.~(\ref{exampleparameterset}), where
the exact numerically calculated values of 
$\delta\lambda_{21} = \lambda_2 - \lambda_1$ and
$|\delta\lambda_{31}| = |\lambda_3 - \lambda_1|$ are plotted against those
obtained from the above approximate expressions.
The vertical axis is plotted using the same log-scale as the horizontal axis
where $|\delta m^2_{31}|$ corresponds to $0$ and
$\delta m^2_{21}$ corresponds to $-2$.
Since a log-scale plot does not reflect the absolute accuracy of each
$\delta\lambda_{ij}$, in Fig.~\ref{errorlambdaXY1} we plot the difference between the
exact and approximate values of $\delta\lambda_{ij}$ normalized to
$\delta\lambda_\mathrm{min,exact}$:
\begin{eqnarray}
\Delta\lambda_{i1} \equiv 
\dfrac{ \delta\lambda_{i1,\mathrm{approx}}-\delta\lambda_{i1,\mathrm{exact}} }
      { \delta\lambda_{\mathrm{min,exact}} }\;,\qquad
(i=2,3)\;,
\label{DeltalambdaXYdef}
\end{eqnarray}
where
\begin{equation}
\delta\lambda_{\mathrm{min,exact}}
\equiv \min(\delta\lambda_{21,\mathrm{exact}},
           |\delta\lambda_{31,\mathrm{exact}}|,
           |\delta\lambda_{32,\mathrm{exact}}|)\;.
\end{equation}
This tells us how large the errors are compared to 
$\delta\lambda_\mathrm{min}$, which is typically used to expand the  
oscillation probabilities in.
As is evident from the figures, the approximation is excellent.
%Unfortunately,
%Eqs.~(\ref{lambdaprimeplusminusdef}) and (\ref{lambdadoubleprimeplusminusdef})
%are too complicated to be of practical use.

When either $\delta m^2_{31}>0$ (normal hierarchy) with $\alpha \alt -1$, \textit{i.e.} $a/|\delta m^2_{31}| \le O(\varepsilon)$, or $\delta m^2_{31}<0$ (inverted hierarchy) with any $a$, the $\lambda$'s can be further approximated by
\begin{eqnarray}
\lambda_1 & \approx & 
\dfrac{ (a + \delta m^2_{21})
         -\sqrt{ (a - \delta m^2_{21})^2 + 4 a\, \delta m^2_{21} s_{12}^2 }
      }
      { 2 }\;,\cr
\lambda_2 & \approx & 
\dfrac{ (a + \delta m^2_{21})
         +\sqrt{ (a - \delta m^2_{21})^2 + 4 a\, \delta m^2_{21} s_{12}^2 }
      }
      { 2 }\;,\cr
\lambda_3 & \approx & \delta m^2_{31}\;.
\label{lambdaapprox2}
\end{eqnarray}
For the $\delta m^2_{31}>0$ (normal hierarchy) case with $\alpha \agt -1$, \textit{i.e.}
$a/|\delta m^2_{31}| \ge O(\varepsilon)$, we can use
\begin{eqnarray}
\lambda_1 & \approx & \delta m^2_{21} c_{12}^2 \;,\cr
\lambda_2 & \approx & 
\dfrac{ (a+\delta m^2_{31})
        -\sqrt{(a-\delta m^2_{31})^2+4a\,\delta m^2_{31} s_{13}^2}
      }{ 2 } \;,\cr
\lambda_3 & \approx & 
\dfrac{ (a+\delta m^2_{31})
        +\sqrt{(a-\delta m^2_{31})^2+4a\,\delta m^2_{31} s_{13}^2}
                   }{ 2 }\;.
\label{lambdaapprox3}
\end{eqnarray}
This second approximation introduces errors of $O(\varepsilon^2 |\delta m^2_{31}|)$ 
in the $\lambda$'s.
However, since $\delta\lambda_{21} \ge O(\varepsilon |\delta m^2_{31}|)$ in this range,
an error of this size is tolerable.
The accuracy of these approximations is illustrated in
Figs.~\ref{deltalambdaXYnormal2} and \ref{deltalambdaXYnormal3} for the 
$\delta m^2_{31}>0$ case, and Fig.~\ref{deltalambdaXYinverted2} for the
$\delta m^2_{31}<0$ case.
Though the accuracy is not as good as when
Eqs.~(\ref{lambdaapprox1normal}) and (\ref{lambdaapprox1inverted}), with 
Eqs.~(\ref{lambdaprimeplusminusdef}) and (\ref{lambdadoubleprimeplusminusdef}),
are used, it is sufficient for most practical purposes.

\subsection{Anti-Neutrino Case}

Matter effects for anti-neutrinos can be treated in an analogous fashion.
We will therefore omit the details and give only an outline of the derivation and
results.

The effective Hamiltonian for anti-neutrinos in matter is
\begin{equation}
\bar{H} = 
\edlit{U}{}^*
\left[ \begin{array}{ccc} \bar{\lambda}_1 & 0 & 0 \\
                          0 & \bar{\lambda}_2 & 0 \\
                          0 & 0 & \bar{\lambda}_3
       \end{array}
\right]
\edlit{U}{}^\mathrm{T}
= U^*
\left[ \begin{array}{ccc} 0 & 0 & 0 \\
                          0 & \delta m^2_{21} & 0 \\
                          0 & 0 & \delta m^2_{31}
       \end{array}
\right]
U^\mathrm{T} +
\left[ \begin{array}{ccc} -a & 0 & 0 \\
                          0 & 0 & 0 \\
                          0 & 0 & 0 
       \end{array}
\right] \;,
\label{Hbar}
\end{equation}
We denote the effective mass-squared eigenvalues as $\bar{\lambda}_{i}$, $(i=1,2,3)$,
and the diagonalization matrix as $\edlit{U}$ to distinguish them from those
for the neutrinos. (Note the mirror image of the tilde on top of $\edlit{U}$.)
This matrix can be partially diagonalized as
\begin{eqnarray}
\bar{H}' & = & \mathcal{Q} U^\mathrm{T} \bar{H} U^* \mathcal{Q}^* \cr
& = & \mathcal{Q} \left\{\;
\left[ \begin{array}{ccc} 0 & 0 & 0 \\
                          0 & \delta m^2_{21} & 0 \\
                          0 & 0 & \delta m^2_{31}
       \end{array}
\right] +
U^\mathrm{T}
\left[ \begin{array}{ccc} -a & 0 & 0 \\
                          0 & 0 & 0 \\
                          0 & 0 & 0 
       \end{array}
\right] 
U^* \;\right\} \mathcal{Q}^* \cr
& = & \mathcal{Q}
\left[ \begin{array}{ccc} 0 & 0 & 0 \\
                          0 & \delta m^2_{21} & 0 \\
                          0 & 0 & \delta m^2_{31}
       \end{array}
\right] \mathcal{Q}^*
- a \,\mathcal{Q}
\left[ \begin{array}{ccc} U_{e1}U^*_{e1} & U_{e1}U^*_{e2} & U_{e1}U^*_{e3} \\
                          U_{e2}U^*_{e1} & U_{e2}U^*_{e2} & U_{e2}U^*_{e3} \\
                          U_{e3}U^*_{e1} & U_{e3}U^*_{e2} & U_{e3}U^*_{e3} 
       \end{array}
\right] \mathcal{Q}^* \cr
& = &
\left[ \begin{array}{ccc} 0 & 0 & 0 \\
                          0 & \delta m^2_{21} & 0 \\
                          0 & 0 & \delta m^2_{31}
       \end{array}
\right] - a 
\left[ \begin{array}{ccc} 
        c_{12}^2 c_{13}^2    & c_{12}s_{12}c_{13}^2 & c_{12}c_{13}s_{13}  \\
        c_{12}s_{12}c_{13}^2 & s_{12}^2c_{13}^2     & s_{12}c_{13}s_{13}  \\
        c_{12}c_{13}s_{13}   & s_{12}c_{13}s_{13}   & s_{13}^2 
       \end{array}
\right] \cr
& = & 
\left[ 
\begin{array}{ccc} 
-a c_{12}^2 c_{13}^2    & -a c_{12}s_{12}c_{13}^2 & -a c_{12}c_{13}s_{13} \\
-a c_{12}s_{12}c_{13}^2 & -a s_{12}^2c_{13}^2 + \delta m^2_{21} & -a s_{12}c_{13}s_{13} \\
-a c_{12}c_{13}s_{13}   & -a s_{12}c_{13}s_{13}   & -a s_{13}^2 + \delta m^2_{31} 
\end{array}
\right] \;.
\label{Hbarprime}
\end{eqnarray}
The only difference from the effective Hamiltonian $H'$ for the neutrinos, Eq.~(\ref{Hprime}),
is in the sign in front of the matter-effect term $a$.

Following the general procedure we employed for the neutrinos,
we begin by diagonalizing the 1-2 submatrix of $\bar{H}'$.
Let
\begin{equation}
\bar{V} = 
\left[ \begin{array}{ccc} \bar{c}_{{\varphi}} & \bar{s}_{{\varphi}} & 0 \\
	                     -\bar{s}_{{\varphi}} & \bar{c}_{{\varphi}} & 0 \\
	                      0 & 0 & 1
	   \end{array}
\right]\;,
\end{equation}
where
\begin{equation}
\bar{c}_{{\varphi}}=\cos\bar{\varphi}\;,\quad
\bar{s}_{{\varphi}}=\sin\bar{\varphi}\;,\quad
\tan 2\bar{\varphi} \equiv 
-\dfrac{a c_{13}^2 \sin2\theta_{12}}
       {\delta m^2_{21} + a c_{13}^2\cos2\theta_{12}}\;,\quad
\left(-\frac{\pi}{2}<\bar{\varphi}\le 0\right)\;.
\label{phi1bardef}
\end{equation}
Using $\bar{V}$, we find
\begin{equation}
\bar{H}'' =
\bar{V}^\dagger \bar{H}' \bar{V} =
\left[ \begin{array}{ccc}
       \bar{\lambda}'_{1} & 0 & -a\bar{c}'_{12}c_{13}s_{13} \\
       0 & \bar{\lambda}'_{2} & -a\bar{s}'_{12}c_{13}s_{13} \\
       -a\bar{c}'_{12}c_{13}s_{13} &
       -a\bar{s}'_{12}c_{13}s_{13} & -a s^2_{13} + \delta m^2_{31} 
       \end{array}
\right] \;,
\end{equation}
where
\begin{equation}
\bar{c}'_{12} = \cos\bar{\theta}'_{12}\;,\quad
\bar{s}'_{12} = \sin\bar{\theta}'_{12}\;,\quad
\bar{\theta}'_{12} = \theta_{12} + \bar{\varphi}\;,\quad
\tan 2\bar{\theta}'_{12} 
= \dfrac{ \delta m^2_{21}\sin 2\theta_{12} }
        { \delta m^2_{21}\cos 2\theta_{12} + a c_{13}^2 }\;,
\label{theta12primebar}
\end{equation}
and 
$\bar{\lambda}'_1 = \bar{\lambda}'_{-}$, $\bar{\lambda}'_2 = \bar{\lambda}'_{+}$, with
\begin{equation}
\bar{\lambda}'_{\pm} \;=\; 
\dfrac{(\delta m^2_{21}-a c_{13}^2)
      \pm \sqrt{(\delta m^2_{21}+a c_{13}^2)^2 - 4a c_{13}^2\,\delta m^2_{21} s_{12}^2}}
      {2} \;.
\label{lambdaprimebardef}
\end{equation}

Next, unlike the neutrino case, we diagonalize the 1-3 submatrix of $\bar{H}''$.
Let
\begin{equation}
\bar{W} = 
\left[ \begin{array}{ccc}
       \bar{c}_{\phi} & 0 & \bar{s}_{\phi} \\
       0 & 1 & 0 \\
      -\bar{s}_{\phi} & 0 & \bar{c}_{\phi}
       \end{array}
\right]
\end{equation}
where
\begin{equation}
\bar{c}_{\phi} = \cos\bar{\phi}\;,\quad
\bar{s}_{\phi} = \sin\bar{\phi}\;,\quad
\tan 2\bar{\phi} \equiv
-\dfrac{a\bar{c}'_{12}\sin 2\theta_{13}}
       {\delta m^2_{31}-a s_{13}^2 -\bar{\lambda}'_{1}}\;.
\label{phi2bardef}
\end{equation}
The angle $\bar{\phi}$ is in the fourth quadrant when $\delta m^2_{31}>0$,
and the first quadrant when $\delta m^2_{31}<0$.
Using $\bar{W}$, we find
\begin{equation}
\bar{H}''' =
\bar{W}^\dagger \bar{H}'' \bar{W} =
\left[ \begin{array}{ccc}
\bar{\lambda}''_1 & a\bar{s}'_{12}c_{13}s_{13}\bar{s}_{\phi} & 0 \\
a\bar{s}'_{12}c_{13}s_{13}\bar{s}_{\phi} & \bar{\lambda}'_2 & -a\bar{s}'_{12}c_{13}s_{13}\bar{c}_{\phi} \\
0 & -a\bar{s}'_{12}c_{13}s_{13}\bar{c}_{\phi} & \bar{\lambda}''_3
\end{array}\right] \;,
\end{equation}
where
\begin{eqnarray}
\bar{\lambda}''_1 = \bar{\lambda}''_{-}\;,\quad
\bar{\lambda}''_3 = \bar{\lambda}''_{+}\;,\qquad \mbox{if $\delta m^2_{31} > 0$}\;, \cr
\bar{\lambda}''_1 = \bar{\lambda}''_{+}\;,\quad
\bar{\lambda}''_3 = \bar{\lambda}''_{-}\;,\qquad \mbox{if $\delta m^2_{31} < 0$}\;.
\end{eqnarray}
with
\begin{equation}
\bar{\lambda}''_{\pm} \equiv
   \dfrac{ [ (\delta m^2_{31}-a s_{13}^2) + \bar{\lambda}'_{1} ]
\pm \sqrt{ [ (\delta m^2_{31}-a s_{13}^2) - \bar{\lambda}'_{1} ]^2 
         + 4 a^2 {\bar{c}_{12}}^{\prime 2} c_{13}^2 s_{13}^2 }
         }
         { 2 } \;.
\label{lambdadoubleprimebardef}
\end{equation}
Evaluation of the off-diagonal terms of $\bar{H}'''$ reveals that
it is approximately diagonalized, and the diagonalization matrix is given approximately by
\begin{equation}
\bar{U}'{}^* \equiv U^*Q^*\bar{V}\bar{W} = U^*\bar{V} Q^* \bar{W}\;,
\end{equation}
or taking the complex conjugate,
\begin{eqnarray}
\lefteqn{\bar{U}'} \cr
& = & U \bar{V} Q\, \bar{W} \cr
& = &
\left[ \begin{array}{ccc} 1 & 0 & 0 \\
                          0 &  c_{23} & s_{23} \\
                          0 & -s_{23} & c_{23}
       \end{array}
\right]
\left[ \begin{array}{ccc} c_{13} & 0 & s_{13} e^{-i\delta} \\
                          0 & 1 & 0 \\
                          -s_{13} e^{i\delta} & 0 & c_{13}
       \end{array}
\right]
\left[ \begin{array}{ccc} c_{12} & s_{12} & 0 \\
                         -s_{12} & c_{12} & 0 \\
                          0 & 0 & 1
       \end{array}
\right]
\left[ \begin{array}{ccc} \bar{c}_{\varphi} & \bar{s}_{\varphi} & 0 \\
	                     -\bar{s}_{\varphi} & \bar{c}_{\varphi} & 0 \\
                          0 & 0 & 1
	   \end{array}
\right]
Q\, \bar{W} \cr
& = & 
\left[ \begin{array}{ccc} 1 & 0 & 0 \\
                          0 &  c_{23} & s_{23} \\
                          0 & -s_{23} & c_{23}
       \end{array}
\right]
\left[ \begin{array}{ccc} c_{13} & 0 & s_{13} e^{-i\delta} \\
                          0 & 1 & 0 \\
                          -s_{13} e^{i\delta} & 0 & c_{13}
       \end{array}
\right]
\left[ \begin{array}{ccc} \bar{c}'_{12} & \bar{s}'_{12} & 0 \\
                         -\bar{s}'_{12} & \bar{c}'_{12} & 0 \\
                          0 & 0 & 1
       \end{array}
\right]
\left[ \begin{array}{ccc} 1 & 0 & 0 \\ 0 & 1 & 0 \\ 0 & 0 & e^{i\delta} 
       \end{array}
\right]
\left[ \begin{array}{ccc} \bar{c}_{\phi} & 0 & \bar{s}_{\phi} \\
                          0 & 1 & 0 \\
                         -\bar{s}_{\phi} & 0 & \bar{c}_{\phi}
       \end{array}
\right] \cr
& = &
\left[ \begin{array}{ccc}
       c_{13}\bar{c}'_{12}\bar{c}_\phi - s_{13}\bar{s}_\phi &
       c_{13}\bar{s}'_{12} & 
       s_{13}\bar{c}_\phi + c_{13}\bar{c}'_{12}\bar{s}_\phi \\
      -c_{23}\bar{s}'_{12}\bar{c}_\phi-s_{23}(c_{13}\bar{s}_\phi+s_{13}\bar{c}'_{12}\bar{c}_\phi)e^{i\delta} &
       c_{23}\bar{c}'_{12}-s_{23}s_{13}\bar{s}'_{12} e^{i\delta} &
      -c_{23}\bar{s}'_{12}\bar{s}_\phi+s_{23}(c_{13}\bar{c}_\phi-s_{13}\bar{c}'_{12}\bar{s}_\phi)e^{i\delta} \\
       s_{23}\bar{s}'_{12}\bar{c}_\phi-c_{23}(c_{13}\bar{s}_\phi+s_{13}\bar{c}'_{12}\bar{c}_\phi)e^{i\delta} &
      -s_{23}\bar{c}'_{12}-c_{23}s_{13}\bar{s}'_{12} e^{i\delta} &
       s_{23}\bar{s}'_{12}\bar{s}_\phi+c_{23}(c_{13}\bar{c}_\phi-s_{13}\bar{c}'_{12}\bar{s}_\phi)e^{i\delta}
      \end{array}
\right]
\cr & &
\label{UQVWbar}
\end{eqnarray}
Identification of this matrix with
\begin{eqnarray}
\edlit{U} & = &
\left[ \begin{array}{ccc} 1 & 0 & 0 \\
                          0 &  \edlit{c}_{23} & \edlit{s}_{23} \\
                          0 & -\edlit{s}_{23} & \edlit{c}_{23}
       \end{array}
\right]
\left[ \begin{array}{ccc} \edlit{c}_{13} & 0 & \edlit{s}_{13} e^{-i\edlit{\delta}} \\
                          0 & 1 & 0 \\
                          -\edlit{s}_{13} e^{i\edlit{\delta}} & 0 & \edlit{c}_{13}
       \end{array}
\right]
\left[ \begin{array}{ccc} \edlit{c}_{12} & \edlit{s}_{12} & 0 \\
                         -\edlit{s}_{12} & \edlit{c}_{12} & 0 \\
                          0 & 0 & 1
       \end{array}
\right]  \cr
& = &
\left[ \begin{array}{ccc}
       \edlit{c}_{12}\edlit{c}_{13} & 
       \edlit{s}_{12}\edlit{c}_{13} & 
       \edlit{s}_{13} e^{-i\edlit{\delta}} \\
      -\edlit{s}_{12}\edlit{c}_{23} - \edlit{c}_{12}\edlit{s}_{13}\edlit{s}_{23}e^{i\edlit{\delta}} &
       \phantom{-}\edlit{c}_{12}\edlit{c}_{23} - \edlit{s}_{12}\edlit{s}_{13}\edlit{s}_{23}e^{i\edlit{\delta}} &
       \edlit{c}_{13}\edlit{s}_{23} \\
       \phantom{-}\edlit{s}_{12}\edlit{s}_{23} - \edlit{c}_{12}\edlit{s}_{13}\edlit{c}_{23}e^{i\edlit{\delta}} &
      -\edlit{c}_{12}\edlit{s}_{23} - \edlit{s}_{12}\edlit{s}_{13}\edlit{c}_{23}e^{i\edlit{\delta}} &
       \edlit{c}_{13}\edlit{c}_{23}
       \end{array} 
\right] \;,
\end{eqnarray}
(up to phases that can be absorbed into redefinitions of the charged lepton 
fields and the Majorana phases of the neutrinos) yields
\begin{eqnarray}
\edlit{\theta}_{13} & \approx & \bar{\theta}_{13}' 
%\equiv \theta_{13} + \bar{\phi}
\;,\cr
\tan\edlit{\theta}_{12} & \approx & \dfrac{c_{13}}{\bar{c}_{13}'}\tan\bar{\theta}_{12}'\;,\cr
\edlit{\theta}_{23} & \approx & \theta_{23} 
-\left(\dfrac{\bar{s}_{12}'\bar{s}_\phi}{\bar{c}_{13}'}\right)\cos\delta\;,\cr
\sin(2\edlit{\theta}_{23})\sin\edlit{\delta} & \approx & \sin(2\theta_{23})\sin\delta\;,
\label{edlittheta0}
\end{eqnarray}
where
\begin{equation}
\bar{s}_{13}' = \sin\bar{\theta}_{13}'\;,\qquad
\bar{c}_{13}' = \cos\bar{\theta}_{13}'\;,\qquad
\bar{\theta}_{13}' \equiv \theta_{13} + \bar{\phi}\;.
\end{equation}
%\begin{eqnarray}
%& & \bar{s}_{13}' = \sin\bar{\theta}_{13}'\;,\qquad
%\bar{c}_{13}' = \cos\bar{\theta}_{13}'\;,\qquad
%\bar{\theta}_{13}' \equiv \theta_{13} + \bar{\phi}\;.\cr
%& & \bar{s}_{12}' = \sin\bar{\theta}_{12}'\;,\qquad
%\bar{c}_{12}' = \cos\bar{\theta}_{12}'\;,\qquad
%\bar{\theta}_{12}' \equiv \theta_{12} + \bar{\varphi}\;.
%\end{eqnarray}
%
%The angles $\bar{\varphi}$ and $\bar{\phi}$ were defined in Eqs.~(\ref{phi1bardef})
%and (\ref{phi2bardef}).
For the $\delta m^2_{31}>0$ case (normal hierarchy) with any $a$, or
the $\delta m^2_{31}<0$ case (inverted hierarchy) with 
$a/|\delta m^2_{31}|\le O(\varepsilon)$, the angle $\bar{\phi}$ is small and
the above relations simplify to
\begin{eqnarray}
\edlit{\theta}_{13} & \approx & \bar{\theta}_{13}' = \theta_{13} + \bar{\phi}\;,\cr
\edlit{\theta}_{12} & \approx & \bar{\theta}_{12}' = \theta_{12} + \bar{\varphi}\;,\cr
\edlit{\theta}_{23} & \approx & \theta_{23}\;,\cr
\edlit{\delta}      & \approx & \delta\;.
\label{edlittheta1}
\end{eqnarray}
For the $\delta m^2_{31}<0$ case (inverted hierarchy) with $a/|\delta m^2_{31}|\ge O(1)$,
we can approximate
\begin{eqnarray}
\edlit{\theta}_{13} & \approx & \bar{\theta}_{13}' \;,\cr
\edlit{\theta}_{12} & \approx & \dfrac{c_{13}}{\bar{c}_{13}'}
\left(\dfrac{\delta m^2_{21}}{2a}\right)\sin(2\theta_{12}) \;,\cr
\edlit{\theta}_{23} & \approx & \theta_{23} - \dfrac{\bar{s}_{\phi}}{\bar{c}_{13}'}
\left(\dfrac{\delta m^2_{21}}{2a}\right)\sin(2\theta_{12})\cos\delta \;,\cr
\edlit{\delta}      & \approx & \delta + \dfrac{\bar{s}_{\phi}}{\bar{c}_{13}'}
\left(\dfrac{\delta m^2_{21}}{a}\right)\dfrac{\sin(2\theta_{12})}{\tan(2\theta_{23})}\sin\delta\;.
\label{edlittheta2}
\end{eqnarray}
%

%%%%%%%%%%%%%%%%%%%
\begin{figure}[p]
\begin{center}
\includegraphics[scale=0.75]{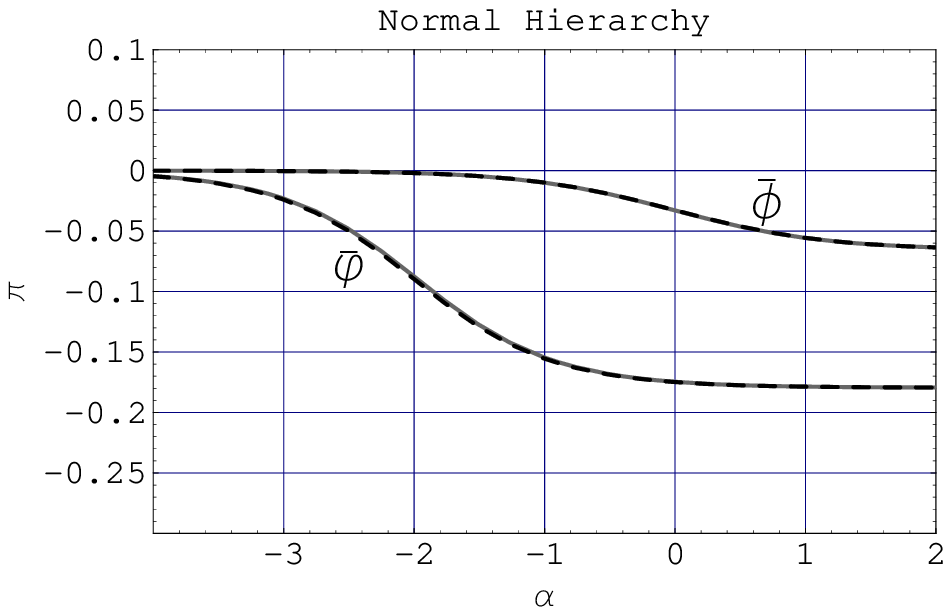}
\includegraphics[scale=0.75]{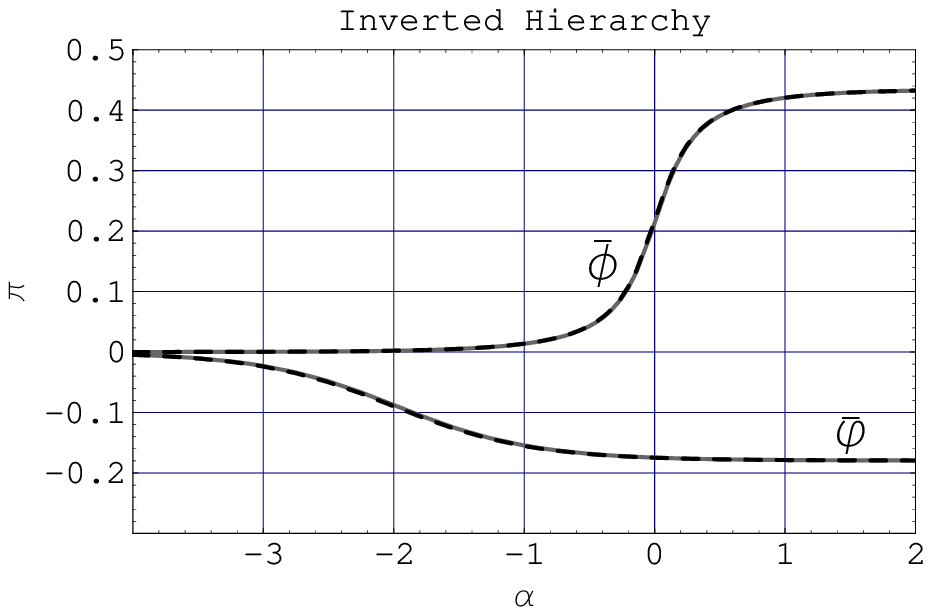}
\caption{The exact (gray solid line) and approximate (black dashed line) values of 
$\bar{\varphi}$ and $\bar{\phi}$ plotted against 
$\alpha = \log_{1/\varepsilon}(a/|\delta m^2_{31}|)$.
The parameter choice was $\tan^2\theta_{12} = 0.4$, $\sin^2(2\theta_{13}) = 0.16$,
$\delta m^2_{21}=8.2\times 10^{-5}\mathrm{eV}^2$ and
$|\delta m^2_{31}|=2.5\times 10^{-3}\mathrm{eV}^2$.}
\label{phibarfigure}
\end{center}
\end{figure}
%%%%%%%%%%%%%%%%%%%
%%%%%%%%%%%%%%%%%%%
\begin{figure}[p]
\begin{center}
\includegraphics[scale=0.75]{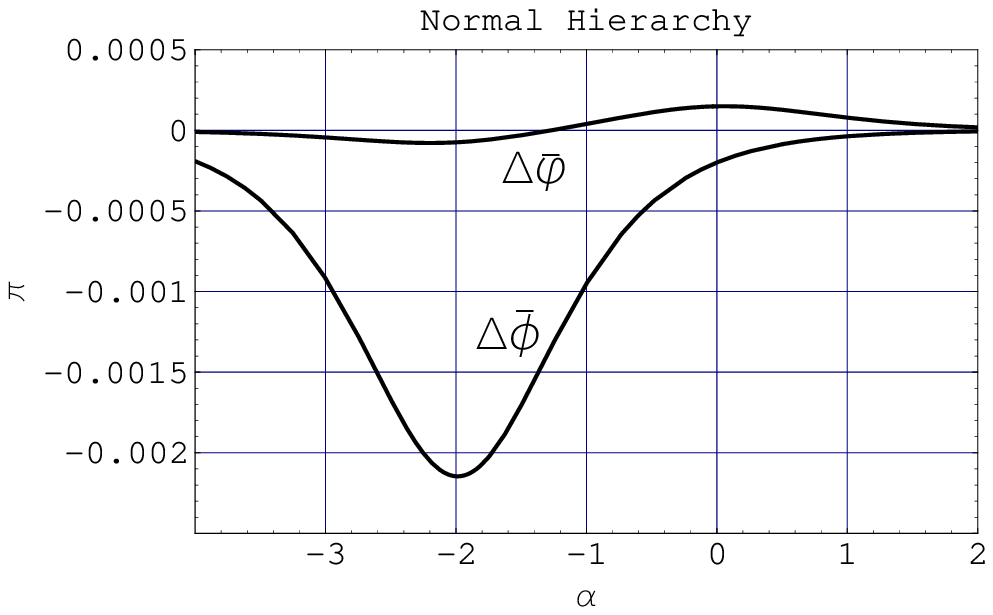}
\includegraphics[scale=0.75]{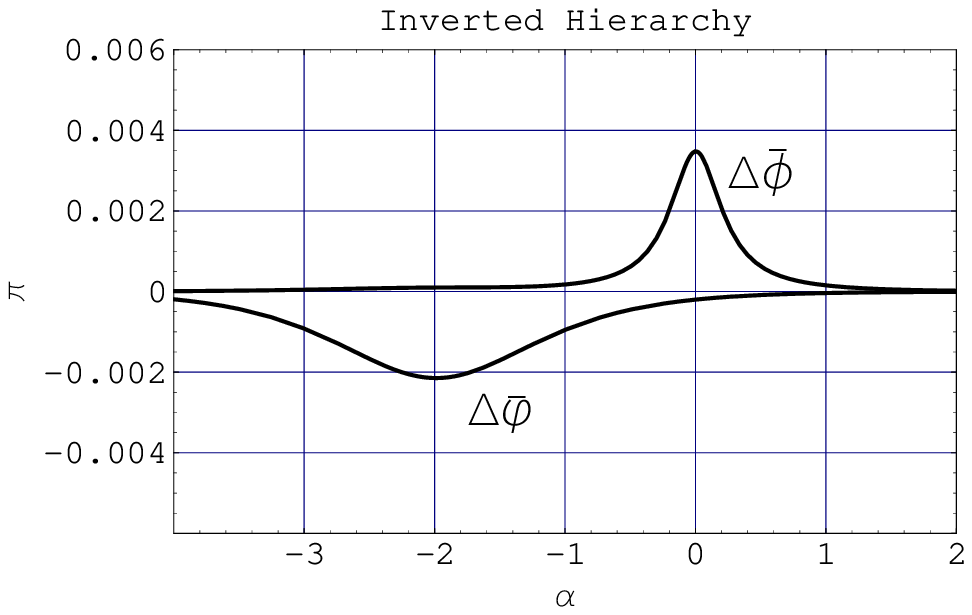}
\caption{$\Delta\bar{\varphi}=\bar{\varphi}_\mathrm{approx}-\bar{\varphi}_\mathrm{exact}$ and 
$\Delta\bar{\phi}=\bar{\phi}_\mathrm{approx}-\bar{\phi}_\mathrm{exact}$ plotted against 
$\alpha = \log_{1/\varepsilon}(a/|\delta m^2_{31}|)$ for the same parameter
choice as Fig.~\protect{\ref{phibarfigure}}.}
\label{Deltaphibarfigure}
\end{center}
\end{figure}
%%%%%%%%%%%%%%%%%%%

\begin{figure}[p]
\begin{center}
\includegraphics[scale=0.75]{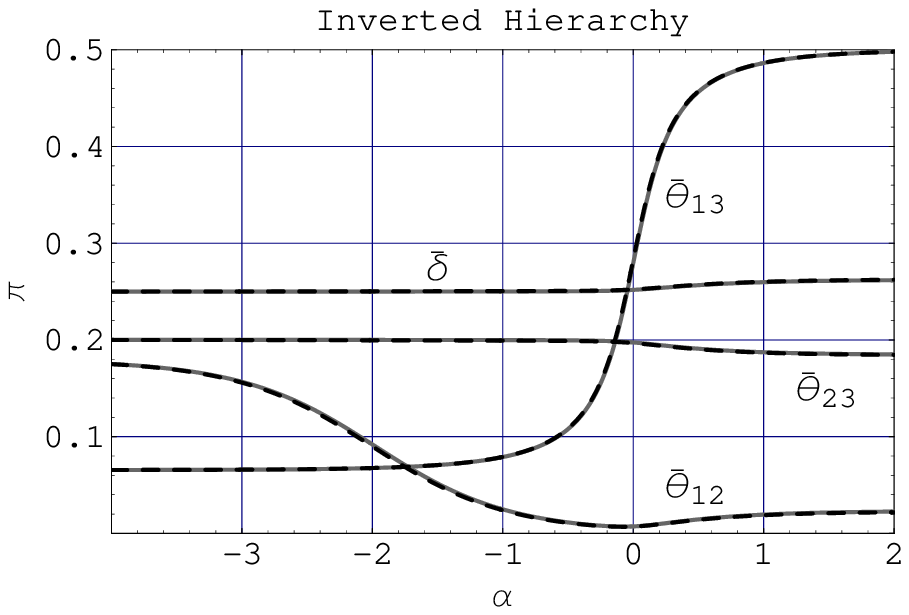}
\includegraphics[scale=0.75]{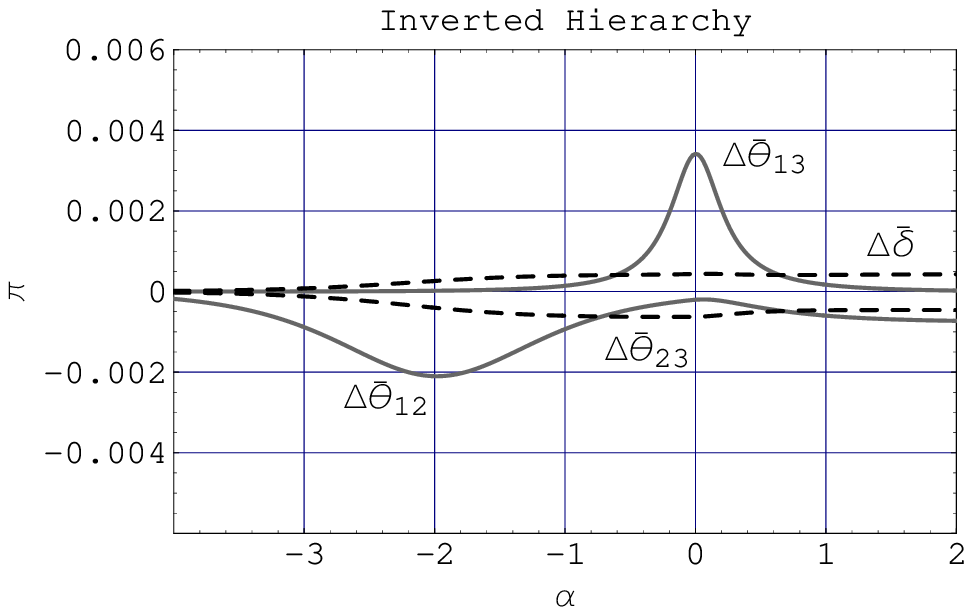}
\caption{(a) The exact values of $\edlit{\theta}_{12}$, $\edlit{\theta}_{13}$, $\edlit{\theta}_{23}$, and $\edlit{\delta}$ (gray solid lines) plotted against their approximate values (black dashed lines) obtained using Eq.~(\protect{\ref{edlittheta0}}), with
Eqs.~(\protect{\ref{phibarapprox}}) and (\protect{\ref{thetaprimebarapprox}}), as functions of
$\alpha=\log_{1/\varepsilon}(a/|\delta m^2_{31}|)$.
(b) The differences 
$\Delta\bar{\theta}_{12} = \edlit{\theta}_{12,\mathrm{approx}}-\edlit{\theta}_{12,\mathrm{exact}}$
and $\Delta\bar{\theta}_{13} = \edlit{\theta}_{13,\mathrm{approx}}-\edlit{\theta}_{13,\mathrm{exact}}$ (solid gray lines), and the differences
$\Delta\bar{\theta}_{23} = \edlit{\theta}_{23,\mathrm{approx}}-\edlit{\theta}_{23,\mathrm{exact}}$
and $\Delta\bar{\delta} = \edlit{\delta}_{\mathrm{approx}}-\edlit{\delta}_{\mathrm{exact}}$ (black dashed lines) of this approximation plotted against $\alpha$.}
\label{thetaTildebarInverted1}
\end{center}
\end{figure}

\begin{figure}[p]
\begin{center}
\includegraphics[scale=0.75]{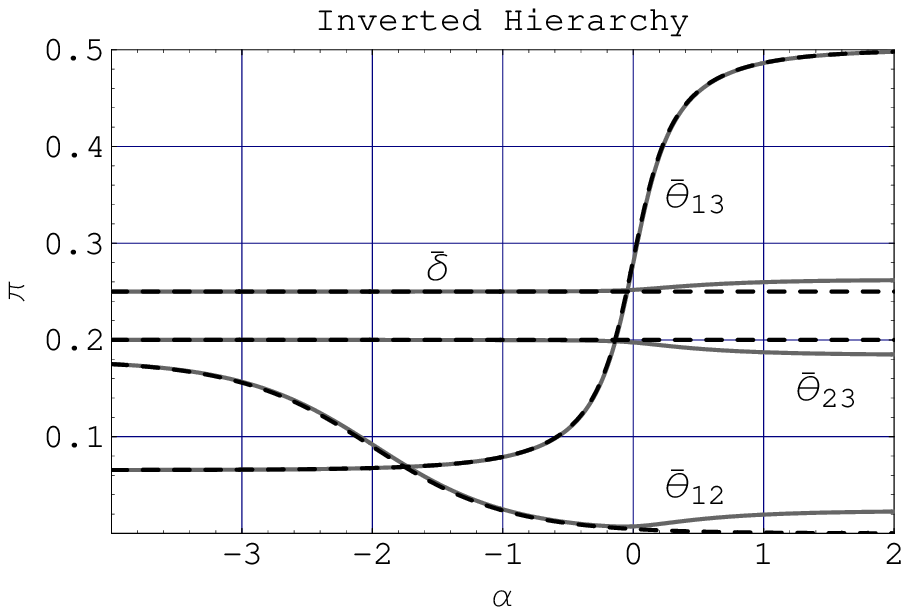}
\includegraphics[scale=0.75]{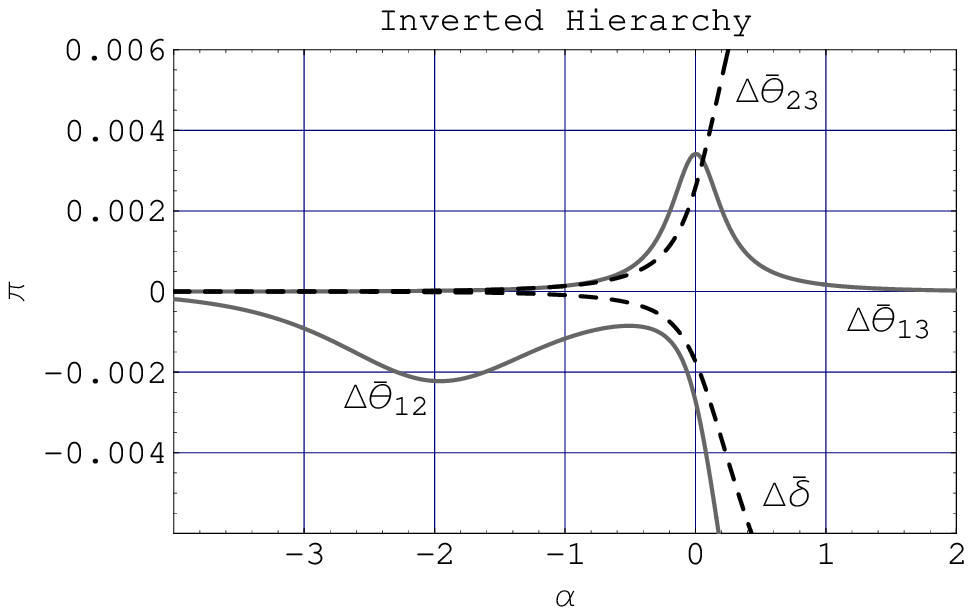}
\caption{(a) The exact values of $\edlit{\theta}_{12}$, $\edlit{\theta}_{13}$, $\edlit{\theta}_{23}$, and $\edlit{\delta}$ (gray solid lines) plotted against their approximate values (black dashed lines) obtained using Eq.~(\protect{\ref{edlittheta1}}) 
as functions of $\alpha=\log_{1/\varepsilon}(a/|\delta m^2_{31}|)$.
(b) $\Delta\bar{\theta}_{12}$ and $\Delta\bar{\theta}_{13}$ (gray solid lines), and
$\Delta\bar{\theta}_{23}$ and $\Delta\bar{\delta}$ (black dashed lines) of this approximation plotted as functions of $\alpha$. This approximation is applicable when $\alpha\alt -1$.}
\label{thetaTildebarInverted2}
\end{center}
\end{figure}

\begin{figure}[p]
\begin{center}
\includegraphics[scale=0.75]{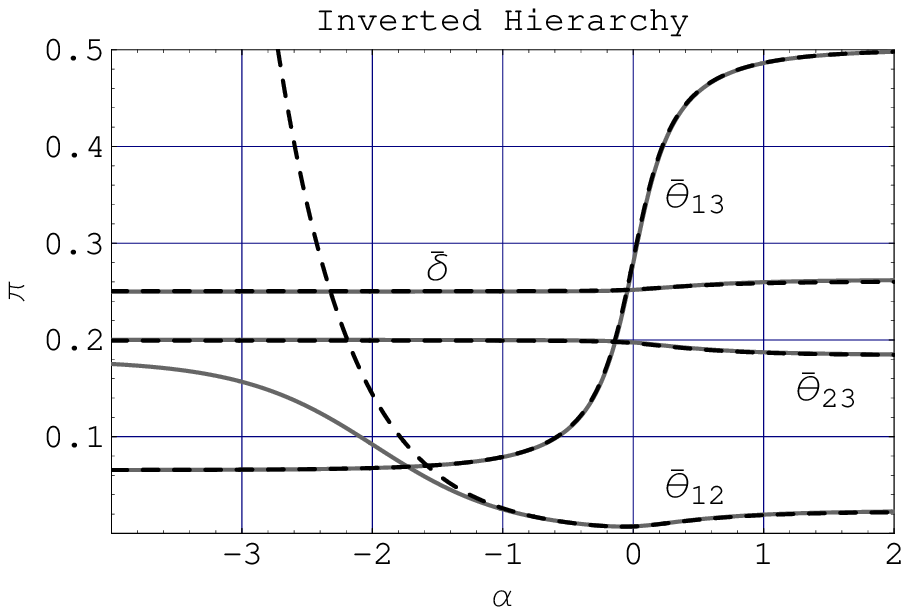}
\includegraphics[scale=0.75]{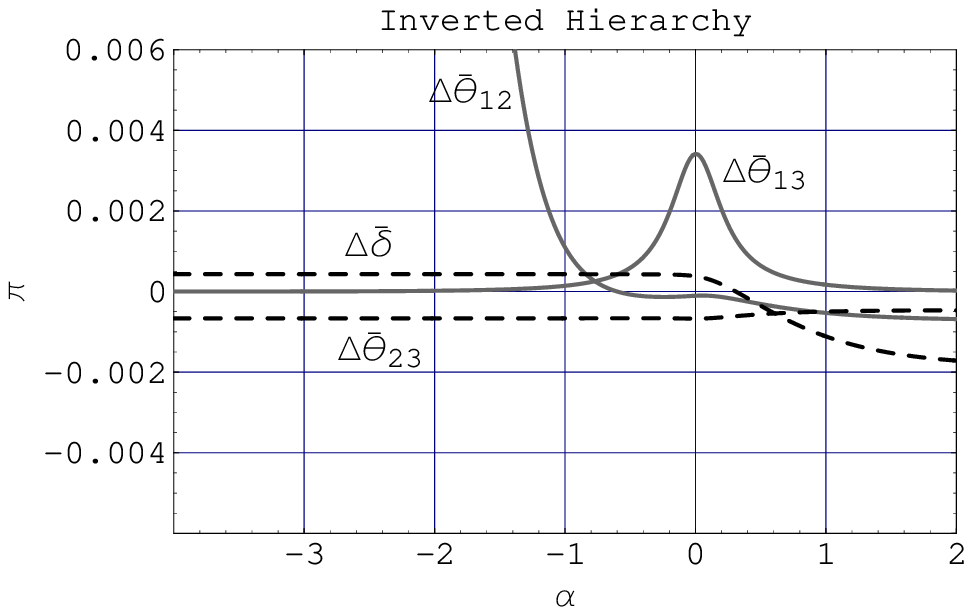}
\caption{(a) The exact values of $\edlit{\theta}_{12}$, $\edlit{\theta}_{13}$, $\edlit{\theta}_{23}$, and $\edlit{\delta}$ (gray solid lines) plotted against their approximate values (black dashed lines) obtained using Eq.~(\protect{\ref{edlittheta2}}) 
as functions of $\alpha=\log_{1/\varepsilon}(a/|\delta m^2_{31}|)$.
(b) $\Delta\bar{\theta}_{12}$ and $\Delta\bar{\theta}_{13}$ (gray solid lines), and
$\Delta\bar{\theta}_{23}$ and $\Delta\bar{\delta}$ (black dashed lines) of this approximation plotted as functions of $\alpha$. This approximation is applicable when $\alpha\agt 0$.}
\label{thetaTildebarInverted3}
\end{center}
\end{figure}

\begin{figure}[p]
\begin{center}
\includegraphics[scale=0.75]{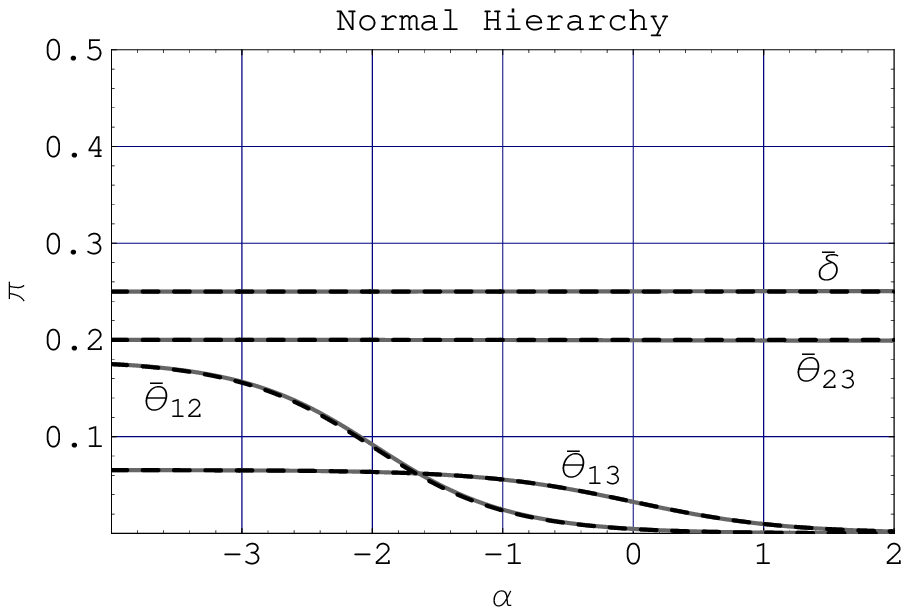}
\includegraphics[scale=0.75]{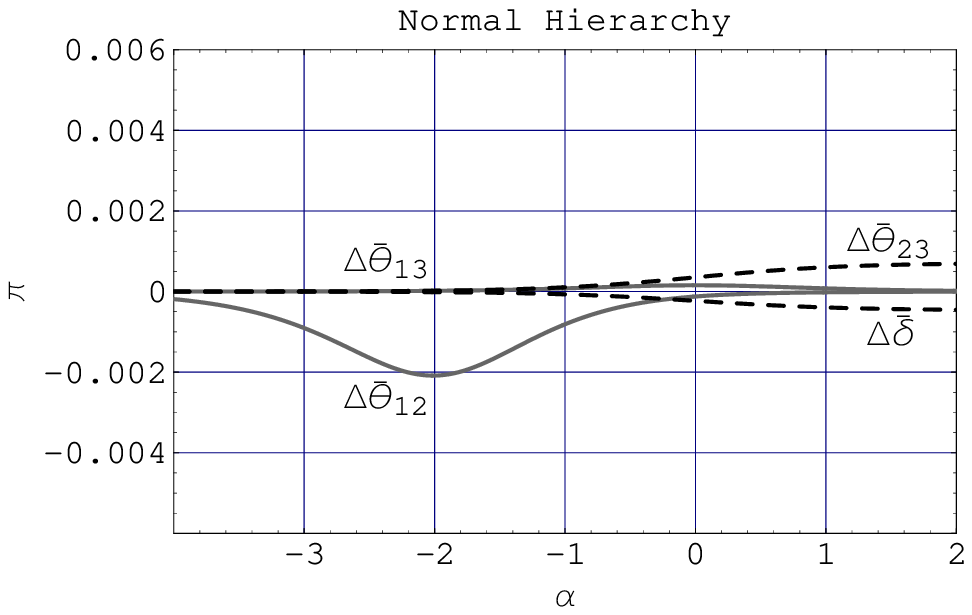}
\caption{(a) The exact values of $\edlit{\theta}_{12}$, $\edlit{\theta}_{13}$, $\edlit{\theta}_{23}$, and $\edlit{\delta}$ (gray solid lines) plotted against their approximate values (black dashed lines) obtained using Eq.~(\protect{\ref{edlittheta1}}) 
as functions of $\alpha=\log_{1/\varepsilon}(a/|\delta m^2_{31}|)$ for the normal
hierarchy case ($\delta m^2_{31}>0$).
(b) $\Delta\bar{\theta}_{12}$ and $\Delta\bar{\theta}_{13}$ (gray solid lines), and
$\Delta\bar{\theta}_{23}$ and $\Delta\bar{\delta}$ (black dashed lines) of this approximation plotted as functions of $\alpha$.}
\label{thetaTildebarNormal}
\end{center}
\end{figure}

The behavior of the angles $\bar{\varphi}$ and $\bar{\phi}$, defined in
Eqs.~(\ref{phi1bardef}) and (\ref{phi2bardef}), are plotted in Fig.~\ref{phibarfigure}
as functions of $\alpha = \log_{1/\varepsilon}(a/|\delta m^2_{31}|)$ 
with gray solid lines for the parameter choice of Eq.~(\ref{exampleparameterset}).
Approximate values can be obtained from
\begin{equation}
\tan 2\bar{\varphi} \approx 
-\dfrac{ a\sin 2\theta_{12} }{ \delta m^2_{21} + a\cos 2\theta_{12} }\;,\qquad
\tan 2\bar{\phi} \approx
-\dfrac{ a\sin 2\theta_{13} }{ \delta m^2_{31} + a\cos 2\theta_{13} }\;,
\label{phibarapprox}
\end{equation}
which are also shown in Fig.~\ref{phibarfigure} with black dashed lines.
The accuracy of this approximation is shown in Fig.~(\ref{Deltaphibarfigure}).
The approximate values of 
$\bar{\theta}_{12}' = \theta_{12}+\bar{\varphi}$ and 
$\bar{\theta}_{13}' = \theta_{13}+\bar{\phi}$ can be calculated from the expressions
\begin{equation}
\tan 2\bar{\theta}_{12}' =
\dfrac{ \delta m^2_{21} \sin 2\theta_{12} }{ \delta m^2_{21}\cos 2\theta_{12} + a }\;,\qquad
\tan 2\bar{\theta}_{13}' =
\dfrac{ \delta m^2_{31} \sin 2\theta_{13} }{ \delta m^2_{31}\cos 2\theta_{13} + a }\;.\qquad
\label{thetaprimebarapprox}
\end{equation}
The accuracy of Eqs.~(\ref{edlittheta0}), (\ref{edlittheta1}), and (\ref{edlittheta2}),
using Eqs.~(\ref{phibarapprox}) and (\ref{thetaprimebarapprox}) as input, is illustrated
in Figures~\ref{thetaTildebarInverted1} through \ref{thetaTildebarNormal} for the
parameter set of Eq.~(\ref{exampleparameterset}).

%%%%%%%%%%%%%%%%%%%
%%%%%%%%%%%%%%%%%%%
%%%%%%%%%%%%%%%%%%%
%%%%%%%%%%%%%%%%%%%
%%%%%%%%%%%%%%%%%%%
%%%%%%%%%%%%%%%%%%%
%%%%%%%%%%%%%%%%%%%
%%%%%%%%%%%%%%%%%%%

\begin{figure}[p]
\begin{center}
\includegraphics[scale=0.75]{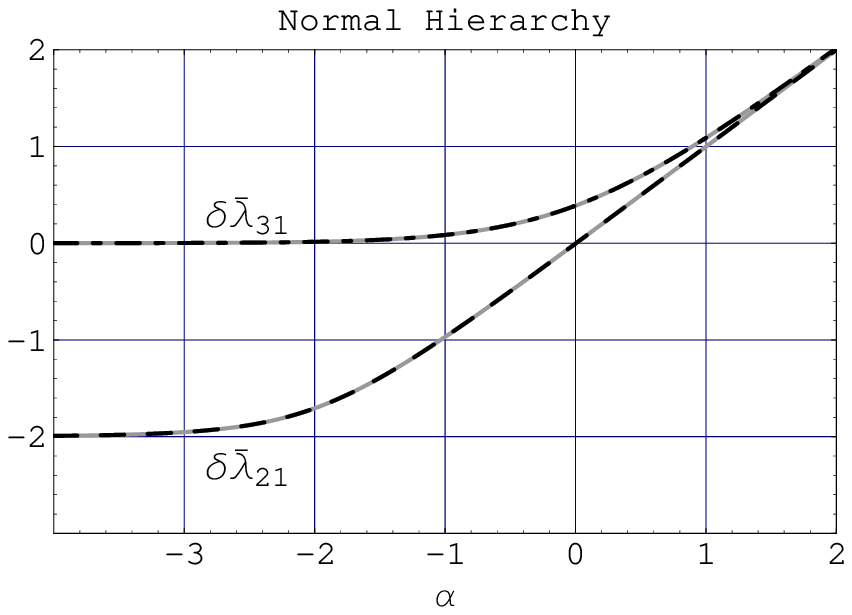}
\includegraphics[scale=0.75]{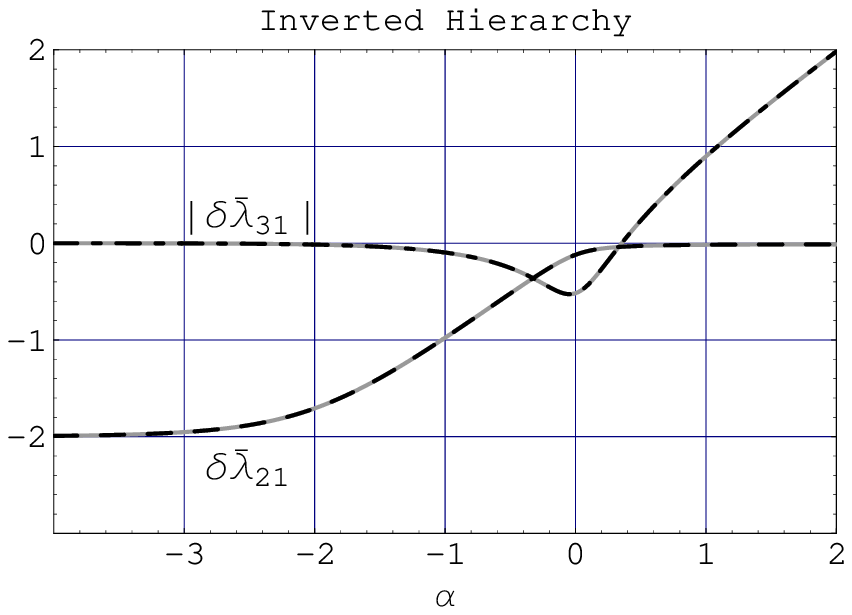}
\caption{The exact and approximate values of 
$\log_{1/\varepsilon}(\delta\bar{\lambda}_{21}/|\delta m^2_{31}|)$ and
$\log_{1/\varepsilon}(|\delta\bar{\lambda}_{31}|/|\delta m^2_{31}|)$ 
for the parameter set of Eq.~(\protect{\ref{exampleparameterset}})
plotted against $\alpha=\log_{1/\varepsilon}(a/|\delta m^2_{31}|)$.
The exact values are in gray solid lines, whereas the approximate values are
in black dashed ($\delta\bar{\lambda}_{21}$) 
and black dot-dashed ($|\delta\bar{\lambda}_{31}|$) lines.
}
\label{deltalambdabarXY1}
\end{center}
\end{figure}

\begin{figure}[p]
\begin{center}
\includegraphics[scale=0.75]{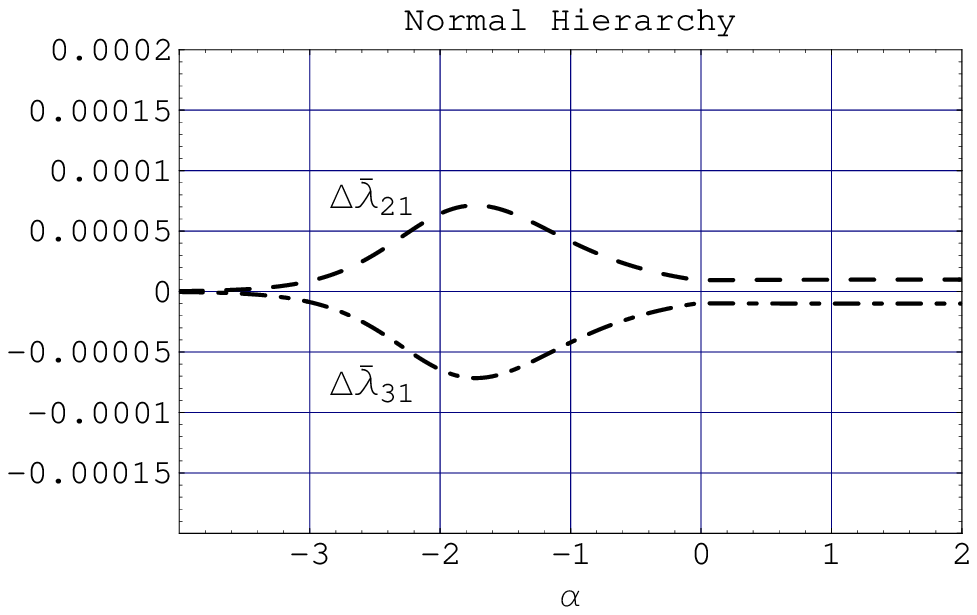}
\includegraphics[scale=0.75]{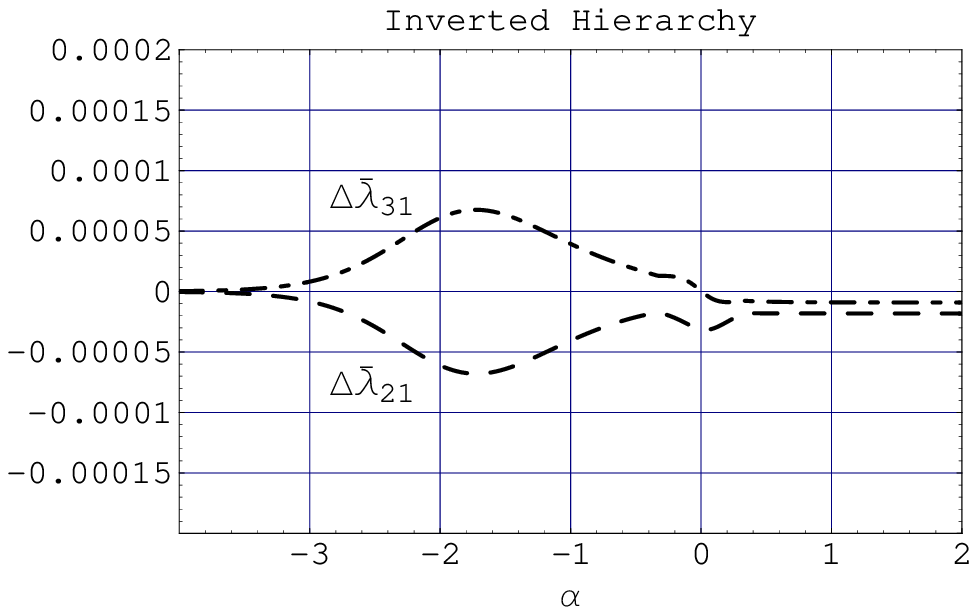}
\caption{The rescaled errors $\Delta\bar{\lambda}_{21}$ (dashed) and 
$\Delta\bar{\lambda}_{31}$ (dot-dashed), as 
defined in Eq.~(\protect{\ref{DeltalambdabarXYdef}}), for the approximation
of Fig.~\protect{\ref{deltalambdabarXY1}}.}
\label{errorlambdabarXY1}
\end{center}
\end{figure}

The approximate values for the effective mass-squared eigenvalues are given by
\begin{eqnarray}
\bar{\lambda}_1 & \approx & \bar{\lambda}''_{-}\;, \cr
\bar{\lambda}_2 & \approx & \bar{\lambda}'_{+} \;, \cr
\bar{\lambda}_3 & \approx & \bar{\lambda}''_{+}\;,
\end{eqnarray}
for the $\delta m^2_{31}>0$ case (normal hierarchy), and
\begin{eqnarray}
\bar{\lambda}_1 & \approx & \bar{\lambda}''_{+}\;, \cr
\bar{\lambda}_2 & \approx & \bar{\lambda}'_{+} \;, \cr
\bar{\lambda}_3 & \approx & \bar{\lambda}''_{-}\;,
\end{eqnarray}
for the $\delta m^2_{31}<0$ case (inverted hierarchy),
where $\bar{\lambda}'_\pm$ and $\bar{\lambda}''_\pm$ are defined in
Eqs.~(\ref{lambdaprimebardef}) and (\ref{lambdadoubleprimebardef}),
respectively.
The accuracy of this approximation is illustrated in 
Fig~\ref{deltalambdabarXY1} using the parameter values of 
Eq.~(\ref{exampleparameterset}), where
the exact numerically calculated values of 
$\delta\bar{\lambda}_{21} = \bar{\lambda}_2 - \bar{\lambda}_1$ and
$|\delta\bar{\lambda}_{31}| = |\bar{\lambda}_3 - \bar{\lambda}_1|$ are plotted against those
obtained from the above approximate expressions.
%The vertical axis is plotted using the same log-scale as the horizontal axis
%where $|\delta m^2_{31}|$ corresponds to $0$ and
%$\delta m^2_{21}$ corresponds to $-2$.
In Fig.~\ref{errorlambdabarXY1}, we plot the difference between the
exact and approximate values of $\delta\bar{\lambda}_{ij}$ normalized to
$\delta\bar{\lambda}_{\mathrm{min,exact}}$:
\begin{eqnarray}
\Delta\bar{\lambda}_{i1} \equiv 
\dfrac{ \delta\bar{\lambda}_{i1,\mathrm{approx}}-\delta\bar{\lambda}_{i1,\mathrm{exact}} }
      { \delta\bar{\lambda}_{\mathrm{min,exact}} }\;,\qquad
(i=2,3)\;,
\label{DeltalambdabarXYdef}
\end{eqnarray}
where
\begin{equation}
\delta\bar{\lambda}_{\mathrm{min,exact}}
\equiv \min(\delta\bar{\lambda}_{21,\mathrm{exact}},
           |\delta\bar{\lambda}_{31,\mathrm{exact}}|,
           |\delta\bar{\lambda}_{32,\mathrm{exact}}|)\;.
\end{equation}

\begin{figure}[p]
\begin{center}
\includegraphics[scale=0.75]{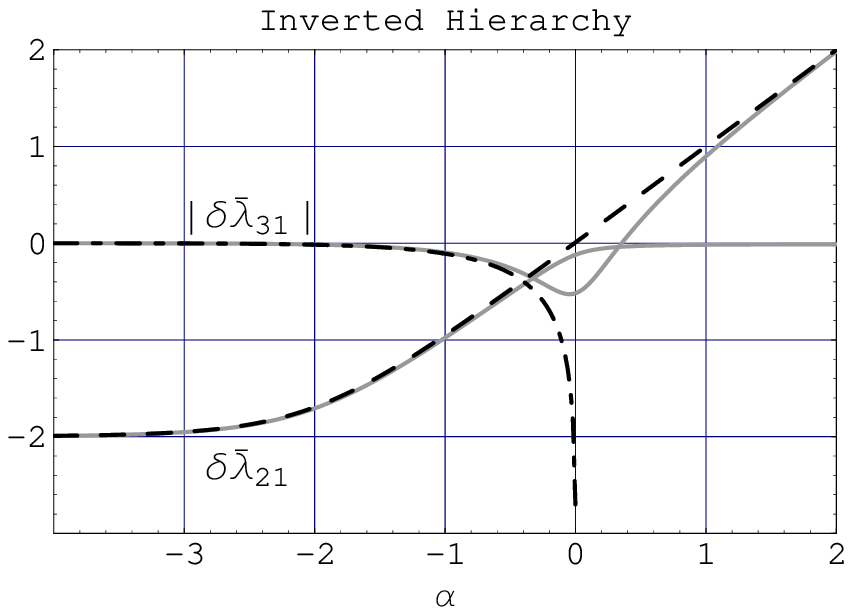}
\includegraphics[scale=0.75]{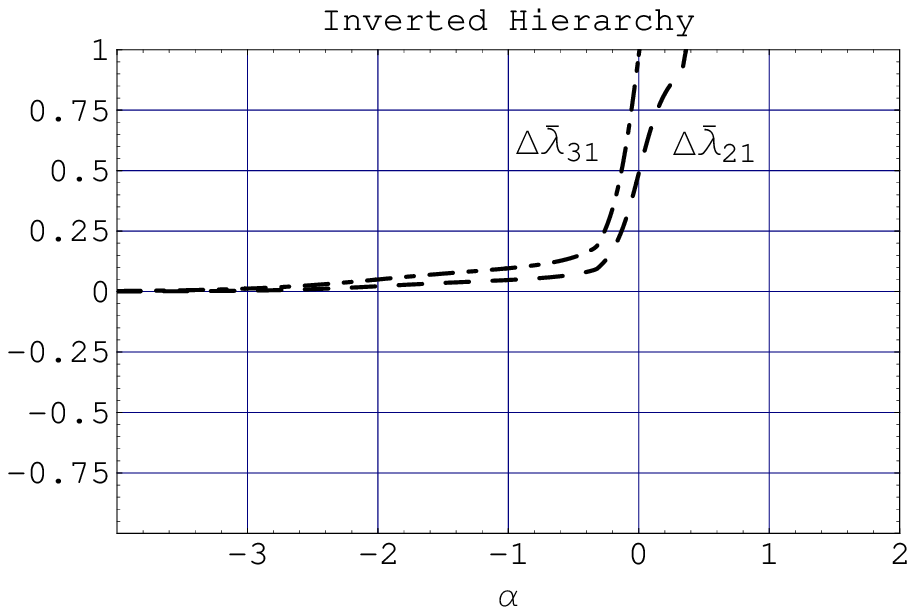}
\caption{Comparison of exact and approximate values using 
Eq.~(\protect{\ref{lambdabarapprox2}}) for the inverted hierarchy case.
The approximation is applicable when $\alpha\alt -1$.}
\label{deltalambdabarXYinverted2}
\end{center}
\end{figure}

\begin{figure}[p]
\begin{center}
\includegraphics[scale=0.75]{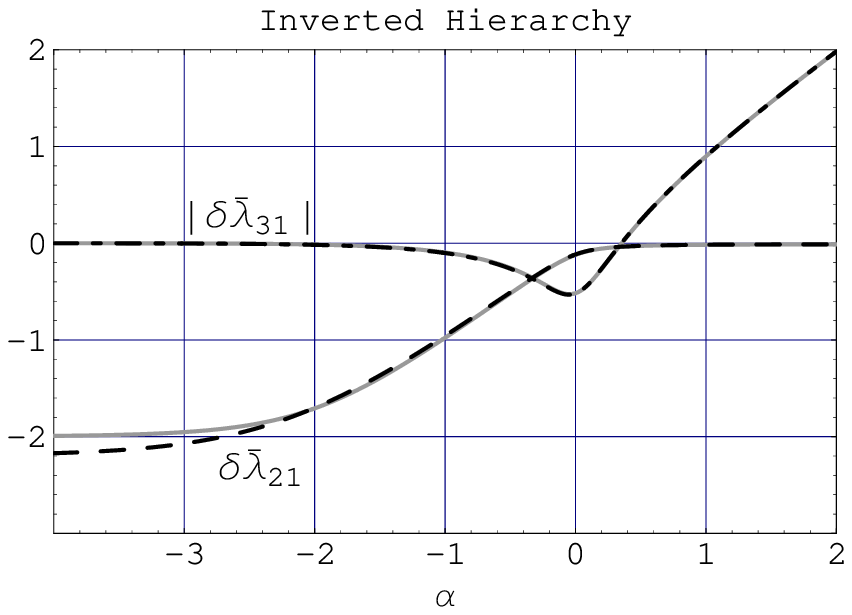}
\includegraphics[scale=0.75]{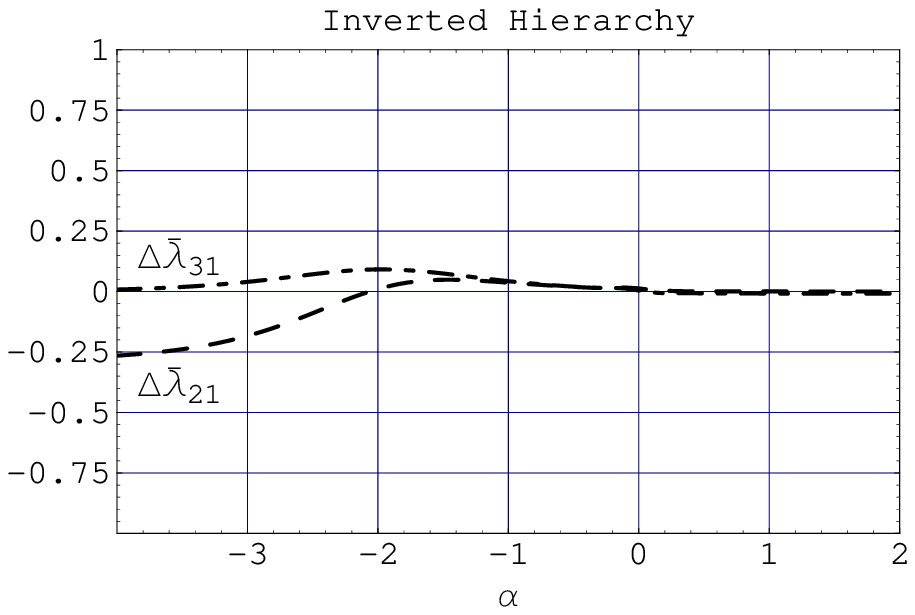}
\caption{Comparison of exact and approximate values using 
Eq.~(\protect{\ref{lambdabarapprox3}}) for the inverted hierarchy case.
The approximation is applicable when $\alpha\agt -1$.}
\label{deltalambdabarXYinverted3}
\end{center}
\end{figure}

\begin{figure}[p]
\begin{center}
\includegraphics[scale=0.75]{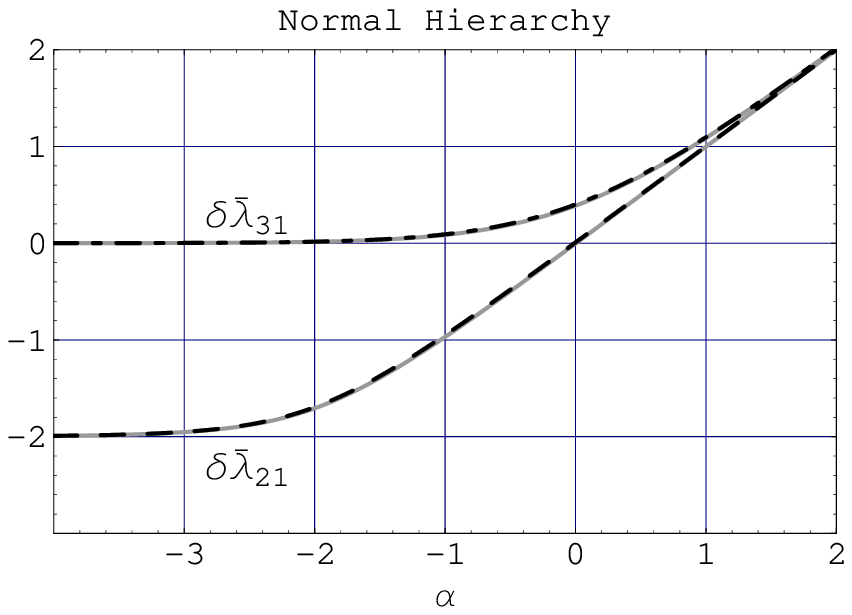}
\includegraphics[scale=0.75]{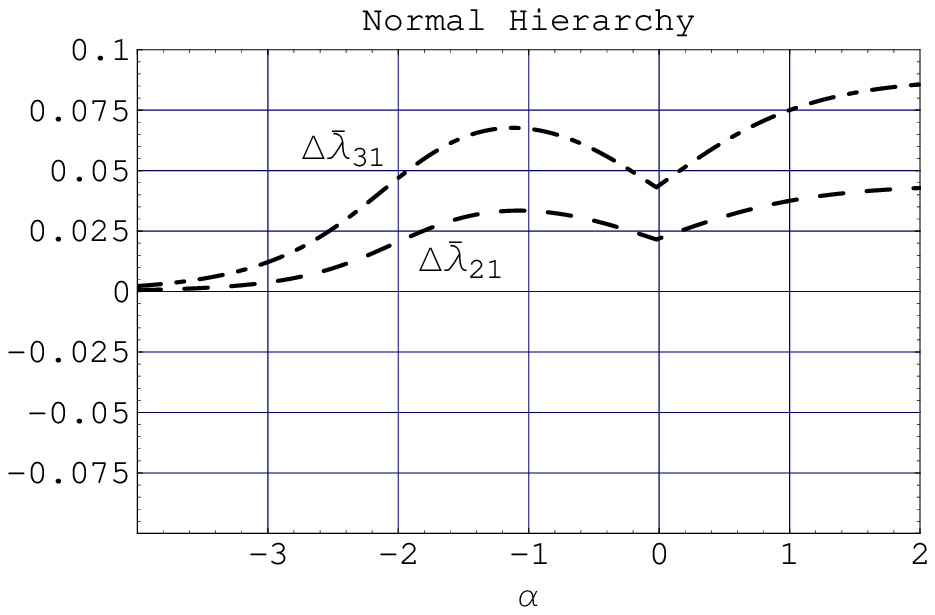}
\caption{Comparison of exact and approximate values using 
Eq.~(\protect{\ref{lambdabarapprox2}}) for the normal hierarchy case.}
\label{deltalambdabarXYnormal2}
\end{center}
\end{figure}

When either $\delta m^2_{31}<0$ (inverted hierarchy) with $\alpha \alt -1$, \textit{i.e.} $a/|\delta m^2_{31}| \le O(\varepsilon)$, or $\delta m^2_{31}>0$ (normal hierarchy) with any $a$, the $\bar{\lambda}$'s can be further approximated by
\begin{eqnarray}
\bar{\lambda}_1 & \approx & 
\dfrac{ (\delta m^2_{21} - a)
         -\sqrt{ (\delta m^2_{21} + a)^2 - 4 a\, \delta m^2_{21} s_{12}^2 }
      }
      { 2 }\;,\cr
\bar{\lambda}_2 & \approx & 
\dfrac{ (\delta m^2_{21} - a)
         +\sqrt{ (\delta m^2_{21} + a)^2 - 4 a\, \delta m^2_{21} s_{12}^2 }
      }
      { 2 }\;,\cr
\bar{\lambda}_3 & \approx & \delta m^2_{31}\;.
\label{lambdabarapprox2}
\end{eqnarray}
For the $\delta m^2_{31}<0$ (inverted hierarchy) case with $\alpha \agt -1$, \textit{i.e.}
$a/|\delta m^2_{31}| \ge O(\varepsilon)$, we can use
\begin{eqnarray}
\bar{\lambda}_1 & \approx & 
\dfrac{ (\delta m^2_{31} - a)
        +\sqrt{(\delta m^2_{31} + a)^2-4a\,\delta m^2_{31} s_{13}^2}
      }{ 2 } \;,\cr
\bar{\lambda}_2 & \approx & \delta m^2_{21} c_{12}^2 \;,\cr
\bar{\lambda}_3 & \approx & 
\dfrac{ (\delta m^2_{31} - a)
        -\sqrt{(\delta m^2_{31} + a)^2-4a\,\delta m^2_{31} s_{13}^2}
                   }{ 2 }\;.
\label{lambdabarapprox3}
\end{eqnarray}
%
%These approximations introduce errors of $O(\varepsilon^2 |\delta m^2_{31}|)$ 
%in the $\bar{\lambda}$'s when $a/|\delta m^2_{31}| \ge O(1)$.
%However, since $\delta\bar{\lambda_{21}} = O(|\delta m^2_{31}|)$ in that range,
%an error of this size is of little consequence.
The accuracy of these approximations is illustrated in
Figs.~\ref{deltalambdabarXYinverted2} and \ref{deltalambdabarXYinverted3} for the 
$\delta m^2_{31}<0$ case, and Fig.~\ref{deltalambdabarXYnormal2} for the
$\delta m^2_{31}>0$ case.

%%%%%%%%%%%%%%%%%%%
%%%%%%%%%%%%%%%%%%%
%%%%%%%%%%%%%%%%%%%
%%%%%%%%%%%%%%%%%%%
%%%%%%%%%%%%%%%%%%%
%%%%%%%%%%%%%%%%%%%
%%%%%%%%%%%%%%%%%%%
%%%%%%%%%%%%%%%%%%%
%%%%%%%%%%%%%%%%%%%
%%%%%%%%%%%%%%%%%%%

%%%%%%%%%%%%%%%%%%%%%%%%%%%%%%%%%%%%%%%%%%%%%%%%%%%%%%%%%%
%%%%%%%%%%%%%%%%%%%%%%%%%%%%%%%%%%%%%%%%%%%%%%%%%%%%%%%%%%
\newpage
\section{Sample Calculation of Oscillation Probabilities}

%%%%%%%%%%%%%%%%%%%%%%%%%%%%%%%%%
\begingroup
\squeezetable
\begin{table}[ht]
\begin{tabular}{|r||c|c|c|c|c|c|}
\hline
& $\;L\;\mathrm{(km)}\;$ 
& $\;\rho\;\mathrm{(g/cm^3)}\;$ 
& $\;E\;\mathrm{(GeV)}\;$ 
& $\alpha = \log_{1/\varepsilon}(a/|\delta m^2_{31}|)$ 
& $|\Delta_{31}|$ 
& Reference \\
\hline\hline
\ T2K (JPARC $\rightarrow$ Super-K)\ \ & $\phantom{0}295$ & $2.6$ & $\;0.25\sim 2\;$ & $\;-2.3\sim -1.1\;$ & $\;(0.3\sim 2.4)\pi\;$ & \protect{\cite{T2K}} \\
\hline
\ JPARC $\rightarrow$ Korea\ \ & $1000$ & $2.7$ & $\;1\sim \phantom{1}6\;$    & $\;-1.5\sim -0.4\;$ & $\;(0.3\sim 2.0)\pi\;$ & \protect{\cite{IKMN,HagiwaraOkamuraSenda}} \\
\hline
\ BNL $\rightarrow$ Home Stake\ \ & $2540$ & $3.4$ & $\;2\sim 10\;$   & $\;-0.9\sim \phantom{-}0\phantom{.0}\;$    & $\;(0.5\sim 2.6)\pi\;$ & \protect{\cite{Diwan}} \\
\hline
\end{tabular}
\caption{The three cases for which we calculate the probabilities
for the processes $\nu_\mu\rightarrow \nu_\mu$ and $\nu_\mu\rightarrow \nu_e$.
The ranges of $\alpha$ and $|\Delta_{31}|$ were calculated assuming 
$\delta m^2_{21}=8.2\times 10^{-5}\,\mathrm{eV^2}$ and
$|\delta m^2_{31}|=2.5\times 10^{-3}\,\mathrm{eV^2}$.
$\rho$ is the average matter density along the baseline calculated using
the Preliminary Earth Reference Model (PREM) \protect{\cite{PREM}}. 
}
\label{ThreeCases}
\end{table}
\endgroup
%%%%%%%%%%%%%%%%%%%%%%%%%%%%%%%%%%%

The accuracy of our approximation in calculating the effective mixing angles
and effective mass-squared differences translates directly into the accuracy in
calculating the oscillation probabilities.
To illustrate this, we calculate the probabilities for
the processes $\nu_\mu\rightarrow \nu_\mu$ and $\nu_\mu\rightarrow \nu_e$
for the three cases listed in Table~\ref{ThreeCases}.
The energy ranges listed include the energies at which $|\Delta_{31}|=\pi$, around
where the first oscillation peak occurs. 
The ranges of the matter effect parameter $\alpha = \log_{1/\varepsilon}(a/|\delta m^2_{31}|)$
for the three cases are roughly $-2\sim -1$, $-1.5\sim -0.5$, and $-1\sim 0$,
so together they cover the range $-2\sim 0$.

Since $\alpha < 0$ for all three cases, the effective mixing angles are well
approximated by Eq.~(\ref{tildetheta1}).
For the effective mass-squared differences, 
we use Eq.~(\ref{lambdaapprox2}) which is applicable to $\alpha\alt -1$ for the $L=295\,\mathrm{km}$ case, and Eq.~(\ref{lambdaapprox3}) which is applicable
to $\alpha\agt -1$ for the $L=2540\,\mathrm{km}$ case.

The $L=1000\,\mathrm{km}$ case is a bit problematic
since neither Eq.~(\ref{lambdaapprox2}) nor (\ref{lambdaapprox3}) can be used
throughout the entire range $\alpha=-1.5\sim -0.5$, 
Eq.~(\ref{lambdaapprox2}) being applicable only
to the low energy end, and Eq.~(\ref{lambdaapprox3}) being applicable only to the high energy end.
Using Eqs.~(\ref{lambdaapprox1normal}) and (\ref{lambdaapprox1inverted}), 
with Eqs.~(\ref{lambdaprimeplusminusdef}) and (\ref{lambdadoubleprimeplusminusdef}),
which are applicable to all energies, would solve our problem and lead to 
approximate oscillation probabilities that are virtually indistinguishable from 
their exact values.
However, we would like to illustrate the power (and limitations) of our
much simpler expressions in Eqs.~(\ref{lambdaapprox2}) and (\ref{lambdaapprox3}).
Here, we will use Eq.~(\ref{lambdaapprox2}) 
since the first oscillation peak occurs towards the lower end of the energy range.

%%%%%%%%%%%%%%%%%%%%%%%%%%%%%%%%%%%%%%%%%%%%%%%%%
\begin{turnpage}
%%%%%%%%%%%%%%%%%%%
\begin{figure}[p]
\begin{center}
\includegraphics[scale=1.0]{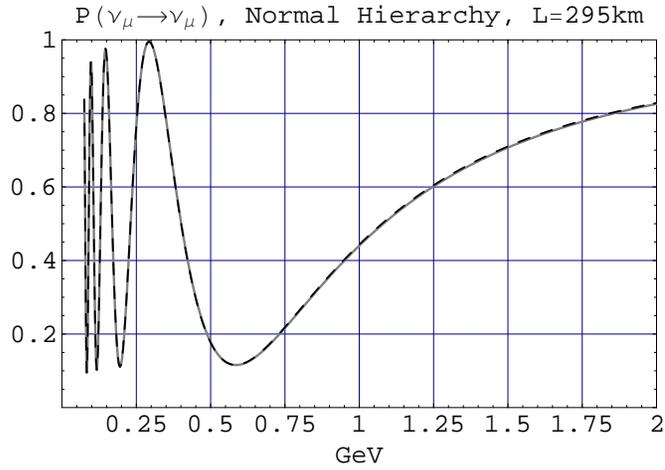}
\includegraphics[scale=1.0]{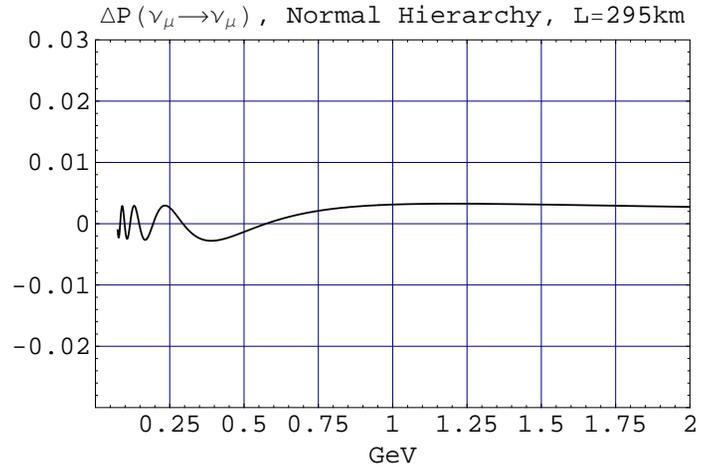}
\includegraphics[scale=1.0]{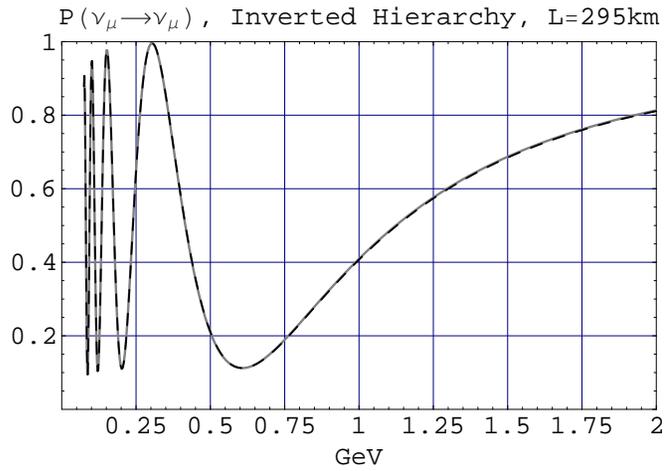}
\includegraphics[scale=1.0]{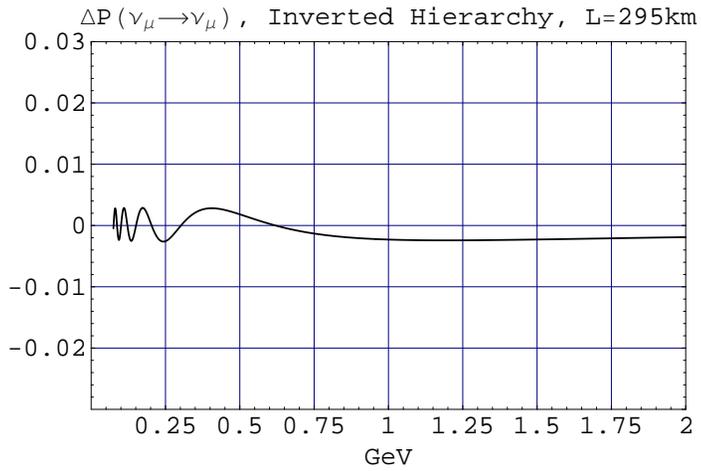}
\caption{Comparison of exact (solid gray line) and approximate (black dashed line) 
values of $P(\nu_\mu\rightarrow\nu_\mu)$ for the $L=295\,\mathrm{km}$ case.
The approximate value was calculated using 
Eq.~(\protect{\ref{tildetheta1}}) for the mixing angles, and
Eq.~(\protect{\ref{lambdaapprox2}}) for the mass-squared differences.
The CP violating phase $\delta$ was set to zero.
The difference $\Delta P \equiv P_\mathrm{approx}-P_\mathrm{exact}$ is plotted on the
right.}
\label{Pmu2mu295}
\end{center}
\end{figure}
%%%%%%%%%%%%%%%%%%%
%%%%%%%%%%%%%%%%%%%
\begin{figure}[p]
\begin{center}
\includegraphics[scale=1.0]{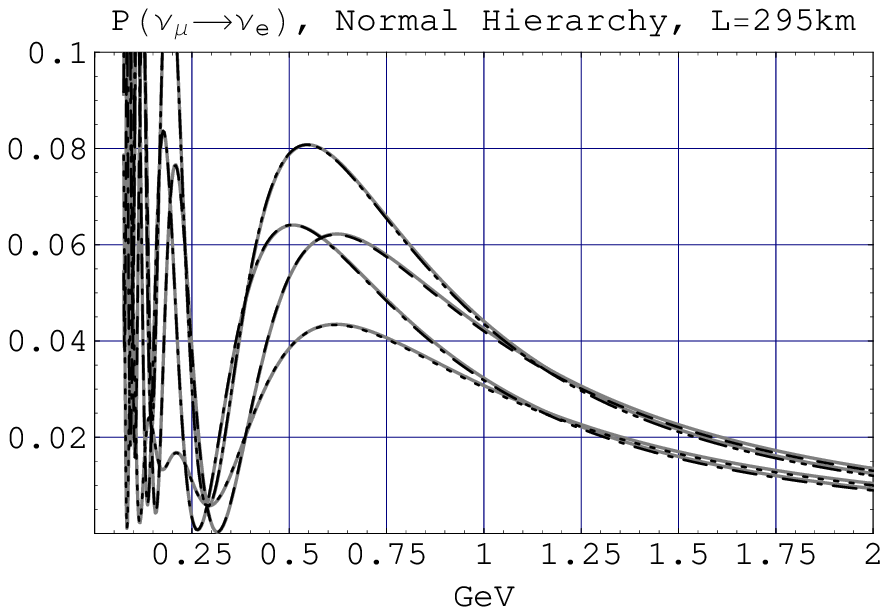}
\includegraphics[scale=1.0]{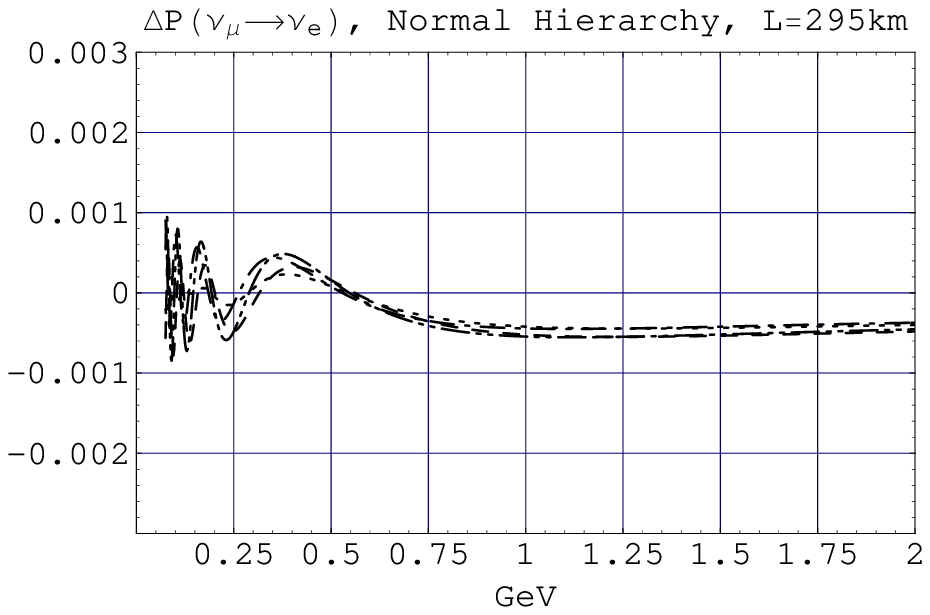}
\includegraphics[scale=1.0]{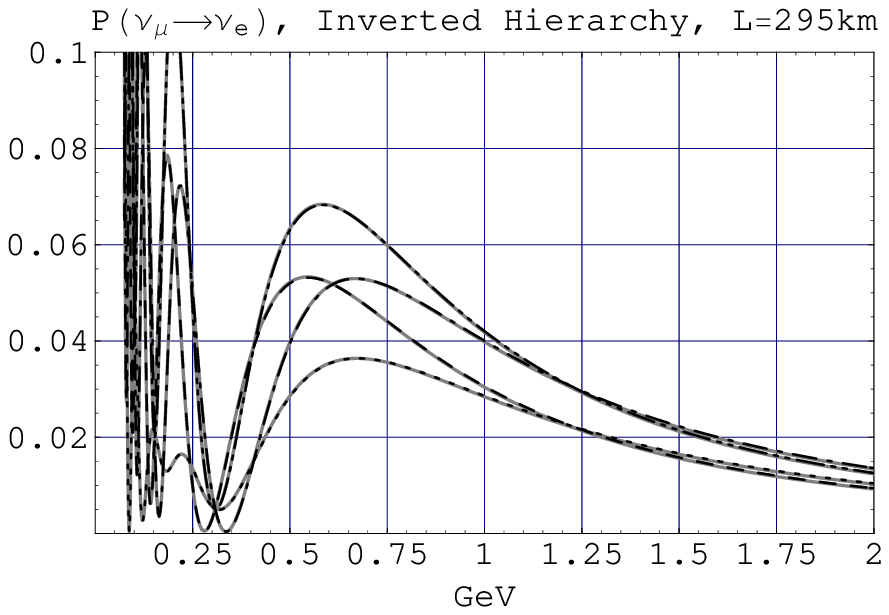}
\includegraphics[scale=1.0]{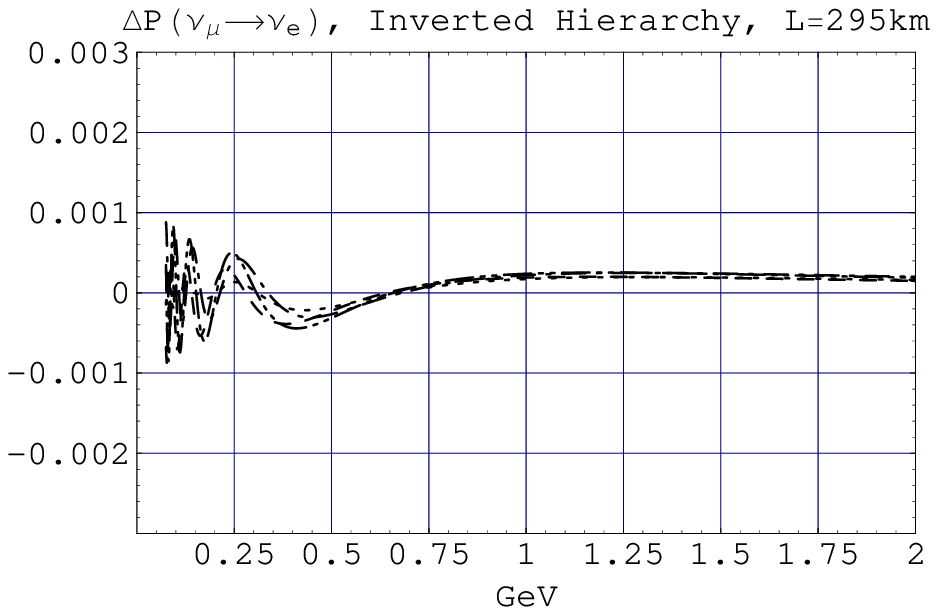}
\caption{Comparison of exact and approximate values of
$P(\nu_\mu\rightarrow\nu_e)$ for the $L=295\,\mathrm{km}$ case 
for several different values of the CP violating phase $\delta$.
The approximate values were calculated using 
Eq.~(\protect{\ref{tildetheta1}}) for the mixing angles, and
Eq.~(\protect{\ref{lambdaapprox2}}) for the mass-squared differences.
The exact values are given by the solid gray lines, while the approximate values are
the black dashed ($\delta=0$), dotted ($\delta=\pi/2$),
dot-dashed ($\delta=\pi$), and double-dot-dashed ($\delta=3\pi/2$) lines.}
\label{Pmu2e295}
\end{center}
\end{figure}
%%%%%%%%%%%%%%%%%%%
%%%%%%%%%%%%%%%%%%%
\begin{figure}[p]
\begin{center}
\includegraphics[scale=1.0]{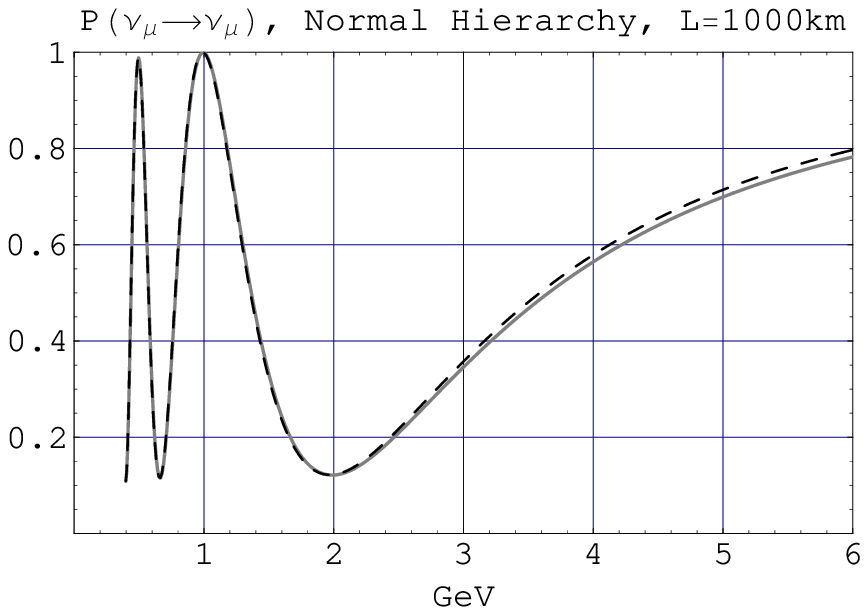}
\includegraphics[scale=1.0]{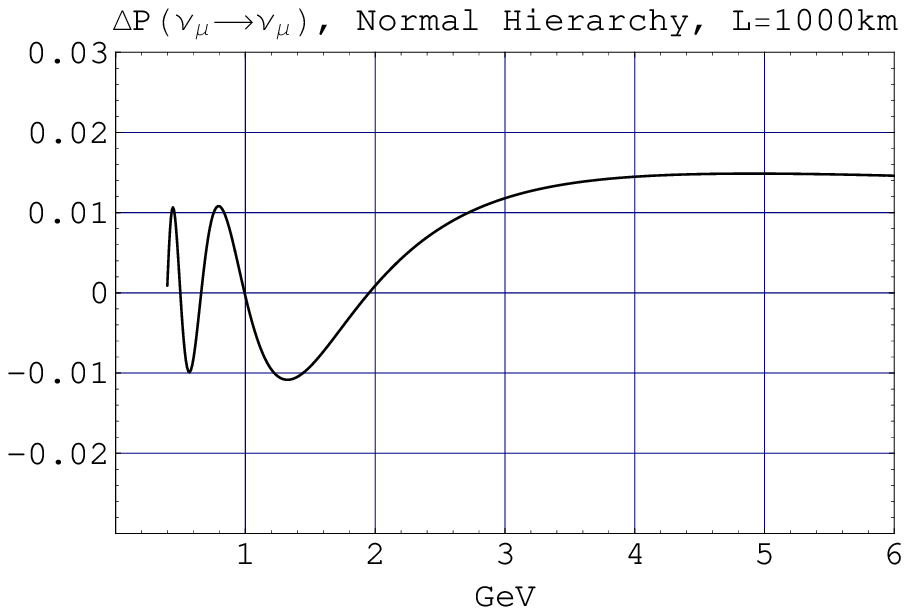}
\includegraphics[scale=1.0]{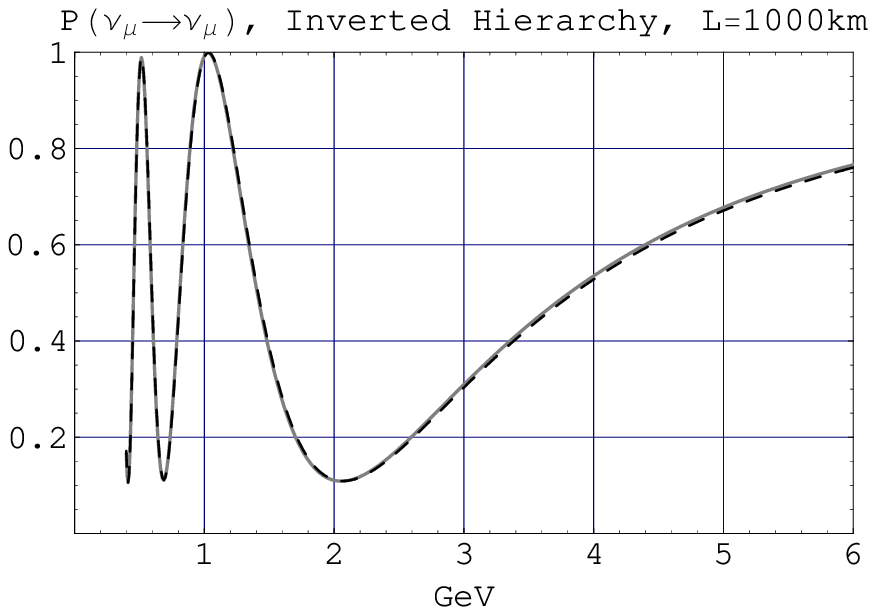}
\includegraphics[scale=1.0]{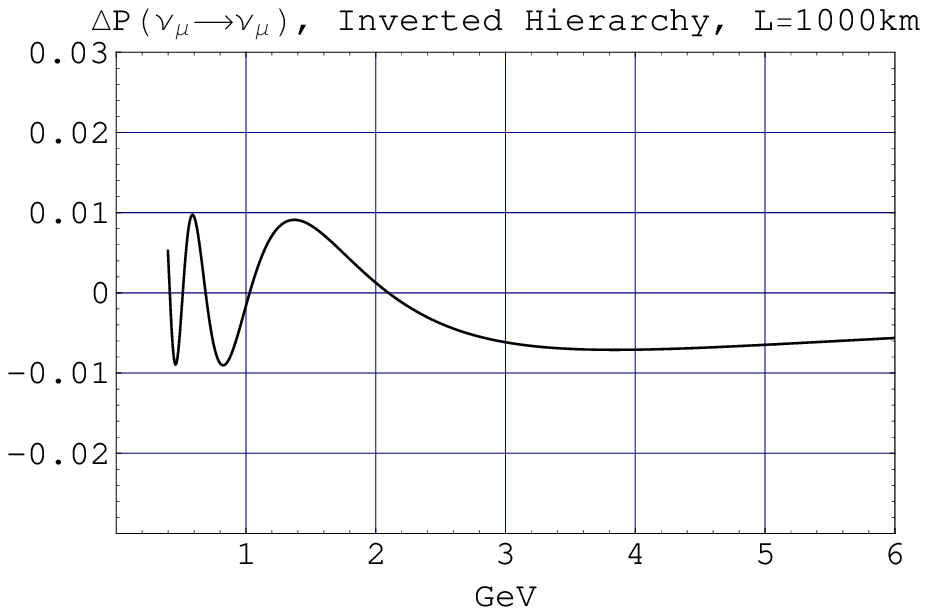}
\caption{Comparison of exact (solid gray line) and approximate (black dashed line) 
values of $P(\nu_\mu\rightarrow\nu_\mu)$ for the $L=1000\,\mathrm{km}$ case.
The approximate value was calculated using 
Eq.~(\protect{\ref{tildetheta1}}) for the mixing angles, and
Eq.~(\protect{\ref{lambdaapprox2}}) for the mass-squared differences.
The CP violating phase $\delta$ was set to zero.
The difference $\Delta P \equiv P_\mathrm{approx}-P_\mathrm{exact}$ is plotted on the
right.}
\label{Pmu2mu1000}
\end{center}
\end{figure}
%%%%%%%%%%%%%%%%%%%
%%%%%%%%%%%%%%%%%%%
\begin{figure}[p]
\begin{center}
\includegraphics[scale=1.0]{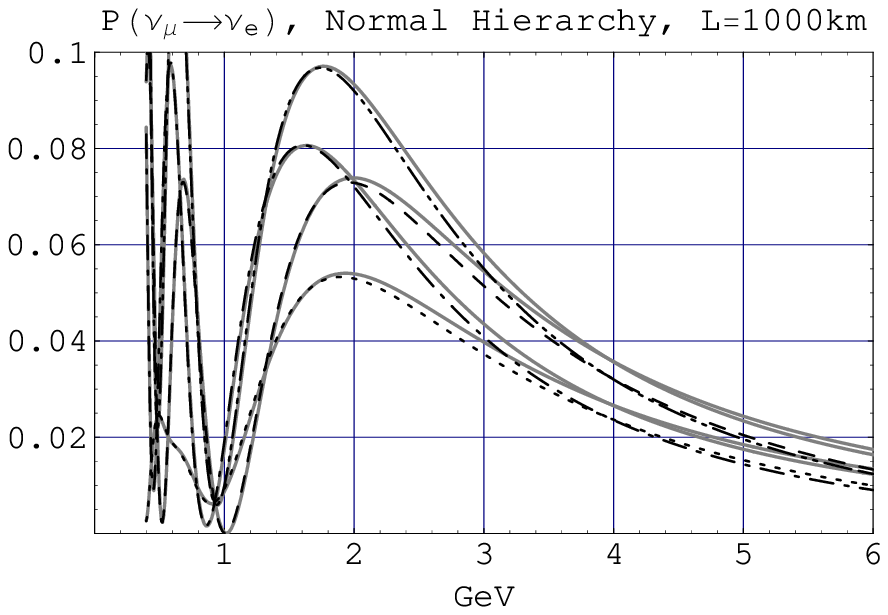}
\includegraphics[scale=1.0]{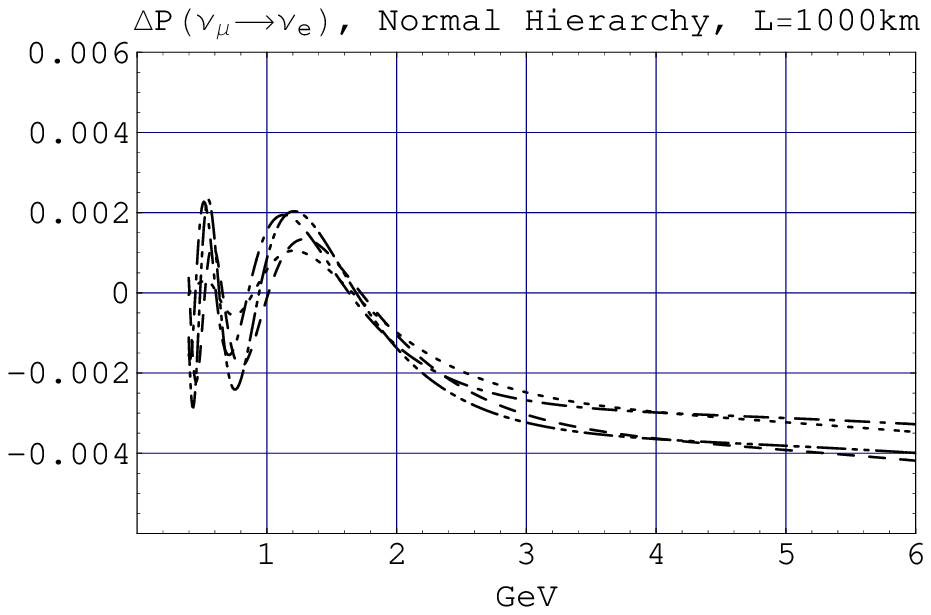}
\includegraphics[scale=1.0]{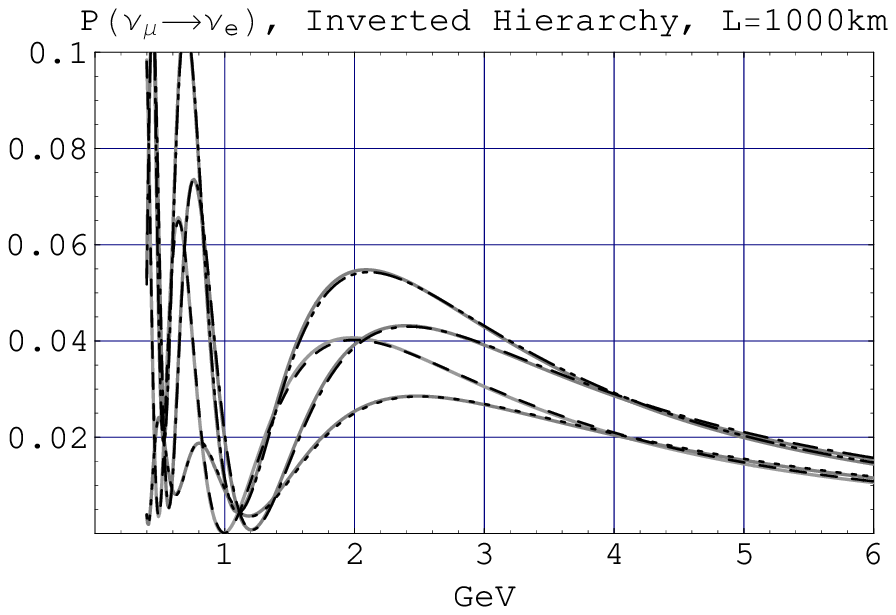}
\includegraphics[scale=1.0]{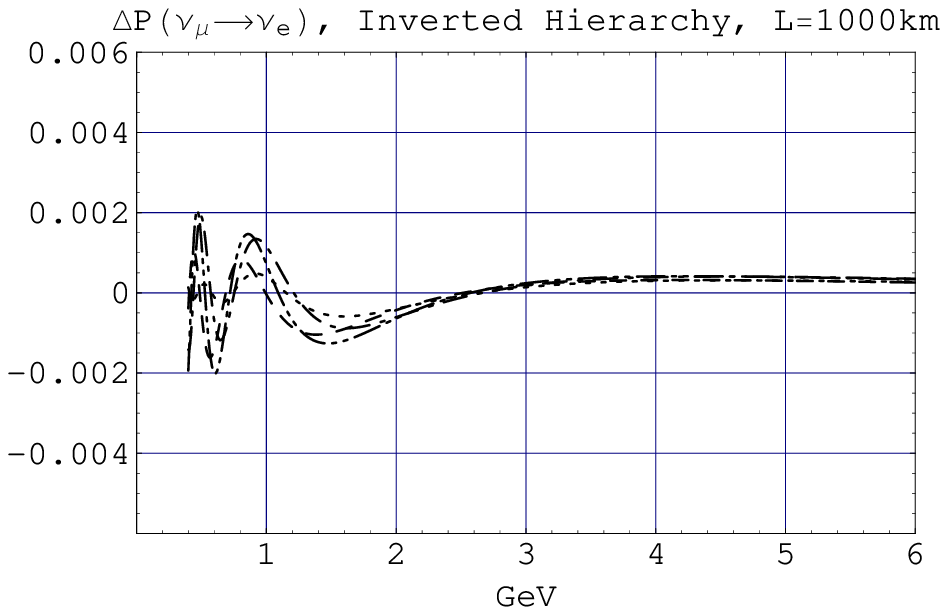}
\caption{Comparison of exact and approximate values of
$P(\nu_\mu\rightarrow\nu_e)$ for the $L=1000\,\mathrm{km}$ case 
for several different values of the CP violating phase $\delta$.
The approximate values were calculated using 
Eq.~(\protect{\ref{tildetheta1}}) for the mixing angles, and
Eq.~(\protect{\ref{lambdaapprox2}}) for the mass-squared differences.
The exact values are given by the solid gray lines, while the approximate values are
the black dashed ($\delta=0$), dotted ($\delta=\pi/2$),
dot-dashed ($\delta=\pi$), and double-dot-dashed ($\delta=3\pi/2$) lines.}
\label{Pmu2e1000}
\end{center}
\end{figure}
%%%%%%%%%%%%%%%%%%%
%%%%%%%%%%%%%%%%%%%
\begin{figure}[p]
\begin{center}
\includegraphics[scale=1.0]{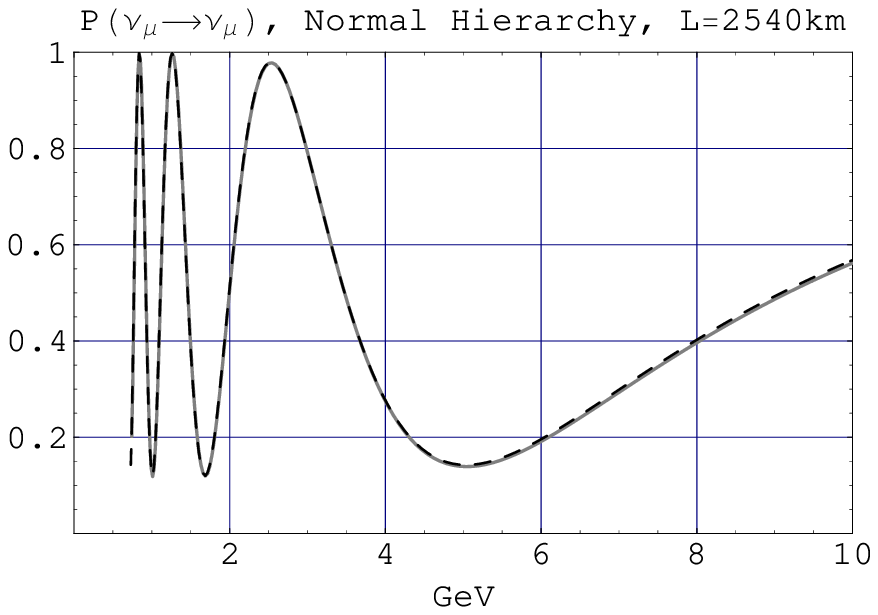}
\includegraphics[scale=1.0]{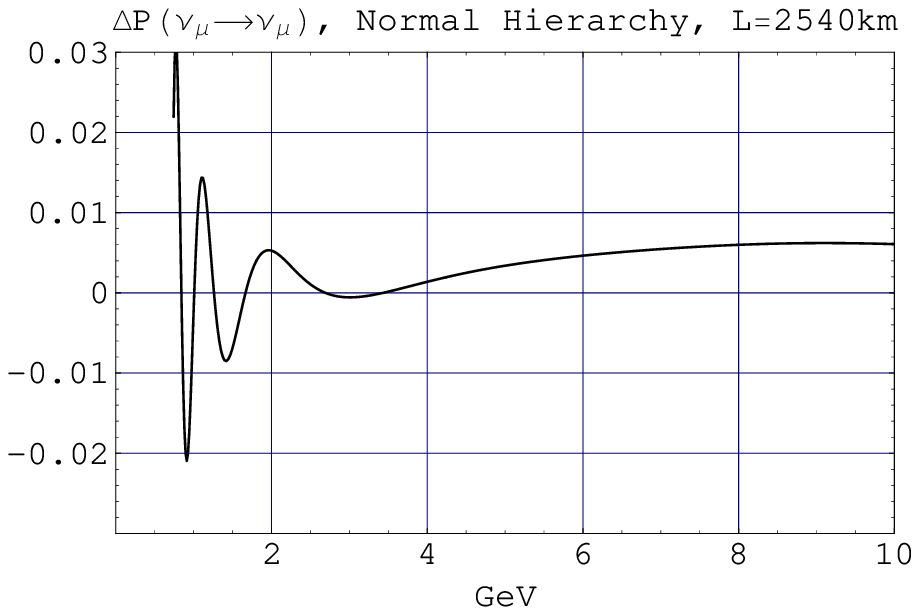}
\includegraphics[scale=1.0]{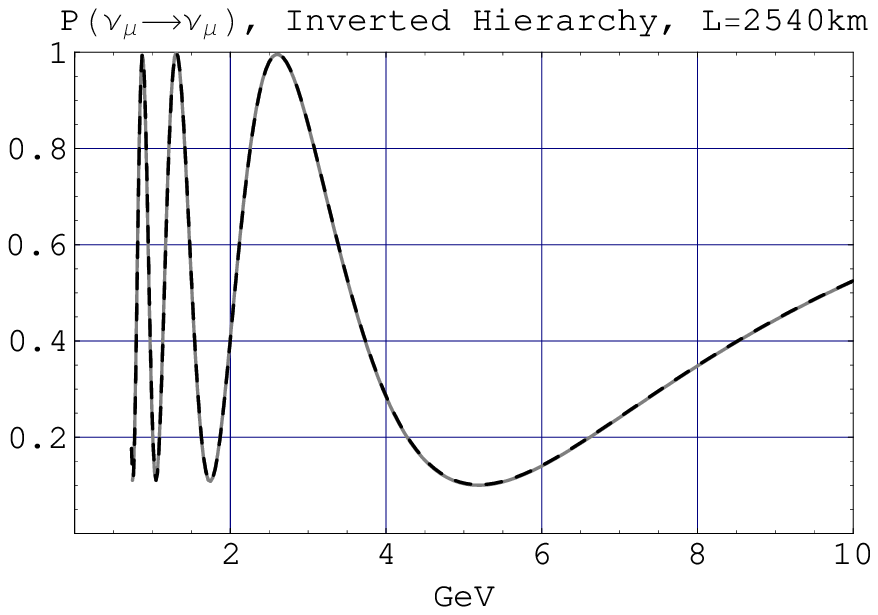}
\includegraphics[scale=1.0]{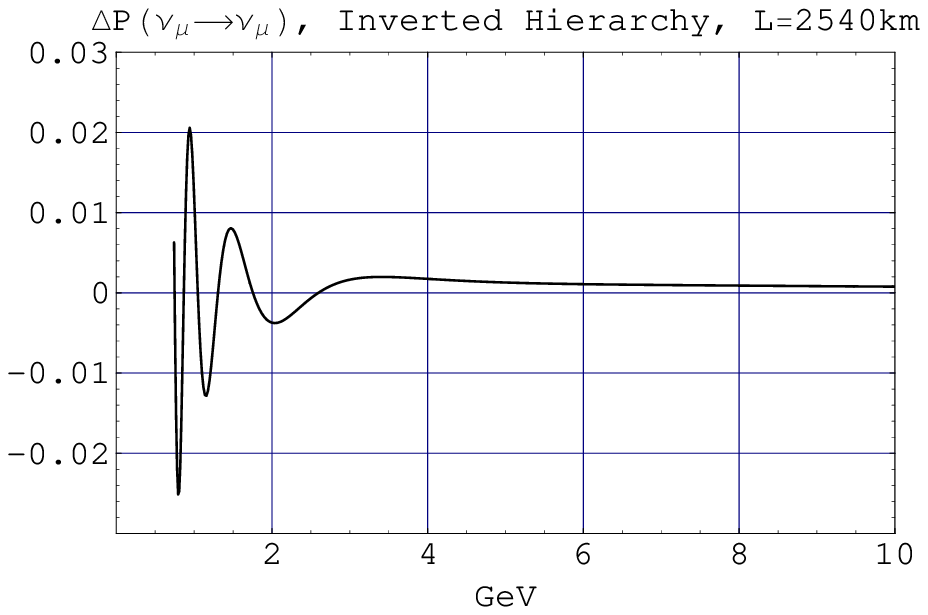}
\caption{Comparison of exact (solid gray line) and approximate (black dashed line) 
values of $P(\nu_\mu\rightarrow\nu_\mu)$ for the $L=2540\,\mathrm{km}$ case.
The approximate value was calculated using 
Eq.~(\protect{\ref{tildetheta1}}) for the mixing angles, and
Eq.~(\protect{\ref{lambdaapprox3}}) for the mass-squared differences.
The CP violating phase $\delta$ was set to zero.}
\label{Pmu2mu2540}
\end{center}
\end{figure}
%%%%%%%%%%%%%%%%%%%
%%%%%%%%%%%%%%%%%%%
\begin{figure}[p]
\begin{center}
\includegraphics[scale=1.0]{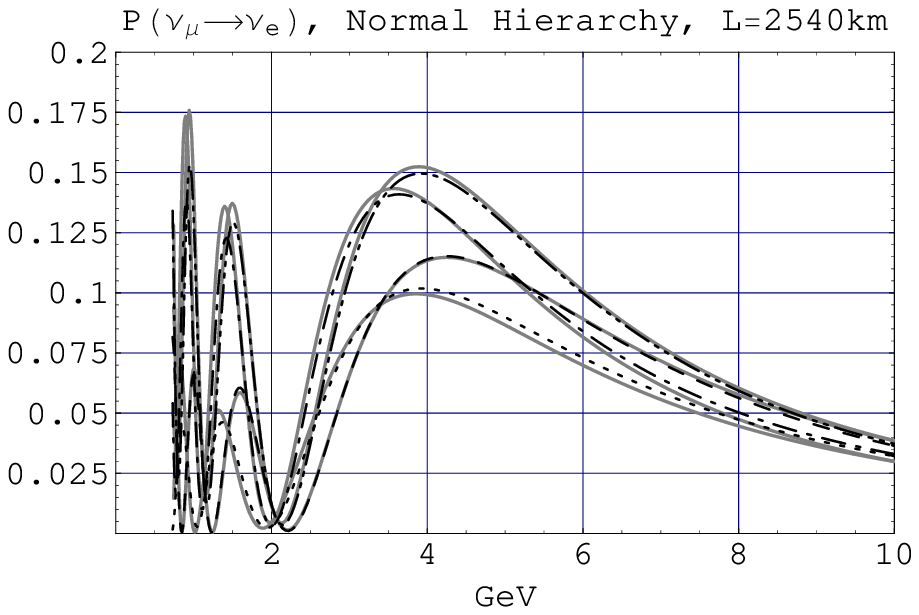}
\includegraphics[scale=1.0]{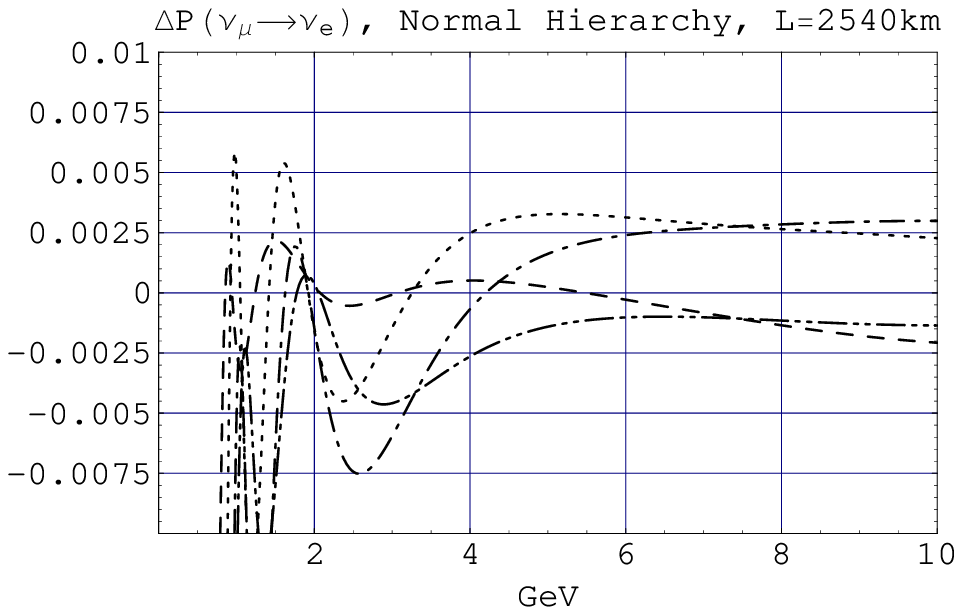}
\includegraphics[scale=1.0]{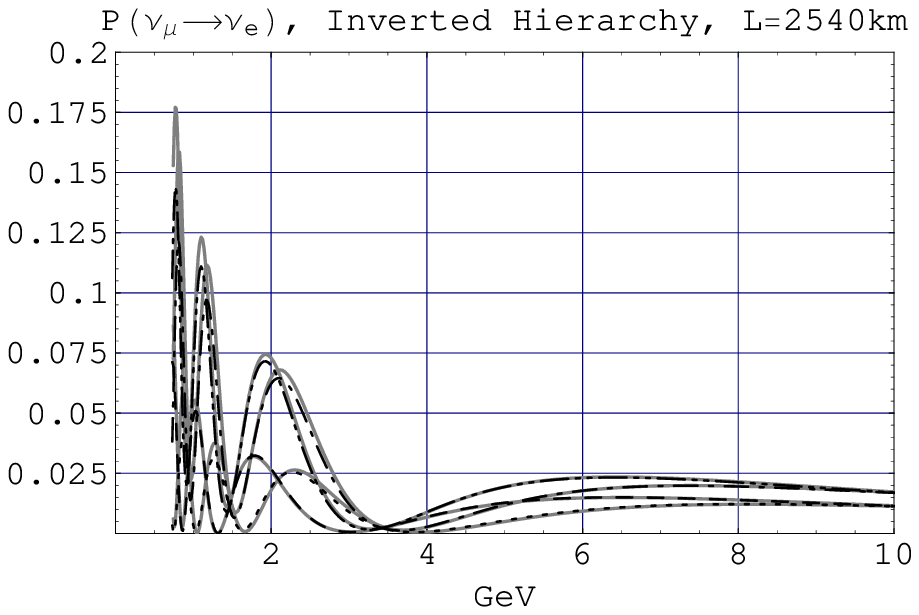}
\includegraphics[scale=1.0]{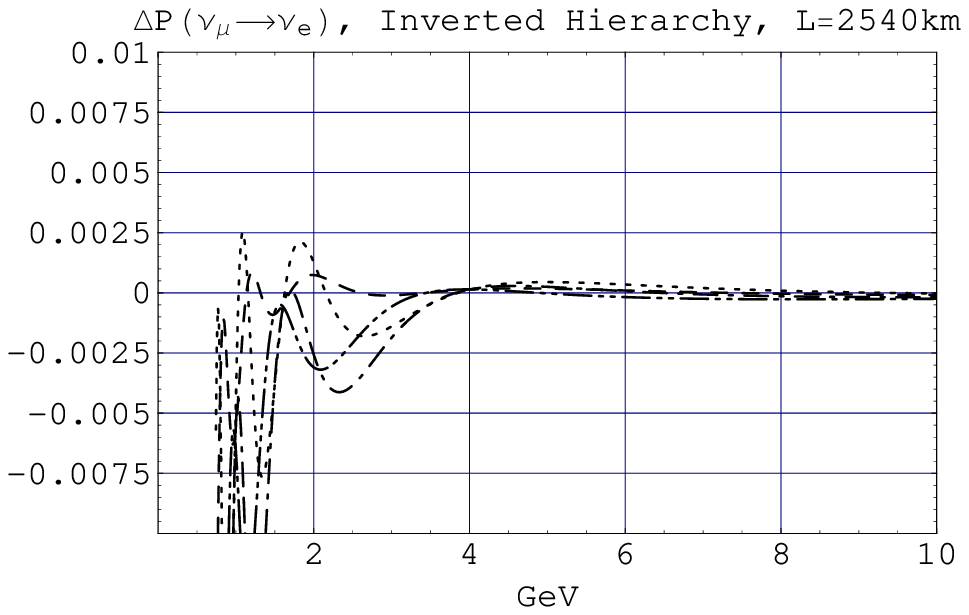}
\caption{Comparison of exact and approximate values of
$P(\nu_\mu\rightarrow\nu_e)$ for the $L=2540\,\mathrm{km}$ case 
for several different values of the CP violating phase $\delta$.
The approximate values were calculated using 
Eq.~(\protect{\ref{tildetheta1}}) for the mixing angles, and
Eq.~(\protect{\ref{lambdaapprox3}}) for the mass-squared differences.
The exact values are given by the solid gray lines, while the approximate values are
the black dashed ($\delta=0$), dotted ($\delta=\pi/2$),
dot-dashed ($\delta=\pi$), and double-dot-dashed ($\delta=3\pi/2$) lines.}
\label{Pmu2e2540}
\end{center}
\end{figure}
%%%%%%%%%%%%%%%%%%%
\end{turnpage}
%%%%%%%%%%%%%%%%%%%%%%%%%%%%%%%%%%%%%%%%%%%%%%%%%%%%%%%%%%%%%%%%%%%%%

In Figs.~\ref{Pmu2mu295} through \ref{Pmu2e2540},
we plot the approximate versus the exact oscillation probabilities for both the
$\delta m^2_{31}>0$ (normal hierarchy) and $\delta m^2_{31}<0$ (inverted hierarchy) cases.
The vacuum parameters were set to the values listed in Eq.~(\ref{exampleparameterset}) 
except for the CP violating phase.
For $\nu_\mu\rightarrow\nu_e$, the four cases 
$\delta=0$, $\frac{\pi}{2}$, $\pi$, and $\frac{3\pi}{2}$ were plotted.
For $\nu_\mu\rightarrow\nu_\mu$, which depends only very weakly on
$\delta$, only the $\delta=0$ case is shown.

The $L=295\,\mathrm{km}$ case is shown in Figs.~\ref{Pmu2mu295} and \ref{Pmu2e295}. 
As is clear from the figures, 
our approximate values are completely indistinguishable from the exact values,
the difference being less than a fraction of a percent throughout the energy range
considered.
For the $L=1000\,\mathrm{km}$ case shown in Figs.~\ref{Pmu2mu1000} and \ref{Pmu2e1000},
deviations can be seen toward the high energy end as expected, but the difference 
is still well under control.
For the $L=2540\,\mathrm{km}$ case shown in Figs.~\ref{Pmu2mu2540} and \ref{Pmu2e2540},
the difference between the exact and approximate values is again less that 
a percent in the range $E=2\sim 10\,\mathrm{GeV}$ where the approximation is applicable,
but becomes large below $E=2\,\mathrm{GeV}$.

%%%%%%%%%%%%%%%%%%%%%%%%%%%%%%%%%%%%%%%%%%%%%%%%%%%%%%%%%%%%%%%%%%%%%
%%%%%%%%%%%%%%%%%%%%%%%%%%%%%%%%%%%%%%%%%%%%%%%%%%%%%%%%%%%%%%%%%%%%%

\section{Qualitative Analysis}

%%%%%%%%%%%%%%%%%%%
\begin{figure}[p]
\begin{center}
\includegraphics[scale=0.75]{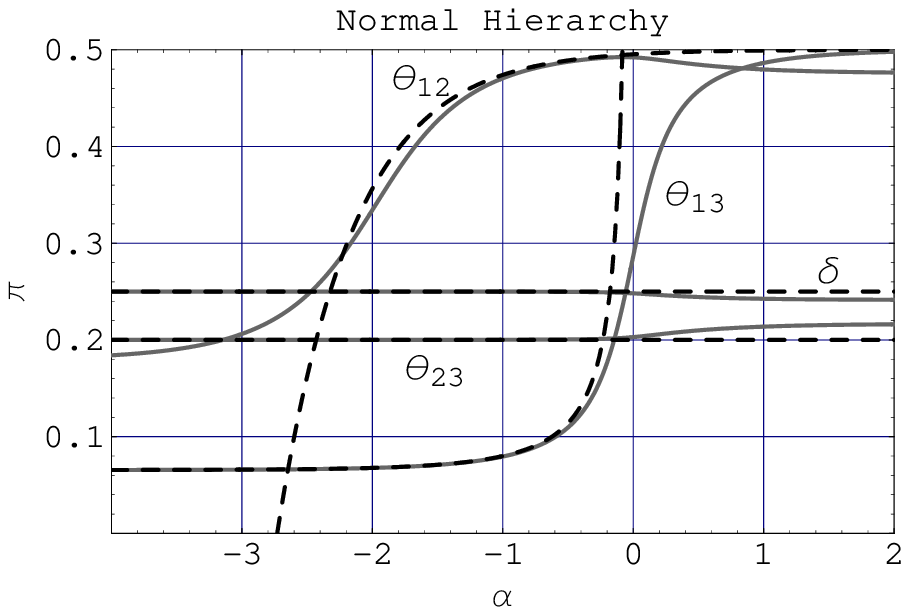}
\includegraphics[scale=0.75]{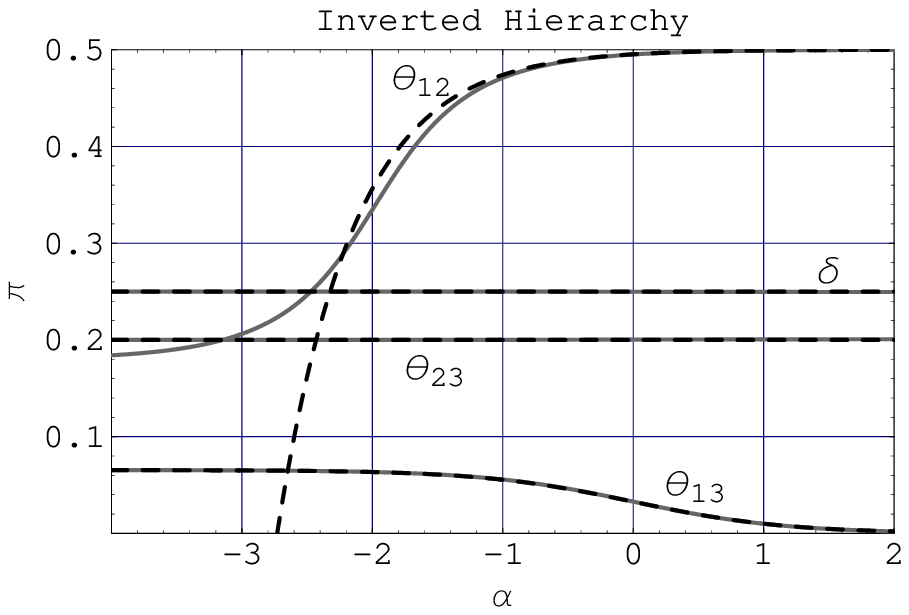}
\caption{The exact values of $\tilde{\theta}_{12}$, $\tilde{\theta}_{13}$, $\tilde{\theta}_{23}$, and $\tilde{\delta}$ (solid gray lines) plotted as functions of
$\alpha=\log_{1/\varepsilon}(a/|\delta m^2_{31}|)$
against their approximate values (black dashed lines) obtained using Eq.~(\protect{\ref{tildetheta4}}).
The $\delta m^2_{31}>0$ (normal hierarchy) case is shown on the left, and 
the $\delta m^2_{31}<0$ (inverted hierarchy) case is shown on the right.
The input parameters are those of Eq.~(\protect{\ref{exampleparameterset}}).
}
\label{thetaTildeApprox4}
\end{center}
\end{figure}
%%%%%%%%%%%%%%%%%%%
%%%%%%%%%%%%%%%%%%%
\begin{figure}[p]
\begin{center}
\includegraphics[scale=0.75]{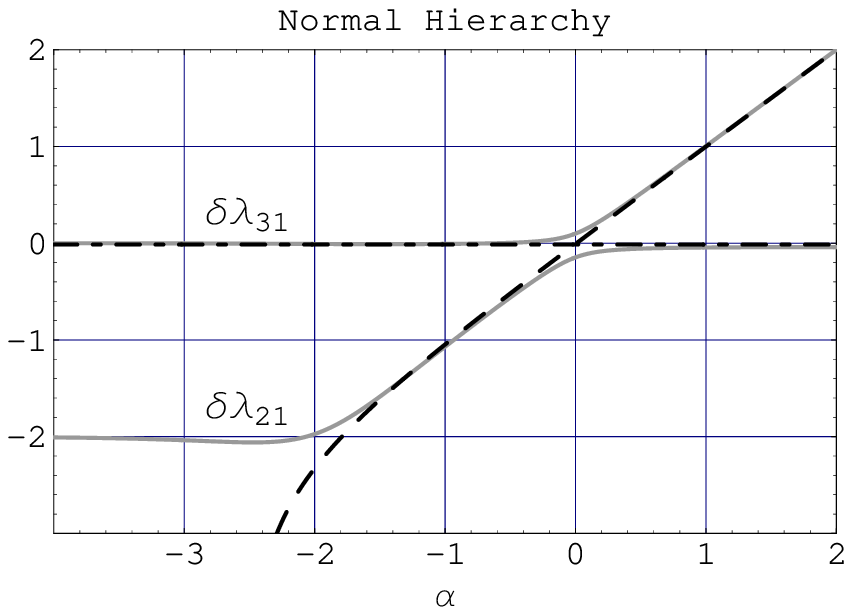}
\includegraphics[scale=0.75]{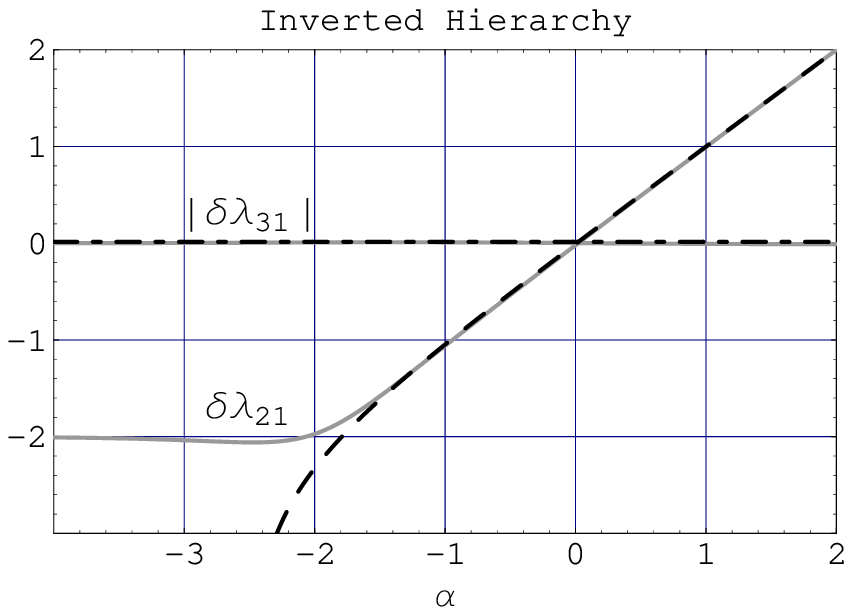}
\caption{The exact values of 
$\log_{1/\varepsilon}(\delta{\lambda}_{21}/|\delta m^2_{31}|)$ and
$\log_{1/\varepsilon}(\delta{\lambda}_{31}/|\delta m^2_{31}|)$ 
(solid gray lines) plotted as
functions of $\alpha=\log_{1/\varepsilon}(a/|\delta m^2_{31}|)$
against their approximate values (black dashed and dot-dashed lines) 
obtained using Eq.~(\protect{\ref{lambda4}}).
The input parameters are those of Eq.~(\protect{\ref{exampleparameterset}}).
}
\label{deltalambdaXY4}
\end{center}
\end{figure}
%%%%%%%%%%%%%%%%%%%

Let us now use our approximation to see whether we can understand various
qualitative features of the oscillation probabilities we plotted.

Since the three cases we considered in the previous section
cover the range $\alpha=-2\sim 0$, let us take the midpoint $\alpha\sim -1$ and
further simplify our expressions to those applicable there. 
The approximations for the effective mixing angles for the neutrinos 
can be obtained from Eqs.~(\ref{phithetaexpand1}), (\ref{alphaminus1phi}), and (\ref{tildetheta1}):
\begin{eqnarray}
\tilde{\theta}_{12} 
& \approx & \frac{\pi}{2} - \frac{\delta m^2_{21}}{2a}\sin(2\theta_{12}) \;, \cr
\tilde{\theta}_{13}
%& \approx & \theta_{13}\left( 1 + \frac{a}{\delta m^2_{31}} \right) 
& \approx & \dfrac{ \theta_{13} }{\left( 1 - \dfrac{a}{\delta m^2_{31}} \right)}, \cr
\tilde{\theta}_{23}
& \approx & \theta_{23} \;,\cr
\tilde{\delta}
& \approx & \delta \;.
\label{tildetheta4}
\end{eqnarray}
In Fig.~\ref{thetaTildeApprox4} we plot these approximations against the exact values.
Though the approximation for $\theta_{13}$ breaks down for the normal hierarchy case
as $\alpha\rightarrow 0$, these expressions nevertheless capture the essential
behavior of the effective mixing angles throughout the range $-2<\alpha< 0$.
They and are also numerically accurate for most of this range.
In fact, for the inverted hierarchy case, the approximations of $\tilde{\theta}_{13}$,
$\tilde{\theta}_{23}$ and $\tilde{\delta}$ are valid for all $\alpha$.
From Eq.~(\ref{alphaminus1lambda}), the approximations for
the effective mass-squared differences are
\begin{eqnarray}
\lambda_1 & \approx & \delta m^2_{21}\,c_{12}^2\;, \cr
\lambda_2 & \approx & a + \delta m^2_{21}\,s_{12}^2\;, \cr
\lambda_3 & \approx & \delta m^2_{31}\;.
\label{lambda4}
\end{eqnarray}
These expressions are compared against the exact values in Fig.~\ref{deltalambdaXY4}, 
and we can again conclude that they are fairly accurate for most of the range $-2<\alpha<0$,
except very near the endpoints.

%from which we find
%
%\begin{eqnarray}
%\delta\lambda_{31}
%& = & \lambda_3 - \lambda_1
%\;\approx\; \delta m^2_{31} - \delta m^2_{21}\,c_{12}^2
%\;\approx\; \delta m^2_{31} \left[ 1 - \frac{1}{2} O(\varepsilon^2) \right] \;,\cr
%\delta\lambda_{21}
%& = & \lambda_2 - \lambda_1
%\;\approx\; a - \delta m^2_{21}\cos(2\theta_{12})
%\;\approx\; a \left[ 1 - 2 O(\varepsilon^2) \right] \;.
%\end{eqnarray}
%

%%%%%%%%%%%%%%%%%%%
\begin{figure}[p]
\begin{center}
\includegraphics[scale=0.75]{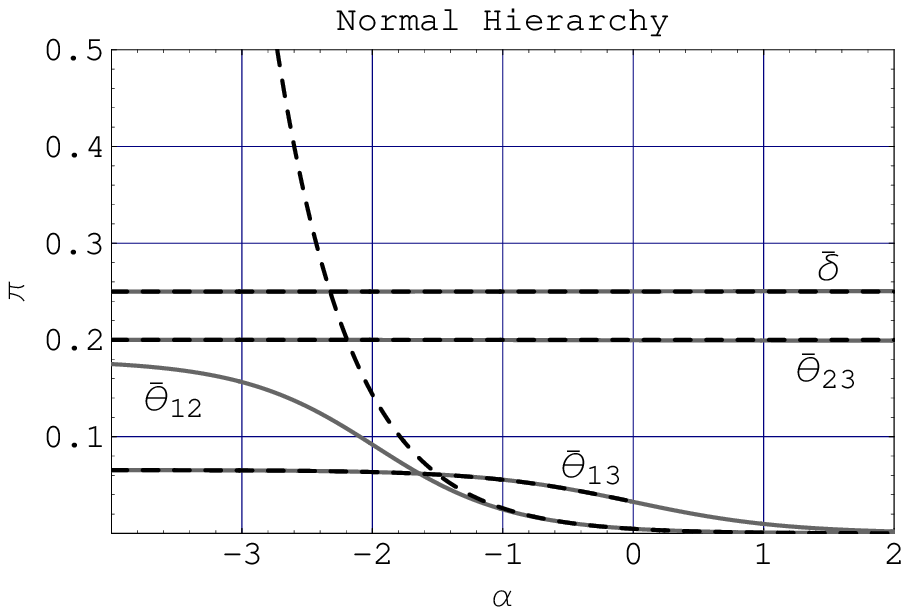}
\includegraphics[scale=0.75]{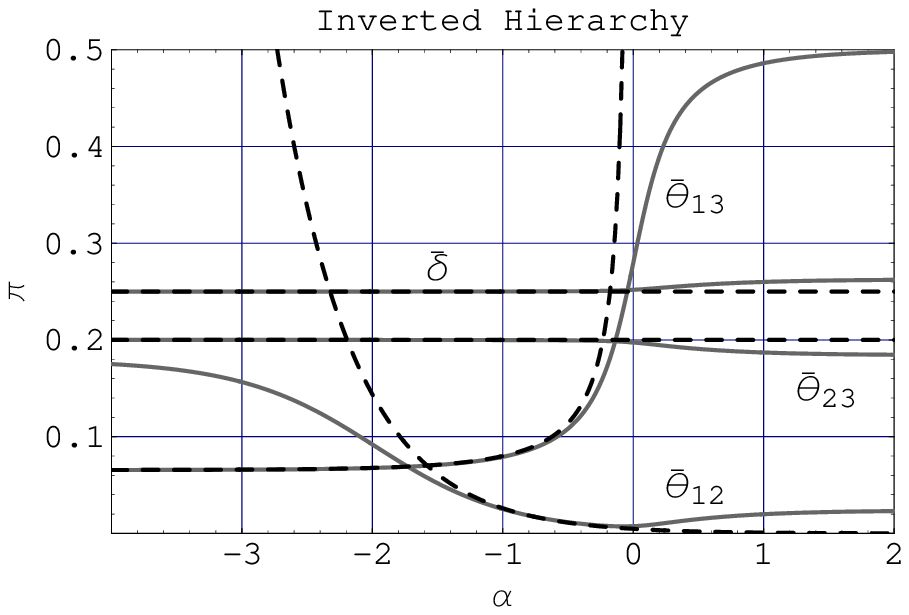}
\caption{The exact values of $\edlit{\theta}_{12}$, $\edlit{\theta}_{13}$, $\edlit{\theta}_{23}$, and $\edlit{\delta}$ (solid gray lines) plotted as functions of
$\alpha=\log_{1/\varepsilon}(a/|\delta m^2_{31}|)$
against their approximate values (black dashed lines) obtained using Eq.~(\protect{\ref{edlittheta4}}).
The $\delta m^2_{31}>0$ (normal hierarchy) case is shown on the left, and 
the $\delta m^2_{31}<0$ (inverted hierarchy) case is shown on the right.
The input parameters are those of Eq.~(\protect{\ref{exampleparameterset}}).
}
\label{thetaTildebarApprox4}
\end{center}
\end{figure}
%%%%%%%%%%%%%%%%%%%
%%%%%%%%%%%%%%%%%%%
\begin{figure}[p]
\begin{center}
\includegraphics[scale=0.75]{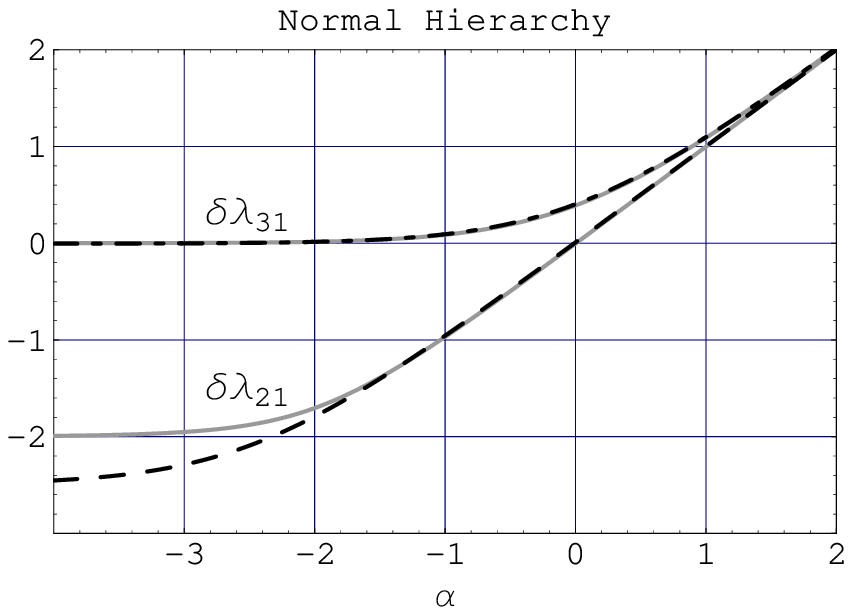}
\includegraphics[scale=0.75]{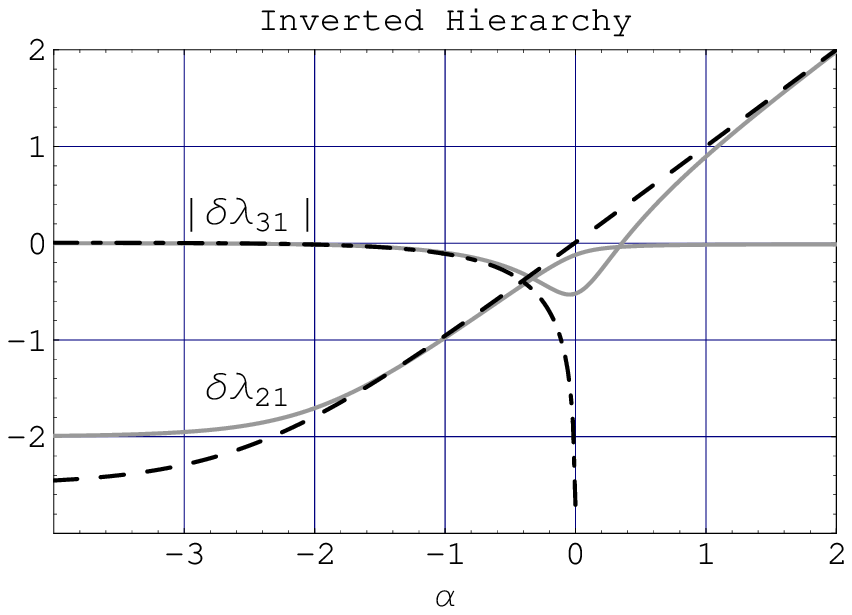}
\caption{The exact values of 
$\log_{1/\varepsilon}(\delta\bar{\lambda}_{21}/|\delta m^2_{31}|)$ and
$\log_{1/\varepsilon}(|\delta\bar{\lambda}_{31}|/|\delta m^2_{31}|)$ 
(solid gray lines) plotted as
functions of $\alpha=\log_{1/\varepsilon}(a/|\delta m^2_{31}|)$
against their approximate values (black dashed and dot-dashed lines),
obtained using Eq.~(\protect{\ref{lambdabar4}}).
The input parameters are those of Eq.~(\protect{\ref{exampleparameterset}}).
}
\label{deltalambdabarXY4}
\end{center}
\end{figure}
%%%%%%%%%%%%%%%%%%%

For the anti-neutrinos, the effective mixing angles when $\alpha\sim -1$ are approximated by
\begin{eqnarray}
\edlit{\theta}_{12} 
& \approx & \frac{\delta m^2_{21}}{2a}\sin(2\theta_{12}) \;, \cr
\edlit{\theta}_{13}
& \approx & \dfrac{ \theta_{13} }{ \left( 1 + \dfrac{a}{\delta m^2_{31}} \right) } \;, \cr
\edlit{\theta}_{23}
& \approx & \theta_{23} \;,\cr
\edlit{\delta}
& \approx & \delta \;.
\label{edlittheta4}
\end{eqnarray}
The accuracy of these expressions is shown in Fig.~\ref{thetaTildebarApprox4}.
In contrast to the neutrino case, the approximations for $\edlit{\theta}_{13}$,
$\edlit{\theta}_{23}$ and $\edlit{\delta}$ are applicable to all $a$ for the
normal hierarchy case, and
the approximation for $\edlit{\theta}_{13}$ breaks down as $\alpha\rightarrow 0$
for the inverted hierarchy case.
The approximations for the effective mass-squared differences are given by
\begin{eqnarray}
\bar{\lambda}_1 & \approx & -a + \delta m^2_{21}\,s_{12}^2\;, \cr
\bar{\lambda}_2 & \approx & \delta m^2_{21}\,c_{12}^2\;, \cr
\bar{\lambda}_3 & \approx & \delta m^2_{31}\;,
\label{lambdabar4}
\end{eqnarray}
with the accuracy shown in Fig.~(\ref{deltalambdabarXY4}).
As in the neutrino case,
we can conclude that these approximations capture the essential behavior
of the effective mixing angles and mass squared differences throughout the
range $-2<\alpha<0$, and are also numerically accurate except near the endpoints.

%%%%%%%%%%%%%%%%%%%%%%%%%%%%%%%%%
Let us now apply these approximations to the
oscillation probabilities $P(\nu_\mu\rightarrow\nu_\mu)$ and $P(\nu_\mu\rightarrow\nu_e)$ 
in matter, and their anti-neutrino counterparts.
First, recall from Eqs.~(\ref{Probabilities}) and (\ref{VacuumParams}) that
these probabilities in vacuum are given by
\begin{eqnarray}
P(\nu_\mu\rightarrow \nu_\mu)
& = & 1 - \sin^2(2\theta_\mathrm{atm})
      \sin^2\left(\frac{\Delta_{31} - \kappa_{\mu\mu}\Delta_{21}}{2}
            \right) 
    + O(\Delta_{21}^2) \;,\cr
P(\nu_\mu\rightarrow \nu_e)
& = & 4 \sin^2\theta_{13}\,\sin^2\theta_\mathrm{atm}
	  \bigl\{ 1
%            - \left( \dfrac{ A\,\Delta_{21}}
%                           { \sin^2\theta_{13}\,\sin^2\theta_\mathrm{atm} }       
%              \right) \sin\delta
	         - \left( B\sin\delta \right) \Delta_{21}
      \bigr\}
%\cr & & \qquad\qquad \times
      \sin^2\left(\frac{\Delta_{31} - \kappa_{\mu e}\Delta_{21}}{2}
            \right)
%\cr & &
    +\; O(\Delta_{21}^2) \;,\cr
& &
\label{Probabilities2}
\end{eqnarray}
where 
\begin{eqnarray}
\sin\theta_\mathrm{atm} & = & s_{23}c_{13} \;, \cr 
A & = & \frac{1}{8}\sin(2\theta_{12})
\sin(2\theta_{13})
\sin(2\theta_\mathrm{atm})
\sqrt{1-\tan^2\theta_{13}\tan^2\theta_\mathrm{atm}}  \cr
& = & \frac{1}{8}\times O(1) \times 2 O(\varepsilon) \times O(1) \times
\sqrt{1-O(\varepsilon^2)}  \cr
& = & \frac{1}{4}O(\varepsilon)\;, \cr
B & \equiv & \frac{A}{\sin^2\theta_{13}\sin^2\theta_\mathrm{atm}}
\;=\; \frac{1}{2}\,O(\varepsilon^{-1}) \;, \cr
\kappa_{\mu\mu}
& = & c_{12}^2 - \cos(2\theta_{12})
      \tan^2\theta_{13}\tan^2\theta_\mathrm{atm}
     -\left(\frac{2A}{\cos^2\theta_{13}\cos^2\theta_{\mathrm{atm}}}\right)
      \cos\delta \cr
& = & \frac{1}{2}\,O(1) - 2 O(\varepsilon^3) - O(\varepsilon) \cos\delta \cr
& = & \frac{1}{2}\,O(1) - O(\varepsilon) \cos\delta 
\;,\cr
\kappa_{\mu e}
& = & s_{12}^2 
%- \left(\frac{A}{\sin^2\theta_{13}\sin^2\theta_\mathrm{atm}}\right)\cos\delta \cr
- B\cos\delta \cr
& = & \frac{1}{2}\,O(1) - \frac{1}{2}\,O(\varepsilon^{-1})\cos\delta \;.
\label{VacuumParams2}
\end{eqnarray}
The oscillation probabilities for the anti-neutrinos are obtained by flipping the sign 
of $\sin\delta$.

Note that the first oscillation peak occurs at a distance/energy of $|\Delta_{31}|\approx\pi$.
Since $\Delta_{21}/|\Delta_{31}|=\delta m^2_{21}/|\delta m^2_{31}|=\varepsilon^2$,
the $O(\Delta_{21}^2)$ terms in Eq.~(\ref{Probabilities2}) 
are of $O(\pi^2\varepsilon^4)$ which justifies our dropping them at those distance/energies.
Note also that the coefficient of $\cos\delta$ in $\kappa_{\mu e}$ is 
$B=\frac{1}{2}O(\varepsilon^{-1})$,
which when multiplied by $\Delta_{21}$ is of $O(\varepsilon)$.
The exact same product of parameters, $B\Delta_{21} = O(\varepsilon)$,
appears in the coefficient of $\sin\delta$ in the 
oscillation envelope of $P(\nu_\mu\rightarrow\nu_e)$.
Therefore, a measurement of $P(\nu_\mu\rightarrow\nu_e)$
can, in principle, constrain $\cos\delta$ from the position of the
peak, and $\sin\delta$ from the height of the peak, provided it is accurate enough
to discern these $O(\varepsilon)$ corrections.
In contrast, the coefficient of $\cos\delta$ in $\kappa_{\mu\mu}$ is $O(\varepsilon)$,
rendering the $\cos\delta$ term in $\kappa_{\mu\mu}\Delta_{21}$ negligible,
and we conclude that $P(\nu_\mu\rightarrow\nu_\mu)$ is insensitive to $\delta$.

Of course, actual long-baseline experiments can only measure oscillation 
probabilities in matter.
The effective $\Delta_{31}$'s in matter for the
neutrinos and anti-neutrinos are
\begin{eqnarray}
\tilde{\Delta}_{31}
& \approx & \left( \Delta_{31} - \Delta_{21}c_{12}^2 \right) \;,\cr
\edlit{\Delta}_{31}
& \approx & \left( \Delta_{31} + \frac{a}{2E}L - \Delta_{21} s_{12}^2 \right) \;,  
\end{eqnarray}
respectively, so they are both the same order as $\Delta_{31}$.
The effective $\Delta_{21}$'s, on the other hand, are
\begin{eqnarray}
\tilde{\Delta}_{21}
& \approx & \left(\frac{a}{2E}L - \Delta_{21}\cos 2\theta_{12}\right)
\;\approx\; \frac{a}{2E}L
\;=\; (\sqrt{2}G_F N_e) L\;, \cr
\edlit{\Delta}_{21}
& \approx & \left(\frac{a}{2E}L + \Delta_{21}\cos 2\theta_{12}\right)
\;\approx\; \frac{a}{2E}L
\;=\; (\sqrt{2}G_F N_e) L\;,
\end{eqnarray}
so they are enhanced by a factor of $a/\delta m^2_{21}$ relative to $\Delta_{21}$.  
Therefore, at the first oscillation peak where $|\Delta_{31}| \approx \pi$,
we can expect $\tilde{\Delta}_{21}$ and $\edlit{\Delta}_{21}$ to be of order
$\pi a/|\delta m^2_{31}|$.  

Note that the value of 
$\tilde{\Delta}_{21} \approx \edlit{\Delta}_{21} \approx (\sqrt{2}G_F N_e)L$ does not
depend on the energy $E$.
It is determined solely by the baseline length $L$, once the matter density $\rho$ is fixed. 
($N_e = N_A\rho/2$ where $N_A$ is the Avogadro number.)
For the three examples we considered in the previous section,
we find
\begin{equation}
(\sqrt{2}G_F N_e) L = 
\left\{ \begin{array}{ll}
0.15 &  \quad(\rho=2.6\,\mathrm{g/cm^3}\;,L=295\,\mathrm{km})\;,  \\
0.5  &  \quad(\rho=2.7\,\mathrm{g/cm^3}\;,L=1000\,\mathrm{km})\;, \\
1.7  &  \quad(\rho=3.4\,\mathrm{g/cm^3}\;,L=2540\,\mathrm{km})\;.
\end{array} \right.
\label{tildeDelta21estimates}
\end{equation}
Since we would like to use the analog of 
Eq.~(\ref{Probabilities2}) to analyze the oscillation probabilities in matter,
we would like maintain the condition
\begin{equation}
\tilde{\Delta}_{21},\; \edlit{\Delta}_{21} < 1\;,
\end{equation}
so that an expansion in $\tilde{\Delta}_{21}$ or $\edlit{\Delta}_{21}$ is justified.
The $L=2540\,\mathrm{km}$ case is clearly problematic and must be treated separately.
We will therefore first restrict our attention to the cases in which the
$|\Delta_{31}|=\pi$ condition occurs in the region $-2<\alpha \alt -1$, so that 
$\tilde{\Delta}_{21} \approx \edlit{\Delta}_{21} \alt O(\pi\varepsilon) = 0.47\sim 0.75$.

Even with this restriction, $\tilde{\Delta}_{21}$ and $\edlit{\Delta}_{21}$ can 
still be enhanced considerably when $|\Delta_{31}|=\pi$ occurs at $\alpha\approx -1$.
One may naively anticipate that this enhancement will enhance the coefficients of $\sin\delta$ and
$\cos\delta$ in $P(\nu_\mu\rightarrow\mu_e)$, thereby facilitate the detection of $\delta$.
At the same time, it could also enhance the 
$O(\tilde{\Delta}_{21}^2)$ and $O(\edlit{\Delta}_{21}^2)$ terms in the oscillation probabilities and
invalidate their complete neglect.
However, it turns out that these are not the case.

\begin{turnpage}
%%%%%%%%%%%%%%%%%%%
\begin{figure}[p]
\begin{center}
\includegraphics[scale=1.0]{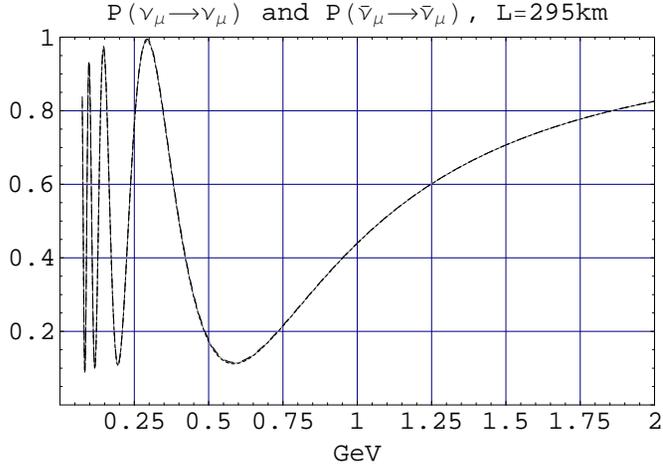}
\includegraphics[scale=1.0]{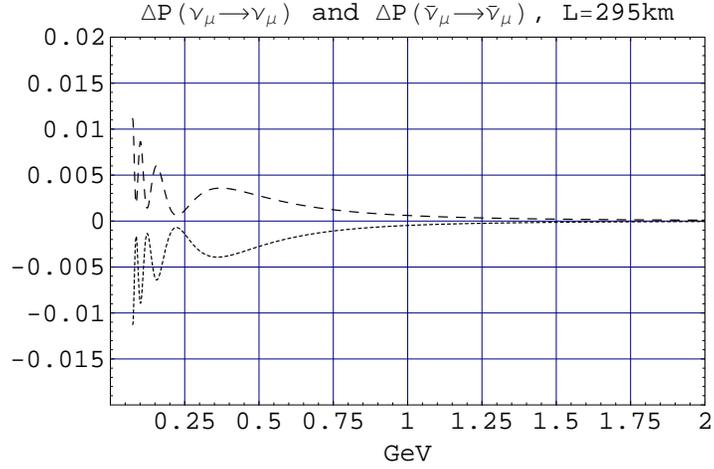}
\includegraphics[scale=1.0]{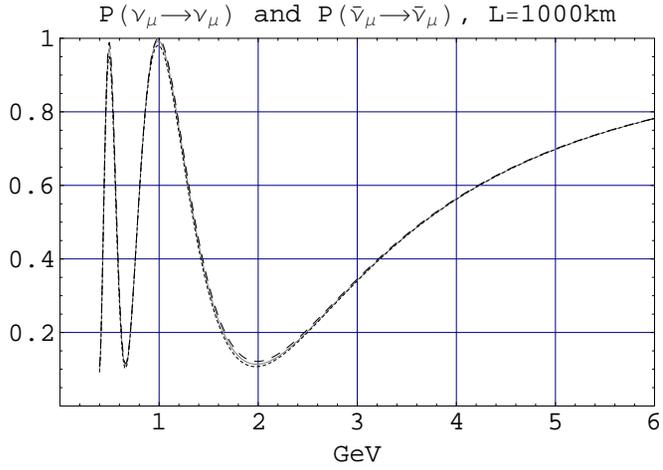}
\includegraphics[scale=1.0]{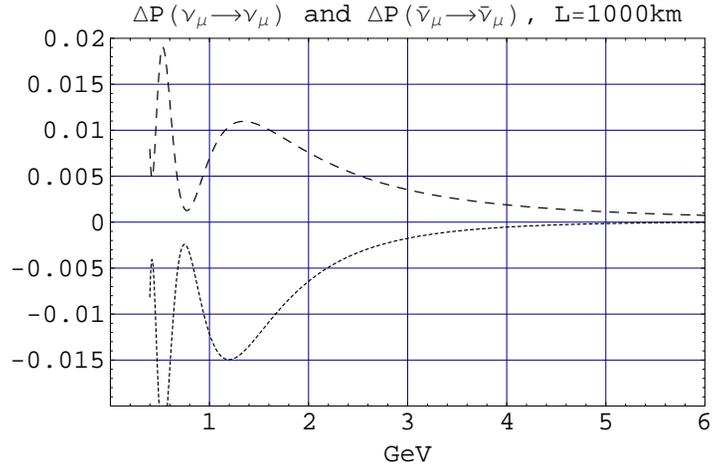}
\caption{Comparison of the $\nu_\mu$ and $\bar{\nu}_\mu$ survival probabilities
in vacuum and in matter. On the left, $P(\nu_\mu\rightarrow\nu_\mu) = P(\bar{\nu}_\mu\rightarrow\bar{\nu}_\mu)$ is the solid gray line,
$\tilde{P}(\nu_\mu\rightarrow\nu_\mu)$ is the dashed black line, and
$\edlit{P}(\bar{\nu}_\mu\rightarrow\bar{\nu}_\mu)$ is the dotted black line.
On the right, the difference 
$\Delta P(\nu_\mu\rightarrow\nu_\mu)
= \tilde{P}(\nu_\mu\rightarrow\nu_\mu)-P(\nu_\mu\rightarrow\nu_\mu)$
is the dashed black line, and the difference
$\Delta P(\bar{\nu}_\mu\rightarrow\bar{\nu}_\mu)
= \edlit{P}(\bar{\nu}_\mu\rightarrow\bar{\nu}_\mu)-P(\bar{\nu}_\mu\rightarrow\bar{\nu}_\mu)$
is the dotted blackline.  The input parameters are those listed in
Eq.~(\protect{\ref{exampleparameterset}}) with $\delta m^2_{31}>0$.}
\label{Pmu2muVacuumMatter}
\end{center}
\end{figure}
%%%%%%%%%%%%%%%%%%%
\end{turnpage}

Let us first look at the $\nu_\mu$ and $\bar{\nu}_\mu$ survival probabilities in matter which 
are obtained by replacing all the quantities in the vacuum probability with their 
tilded and anti-tilded counterparts:
\begin{eqnarray}
\lefteqn{\tilde{P}(\nu_\mu \rightarrow \nu_\mu)} \cr
& = & 1
%-4\;|\tilde{U}_{\mu 3}|^2 \left( 1 - |\tilde{U}_{\mu 3}|^2 \right)
-\sin^2(2\tilde{\theta}_\mathrm{atm})
\sin^2\left( \frac{ \tilde{\Delta}_{31} - \tilde{\kappa}_{\mu\mu}\tilde{\Delta}_{21} }{ 2 }
      \right)
\cr
& & \phantom{1}
-|\tilde{U}_{\mu 1}|^2 |\tilde{U}_{\mu 2}|^2 
 \left( 1 + \dfrac{|\tilde{U}_{\mu 3}|^2}{1-|\tilde{U}_{\mu 3}|^2}\cos\tilde{\Delta}_{31} \right) \tilde{\Delta}_{21}^2
\cr
& & \phantom{1}
-|\tilde{U}_{\mu 1}|^2 |\tilde{U}_{\mu 2}|^2 |\tilde{U}_{\mu 3}|^2
     \left\{\dfrac{ 1 + |\tilde{U}_{\mu 2}|^2 - |\tilde{U}_{\mu 3}|^2 }{ 3(1-|\tilde{U}_{\mu 3}|^2)^2 }
            \sin\tilde{\Delta}_{31}
     \right\} \tilde{\Delta}_{21}^3    
+ O(\tilde{\Delta}_{21}^4) \;, \label{Pmu2muExpand} \\
& & \cr
\lefteqn{\edlit{P}(\bar{\nu}_\mu \rightarrow \bar{\nu}_\mu)} \cr
& = & 1
%-4\;|\edlit{U}_{\mu 3}|^2 \left( 1 - |\edlit{U}_{\mu 3}|^2 \right)
-\sin^2(2\edlit{\theta}_\mathrm{atm})
\sin^2\left( \frac{ \edlit{\Delta}_{31} - \edlit{\kappa}_{\mu\mu}\edlit{\Delta}_{21} }{ 2 }
      \right)
\cr
& & \phantom{1}
-|\edlit{U}_{\mu 1}|^2 |\edlit{U}_{\mu 2}|^2 
 \left( 1 + \dfrac{|\edlit{U}_{\mu 3}|^2}{1-|\edlit{U}_{\mu 3}|^2}\cos\edlit{\Delta}_{31} \right) \edlit{\Delta}_{21}^2
\cr
& & \phantom{1}
-|\edlit{U}_{\mu 1}|^2 |\edlit{U}_{\mu 2}|^2 |\edlit{U}_{\mu 3}|^2
     \left\{\dfrac{ 1 + |\edlit{U}_{\mu 2}|^2 - |\edlit{U}_{\mu 3}|^2 }{ 3(1-|\edlit{U}_{\mu 3}|^2)^2 }
            \sin\edlit{\Delta}_{31}
     \right\} \edlit{\Delta}_{21}^3    
+ O(\edlit{\Delta}_{21}^4) \;.  \label{Pmubar2mubarExpand}
\end{eqnarray}
We have kept terms up to $\tilde{\Delta}_{21}^3$ and $\edlit{\Delta}_{21}^3$ 
explicitly to evaluate their sizes.
Using the approximations of Eqs.~(\ref{tildetheta4}) and (\ref{edlittheta4}),
the effective MNS matrix elements that appear in these expressions can be evaluated to be
\begin{eqnarray}
|\tilde{U}_{\mu 1}|^2
& = & |\tilde{s}_{12}\tilde{c}_{23} + \tilde{c}_{12}\tilde{s}_{13}\tilde{s}_{23}e^{i\tilde{\delta}}|^2
\;=\; \frac{1}{2} O(1) \;, \cr
|\tilde{U}_{\mu 2}|^2
& = & |\tilde{c}_{12}\tilde{c}_{23} - \tilde{s}_{12}\tilde{s}_{13}\tilde{s}_{23}e^{i\tilde{\delta}}|^2
\;\approx\; O(\varepsilon^2) \;, \cr
|\tilde{U}_{\mu 3}|^2
& = & (\tilde{c}_{13}\tilde{s}_{23})^2 
\;=\; \frac{1}{2}O(1) \;.
\label{MNSmatrixelements}
\end{eqnarray}
for the neutrinos, and
\begin{eqnarray}
|\edlit{U}_{\mu 1}|^2
& = & |\edlit{s}_{12}\edlit{c}_{23} + \edlit{c}_{12}\edlit{s}_{13}\edlit{s}_{23}e^{i\edlit{\delta}}|^2
\;\approx\; O(\varepsilon^2) \;,\cr
|\edlit{U}_{\mu 2}|^2
& = & |\edlit{c}_{12}\edlit{c}_{23} - \edlit{s}_{12}\edlit{s}_{13}\edlit{s}_{23}e^{i\edlit{\delta}}|^2
\;=\; \frac{1}{2} O(1) \;, \cr
|\edlit{U}_{\mu 3}|^2
& = & (\edlit{c}_{13}\edlit{s}_{23})^2 
\;=\; \frac{1}{2}O(1) \;,
\label{antiMNSmatrixelements}
\end{eqnarray}
for the anti-neutrinos. This shows that the 
$\tilde{\Delta}_{21}^2$, $\edlit{\Delta}_{21}^2$
and $\tilde{\Delta}_{21}^3$, $\edlit{\Delta}_{21}^3$ terms
are suppressed by $|\tilde{U}_{\mu 2}|^2 = O(\varepsilon^2)$ for the neutrinos, and by
$|\edlit{U}_{\mu 1}|^2 = O(\varepsilon^2)$ for the anti-neutrinos, 
cancelling out the enhancements of $\tilde{\Delta}_{21}$, $\edlit{\Delta}_{21}$ over
$\Delta_{21}$, and reducing these
higher order terms to $O(\pi^2\varepsilon^4)$,
which is the same order as the $O(\Delta_{21}^2)$ terms that were neglected for the vacuum case.  
The fact that $\cos\tilde{\Delta}_{31}\approx -1$ and
$\sin\tilde{\Delta}_{31}\approx 0$ near the first oscillation peak also helps in suppressing
these terms. 
Therefore, to a very good approximation, we can use the expressions
\begin{eqnarray}
\tilde{P}(\nu_\mu\rightarrow \nu_\mu) & = &
1 - \sin^2(2\tilde{\theta}_\mathrm{atm})
\sin^2\left(\dfrac{\tilde{\Delta}_{31}-\tilde{\kappa}_{\mu\mu}\tilde{\Delta}_{21}}{2}\right) \;, \cr
\edlit{P}(\bar{\nu}_\mu\rightarrow \bar{\nu}_\mu) & = &
1 - \sin^2(2\edlit{\theta}_\mathrm{atm})
\sin^2\left(\dfrac{\edlit{\Delta}_{31}-\edlit{\kappa}_{\mu\mu}\edlit{\Delta}_{21}}{2}\right) \;.
\end{eqnarray}
Looking at the remaining parameters in these expressions, we find
\begin{eqnarray}
\sin\tilde{\theta}_{\mathrm{atm}} 
& = & \tilde{c}_{13}\tilde{s}_{23}
\;=\; c_{13}s_{23} \left[ 1 + O(\varepsilon^3) \right]
\;\approx\; \sin\theta_{\mathrm{atm}} \;,\cr
\tilde{A} 
& = & \frac{1}{8}\sin(2\tilde\theta_{12})
\sin(2\tilde{\theta}_\mathrm{13})
\sin(2\tilde{\theta}_\mathrm{atm}) 
\sqrt{ 1 - \tan^2\tilde{\theta}_{13}\tan^2\tilde{\theta}_\mathrm{atm} } \cr
& \approx & \frac{1}{4}\left(\frac{\delta m^2_{21}}{a}\right)
	        \left( 1 + \frac{a}{\delta m^2_{31}} \right)\theta_{13} \cr
& = & \frac{1}{4}\,O(\varepsilon^2)
\;,\cr
\tilde\kappa_{\mu\mu}
& = & \tilde{c}^2_{12} 
- \cos(2\tilde{\theta}_{12})\tan^2\tilde{\theta}_{13}\tan^2\tilde{\theta}_\mathrm{atm}
- \left( \dfrac{2\tilde{A}}{\cos^2\tilde{\theta}_{13}\cos^2\tilde{\theta}_\mathrm{atm}} \right)
  \cos\tilde{\delta} \cr
& = &
O(\varepsilon^2) - O(\varepsilon^2)\cos\delta \;,
%\tilde\kappa_{\mu e} 
%& \approx & \tilde{s}^2_\mathrm{12} 
%- %\left(\frac{\tilde{A}}{\sin^2\tilde{\theta}_\mathrm{13}\sin^2\tilde{\theta}_{atm}}\right)\cos\tilde{\delta%}  \cr
%& = & O(1) - \frac{1}{2}\, O(1)\cos\delta \;, \cr
%
\label{NeutrinoMatterParameters} 
\end{eqnarray}
for the neutrinos, and
\begin{eqnarray}
\sin\edlit{\theta}_{\mathrm{atm}}
& = & \edlit{c}_{13}\edlit{s}_{23}
\;=\; c_{13}s_{23} \left[ 1 + O(\varepsilon^3) \right]
\;\approx\; \sin\theta_{\mathrm{atm}} \;,\cr
\edlit{A} 
& \approx & \frac{1}{8}\sin(2\edlit{\theta}_{12})
\sin(2\edlit{\theta}_\mathrm{13})
\sin(2\edlit{\theta}_\mathrm{atm})
\sqrt{ 1 - \tan^2\edlit{\theta}_{13}\tan^2\edlit{\theta}_\mathrm{atm} } \cr
& \approx & \frac{1}{4}\left(\frac{\delta m^2_{21}}{a}\right)
            \left( 1-\frac{a}{\delta m^2_{31}} \right) \theta_{13}
\;=\; \frac{1}{4}\,O(\varepsilon^2)
\;,\cr
\edlit{\kappa}_{\mu\mu}
& = & \cos^2\edlit{\theta}_\mathrm{12}
- \cos(2\edlit{\theta}_{12})\tan^2\edlit{\theta}_{13}\tan^2\edlit{\theta}_\mathrm{atm}
- \left( \dfrac{2\edlit{A}}{\cos^2\edlit{\theta}_{13}\cos^2\edlit{\theta}_\mathrm{atm}} \right)
  \cos\edlit{\delta} \cr
& = & O(1) - O(\varepsilon^2)\cos\delta \;,
%\edlit{\kappa}_{\mu e} 
%& \approx & \sin^2\edlit{\theta}_\mathrm{12} 
%- \left(\frac{ \edlit{A} }
%             { \sin^2\edlit{\theta}_\mathrm{13}
%               \sin^2\edlit{\theta}_{atm} }
%  \right)\cos\delta \cr
%& \approx & \sin^2\edlit{\theta}_\mathrm{12}
%\Bigl[\, 1 - \cot\edlit{\theta}_\mathrm{12}
%             \cot\edlit{\theta}_\mathrm{13}
%             \cot\edlit{\theta}_\mathrm{atm}\cos\delta
%\,\Bigr] \cr 
%& = & \frac{1}{4}\,O(\varepsilon^2) - \frac{1}{2}\,O(1)\cos\delta
%\;\approx\; - \frac{1}{2}\,O(1)\cos\delta \;. 
\label{AntiNeutrinoMatterParameters}
\end{eqnarray}
for the anti-neutrinos. 
(Recall that $\sin(2\theta_{12}) = 1-2O(\varepsilon^2)$ and 
$\sin(2\theta_\mathrm{atm}) = 1-\frac{1}{2}O(\varepsilon^2)$.)
This shows that the coefficient of $\cos\delta$ in $\tilde{\kappa}_{\mu\mu}$ 
($\edlit{\kappa}_{\mu\mu}$) is suppressed by one power of $\varepsilon$ relative to 
$\kappa_{\mu\mu}$.
Consequently, the $\nu_\mu$ and $\bar{\nu}_\mu$ survival probabilities
remain insensitive to the CP violating phase $\delta$, despite the enhancements of
$\tilde{\Delta}_{21}$, $\edlit{\Delta}_{21}$ over $\Delta_{21}$.
Looking at the arguments of the sines more carefully, we find
\begin{eqnarray}
\Delta_{31} - \kappa_{\mu\mu}\Delta_{21}
& = & \left(\Delta_{31} - c_{12}^2\Delta_{21}\right) + O(\varepsilon^3\Delta_{31}) \;,\cr
\tilde{\Delta}_{31} - \tilde{\kappa}_{\mu\mu} \tilde{\Delta}_{21}
& \approx & \left(\Delta_{31} - c_{12}^2 \Delta_{21}\right)
    - \tilde{c}^2_{12}\left( \frac{a}{2E}L - \Delta_{21}\cos 2\theta_{12} \right) \cr
& = & \left(\Delta_{31} - c_{12}^2 \Delta_{21}\right) + O(\varepsilon^3 \Delta_{31}) \;,\cr
%& = & \left(\Delta_{31} - \kappa_{\mu\mu} \Delta_{21}\right)
%    + O(\varepsilon^3 \Delta_{31}) \;, \cr
\edlit{\Delta}_{31} - \edlit{\kappa}_{\mu\mu} \edlit{\Delta}_{21}
& \approx & \left(\Delta_{31} + \frac{a}{2E}L - \Delta_{21} s_{12}^2 \right)
    - \edlit{c}_{12}^2\left(\frac{a}{2E}L + \Delta_{21}\cos 2\theta_{12}\right) \cr
& = & \left(\Delta_{31} + \frac{a}{2E}L - \Delta_{21} s_{12}^2 \right)
    - \left(\frac{a}{2E}L + \Delta_{21}\cos 2\theta_{12}\right)
    + O(\varepsilon^3\Delta_{31}) \cr
& = & \left(\Delta_{31} - c_{12}^2\Delta_{21}\right)
    + O(\varepsilon^3\Delta_{31}) \;.
%& = & \left(\Delta_{31} - \kappa_{\mu\mu} \Delta_{21}\right)
%    + O(\varepsilon^3 \Delta_{31}) \;.
\label{Pmu2muArgument}
\end{eqnarray}
Since 
$\sin(2\theta_\mathrm{atm}) 
\approx \sin(2\tilde{\theta}_\mathrm{atm}) 
\approx \sin(2\edlit{\theta}_\mathrm{atm})$,
we conclude
\begin{equation}
\tilde{P}(\nu_\mu\rightarrow \nu_\mu) 
\approx P(\nu_\mu\rightarrow \nu_\mu)
= P(\bar{\nu}_\mu\rightarrow \bar{\nu}_\mu)
\approx \edlit{P}(\bar{\nu}_\mu\rightarrow \bar{\nu}_\mu)\;.
\end{equation}
That is, the $\nu_\mu$ and $\bar{\nu}_\mu$ survival probabilities are 
insensitive to matter effects and their values in matter are the same as their
values in vacuum.  In Fig.~\ref{Pmu2muVacuumMatter} we compare the 
exact numerical values of these probabilities calculated for our example
parameter set Eq.~(\ref{exampleparameterset}) with $\delta m^2_{31}>0$.
As can be seen, the differences among these probabilities are extremely small,
and our approximation has allowed us to understand this analytically.

%%%%%%%%%%%%%%%%%%%%%%%%%%%%%%%%
%%%%%%%%%%%%%%%%%%%%%%%%%%%%%%%%

Next, let us consider the $\nu_\mu\rightarrow \nu_e$ and
$\bar{\nu}_\mu\rightarrow \bar{\nu}_e$ oscillation probabilities.
They are
\begin{eqnarray}
\lefteqn{\tilde{P}(\nu_\mu \rightarrow \nu_e)} \cr
& = &
4\sin^2\tilde{\theta}_{13}\sin^2\tilde{\theta}_\mathrm{atm}
\Biggl[
\left\{ 1 - ( \tilde{B}\sin\tilde{\delta} ) \tilde{\Delta}_{21} \right\}
\sin^2
\left(\frac{\tilde{\Delta}_{31} - \tilde{\kappa}_{\mu e}\tilde{\Delta}_{21}}{2}
\right)
\Biggr.
\cr
& &
+ \frac{1}{4}
  \left\{ ( \tilde{B}\sin\tilde{\delta})^2
        - 2 \tilde{\kappa}_{\mu e}
          ( 1 - \tilde{\kappa}_{\mu e} ) \sin^2\frac{\tilde{\Delta}_{31}}{2}
        - ( \tilde{B}\sin\tilde{\delta} )
          ( 1 - 2\tilde{\kappa}_{\mu e} ) \sin\tilde{\Delta}_{31}
  \right\} \tilde{\Delta}_{21}^2
\cr
& & 
- \frac{1}{12}
  \left\{ 3 ( \tilde{B}\sin\tilde{\delta} )\tilde{\kappa}_{\mu e}^2 
        + 2 ( \tilde{B}\sin\tilde{\delta} )
            ( 1 - 3\tilde{\kappa}_{\mu e}^2 )\sin^2\frac{\tilde{\Delta}_{31}}{2}
        - \tilde{\kappa}_{\mu e} 
            ( 1 - \tilde{\kappa}_{\mu e}^2 )\sin\tilde{\Delta}_{31}
  \right\} \tilde{\Delta}_{21}^3 \cr
& & \Biggl. +\,O(\tilde{\Delta}_{21}^4)\;  \Biggr] \;,  
\label{Pmu2eExpand} \\
& & \cr
\lefteqn{\edlit{P}(\bar{\nu}_\mu \rightarrow \bar{\nu}_e)} \cr
& = &
4\sin^2\edlit{\theta}_{13}\sin^2\edlit{\theta}_\mathrm{atm}
\Biggl[
\left\{ 1 + ( \edlit{B}\sin\edlit{\delta} ) \edlit{\Delta}_{21} \right\}
\sin^2
\left(\frac{\edlit{\Delta}_{31} - \edlit{\kappa}_{\mu e}\edlit{\Delta}_{21}}{2}
\right)
\Biggr.
\cr
& &
+ \frac{1}{4}
  \left\{ ( \edlit{B}\sin\edlit{\delta})^2
        - 2 \edlit{\kappa}_{\mu e}
          ( 1 - \edlit{\kappa}_{\mu e} ) \sin^2\frac{\edlit{\Delta}_{31}}{2}
        + ( \edlit{B}\sin\edlit{\delta} )
          ( 1 - 2\edlit{\kappa}_{\mu e} ) \sin\edlit{\Delta}_{31}
  \right\} \edlit{\Delta}_{21}^2
\cr
& & 
+ \frac{1}{12}
  \left\{ 3 ( \edlit{B}\sin\edlit{\delta} )\edlit{\kappa}_{\mu e}^2 
        + 2 ( \edlit{B}\sin\edlit{\delta} )
            ( 1 - 3\edlit{\kappa}_{\mu e}^2 )\sin^2\frac{\edlit{\Delta}_{31}}{2}
        + \edlit{\kappa}_{\mu e} 
            ( 1 - \edlit{\kappa}_{\mu e}^2 )\sin\edlit{\Delta}_{31}
  \right\} \edlit{\Delta}_{21}^3 \cr
& & \Biggl. +\,O(\edlit{\Delta}_{21}^4)\;  \Biggr] \;,  
\label{Pmubar2ebarExpand} 
\end{eqnarray} 
From Eq.~(\ref{NeutrinoMatterParameters}) we obtain
\begin{eqnarray*}
\tilde{B}
& = & \frac{\tilde{A}}{\sin^2\tilde{\theta}_\mathrm{13}\sin^2\tilde{\theta}_{atm}}
\;\approx\; \frac{1}{2}\left( \dfrac{\delta m^2_{21}}{a} \right)
\left( 1 - \dfrac{a}{\delta m^2_{31}} \right) \frac{1}{\theta_{13}}
\;=\; \frac{1}{2}\,O(1)\;, \cr
\tilde\kappa_{\mu e} 
& = &  \tilde{s}^2_\mathrm{12} - \tilde{B}\cos\tilde{\delta} 
\;\approx\; 1 - \frac{1}{2}\, O(1)\cos\delta \;, 
\end{eqnarray*}
and from Eq.~(\ref{AntiNeutrinoMatterParameters}) we obtain
\begin{eqnarray*}
\edlit{B}
& = & \frac{\edlit{A}}{\sin^2\edlit{\theta}_\mathrm{13}\sin^2\edlit{\theta}_{atm}} 
\;\approx\; \frac{1}{2}\left( \dfrac{\delta m^2_{21}}{a} \right)
\left( 1 + \dfrac{a}{\delta m^2_{31}} \right) \frac{1}{\theta_{13}} 
\;=\; \frac{1}{2}\,O(1)\;, \cr
\edlit\kappa_{\mu e} 
& = &  \edlit{s}^2_\mathrm{12} - \edlit{B}\cos\edlit{\delta} 
\;=\; \frac{1}{4}\,O(\varepsilon^2) - \frac{1}{2}\, O(1)\cos\delta
\;\approx\; -\frac{1}{2}\,O(1)\cos\delta \;.
\end{eqnarray*}
Therefore, the expressions inside the curly brackets in the 
$\tilde{\Delta}_{21}^2$, $\edlit{\Delta}_{21}^2$ and 
$\tilde{\Delta}_{21}^3$, $\edlit{\Delta}_{21}^3$ terms are roughly of order one,
with factors of $1/4$ and $1/12$ in front suppressing them.
And since the entire expression is multiplied by 
$\sin^2\tilde{\theta}_{13} = O(\varepsilon^2)$, 
$\sin^2\edlit{\theta}_{13} = O(\varepsilon^2)$ from outside the 
square brackets, these terms are of $O(\pi^2\varepsilon^4)$ and are negligible.

Therefore, to a good approximation, we can use the expressions
\begin{eqnarray}
\tilde{P}(\nu_\mu\rightarrow \nu_e)
& = & 4\sin^2\tilde{\theta}_{13}\sin^2\tilde{\theta}_\mathrm{atm}
\left\{ 1 - (\tilde{B}\sin\tilde{\delta})\tilde{\Delta}_{21} \right\}
\sin^2\left( \dfrac{ \tilde{\Delta}_{31} - \tilde{\kappa}_{\mu e}\tilde{\Delta}_{21} }{ 2 }
      \right) \;, \cr
\edlit{P}(\bar{\nu}_\mu\rightarrow \bar{\nu}_e)
& = & 4\sin^2\edlit{\theta}_{13}\sin^2\edlit{\theta}_\mathrm{atm}
\left\{ 1 + (\edlit{B}\sin\edlit{\delta})\edlit{\Delta}_{21} \right\}
\sin^2\left( \dfrac{ \edlit{\Delta}_{31} - \edlit{\kappa}_{\mu e}\edlit{\Delta}_{21} }{ 2 }
      \right) \;,
\end{eqnarray}
where the arguments of the sine functions are given by
\begin{eqnarray}
\lefteqn{\tilde{\Delta}_{31} - \tilde{\kappa}_{\mu e}\tilde{\Delta}_{21}} \cr
& \approx & \tilde{\Delta}_{31}
          - \left( \tilde{s}^2_{12} - \tilde{B}\cos\tilde{\delta} \right)\tilde{\Delta}_{21} \cr
& \approx & (\Delta_{31} - c_{12}^2 \Delta_{21})
          - \left\{ 1
                  - \frac{1}{2}\left(\dfrac{\delta m^2_{21}}{a\theta_{13}}\right)
                               \left(1-\dfrac{a}{\delta m^2_{31}}\right)\cos\delta
            \right\}
            \left( \frac{a}{2E}L - \Delta_{21}\cos 2\theta_{12} \right) \cr
%& = & \Delta_{31}
%\left[ \left\{ 1 - c_{12}^2\left(\dfrac{\delta m^2_{21}}{\delta m^2_{31}}\right)
%       \right\}
%     - \left(\frac{a}{\delta m^2_{31}}\right)
%       \left\{ 1
%             - \frac{1}{2}\left(\dfrac{\delta m^2_{21}}{a\theta_{13}}\right)
%                          \left(1-\dfrac{a}{\delta m^2_{31}}\right)\cos\delta
%       \right\}
%\right] \cr
& = & \Delta_{31}
\left[ 1 - \left(\dfrac{a}{\delta m^2_{31}}\right)
         + \dfrac{1}{2\theta_{13}}\left(\dfrac{\delta m^2_{21}}{\delta m^2_{31}}\right)\cos\delta
	     + O(\varepsilon^2)
\right] \cr
& = & \Delta_{31}
\left[ 1 + \mathrm{sign}(\delta m^2_{31})
       \left( -\dfrac{a}{|\delta m^2_{31}|} + \dfrac{\varepsilon^2}{2\theta_{13}}\cos\delta \right)
       + O(\varepsilon^2)
\right] \;, \cr
& & \cr
\lefteqn{\edlit{\Delta}_{31} - \edlit{\kappa}_{\mu e}\edlit{\Delta}_{21}} \cr
& \approx & \edlit{\Delta}_{31}
          - \left( \edlit{s}^2_{12} - \edlit{B}\cos\edlit{\delta} \right)\edlit{\Delta}_{21} \cr
& \approx & \left(\Delta_{31} + \dfrac{a}{2E}L - s_{12}^2 \Delta_{21}\right)
          - \left\{ 
                  - \frac{1}{2}\left(\dfrac{\delta m^2_{21}}{a\theta_{13}}\right)
                               \left(1+\dfrac{a}{\delta m^2_{31}}\right)\cos\delta
            \right\}\left( \frac{a}{2E}L + \Delta_{21}\cos 2\theta_{12} \right) \cr
%& = & \Delta_{31}
%\left[ \left\{ 1 + \left(\dfrac{a}{\delta m^2_{31}}\right)
%                 - s_{12}^2\left(\dfrac{\delta m^2_{21}}{\delta m^2_{31}}\right)
%       \right\}
%     - \left(\frac{a}{\delta m^2_{31}}\right)
%       \left\{ 
%             - \frac{1}{2}\left(\dfrac{\delta m^2_{21}}{a\theta_{13}}\right)
%                          \left(1-\dfrac{a}{\delta m^2_{31}}\right)\cos\delta
%       \right\}
%\right] \cr
& = & \Delta_{31}
\left[ 1 + \left(\dfrac{a}{\delta m^2_{31}}\right)
         + \dfrac{1}{2\theta_{13}}\left(\dfrac{\delta m^2_{21}}{\delta m^2_{31}}\right)\cos\delta
	     + O(\varepsilon^2)
\right] \cr
& = & \Delta_{31}
\left[ 1 + \mathrm{sign}(\delta m^2_{31})
       \left( \dfrac{a}{|\delta m^2_{31}|} + \dfrac{\varepsilon^2}{2\theta_{13}}\cos\delta \right)
       + O(\varepsilon^2)
\right] \;,
\end{eqnarray}
while the amplitudes are given by
\begin{eqnarray}
\lefteqn{ 4\sin^2\tilde{\theta}_{13}\sin^2\tilde{\theta}_\mathrm{atm}
          \left\{ 1 - (\tilde{B}\sin\tilde{\delta}) \tilde{\Delta}_{21} \right\} } \cr
& \approx & 2\theta_{13}^2
\left( 1 + \dfrac{2a}{\delta m^2_{31}} \right) 
\left[ 1 
     - \left\{ \frac{1}{2}\left(\dfrac{\delta m^2_{21}}{a\theta_{13}}\right)
               \left( 1 - \dfrac{a}{\delta m^2_{31}} \right)\sin\delta
       \right\}
       \left( \dfrac{a}{2E}L - \Delta_{21}\cos 2\theta_{12} \right)
\right] \cr
& = & 2\theta_{13}^2
\left[ 1 + \dfrac{2a}{\delta m^2_{31}}
         - \dfrac{1}{2\theta_{13}}\left( \dfrac{\delta m^2_{21}}{\delta m^2_{31}} \right)
           \Delta_{31} \sin\delta + O(\varepsilon^2) 
\right] \cr
& = & 2\theta_{13}^2
\left[ 1 + \dfrac{2a}{\delta m^2_{31}}
         - \dfrac{\varepsilon^2}{2\theta_{13}}|\Delta_{31}| \sin\delta
	     + O(\varepsilon^2)
\right] \;,\cr
& & \cr
\lefteqn{ 4\sin^2\edlit{\theta}_{13}\sin^2\edlit{\theta}_\mathrm{atm}
          \left\{ 1 + (\edlit{B}\sin\edlit{\delta}) \edlit{\Delta}_{21} \right\} } \cr
& \approx & 2\theta_{13}^2
\left( 1 - \dfrac{2a}{\delta m^2_{31}} \right) 
\left[ 1 
     + \left\{ \frac{1}{2}\left(\dfrac{\delta m^2_{21}}{a\theta_{13}}\right)
               \left( 1 + \dfrac{a}{\delta m^2_{31}} \right)\sin\delta
       \right\}
       \left( \dfrac{a}{2E}L + \Delta_{21}\cos 2\theta_{12} \right)
\right] \cr
& = & 2\theta_{13}^2
\left[ 1 - \dfrac{2a}{\delta m^2_{31}}
         + \dfrac{1}{2\theta_{13}}\left( \dfrac{\delta m^2_{21}}{\delta m^2_{31}} \right)
           \Delta_{31} \sin\delta + O(\varepsilon^2) 
\right] \cr
& = & 2\theta_{13}^2
\left[ 1 - \dfrac{2a}{\delta m^2_{31}}
         + \dfrac{\varepsilon^2}{2\theta_{13}}|\Delta_{31}| \sin\delta
	     + O(\varepsilon^2)
\right] \;.
\end{eqnarray}
From these simple expressions, we can discern a few facts about 
$\nu_e$ ($\bar{\nu}_e$) appearance experiments.

Roughly speaking, the positions of the oscillation peaks will provide information on the combination
\begin{equation}
\mathrm{sign}(\delta m^2_{31})\left(\mp\dfrac{a}{|\delta m^2_{31}|}
                           + \dfrac{\varepsilon^2}{2\theta_{13}}\cos\delta
                      \right)\;,
\label{PeakPositionInfo}
\end{equation}
while the heights of the oscillation peaks will provide information on the combination
\begin{equation}
2\theta_{13}^2
\left[ 1 \pm
\left\{ \mathrm{sign}(\delta m^2_{31}) \dfrac{2a}{|\delta m^2_{31}|} 
        - \frac{\varepsilon^2}{2\theta_{13}}|\Delta_{31}|\sin\delta
\right\} 
\right] \;,
\label{PeakHeightInfo}
\end{equation}
where the upper signs are for the neutrinos, and the lower signs are for the
anti-neutrinos.
If the oscillation peaks occur in an energy region in which
\begin{equation}
\frac{a}{|\delta m^2_{31}|} \ll \frac{\varepsilon^2}{2\theta_{13}}\;,
\end{equation}
($\alpha \sim -2$) then the $a/|\delta m^2_{31}|$ terms in these expressions can be neglected.
Then, measuring the height of the peak (or just the total $\nu_e$ or $\bar{\nu}_e$ flux using a narrow band beam) will allow us to constrain $\sin\delta$, provided that $\theta_{13}$ is well known from future reactor experiments \cite{DoubleChooz,kaska}.
If $\theta_{13}$ is not well-known, then measuring the peak heights for
both the neutrino and anti-neutrino will allow us to constrain both $\theta_{13}$ and $\sin\delta$.
If one also measures the position of the peaks, either using a wide band beam or by changing
the beam energy, then one can also extract information on the product
\begin{equation}
\mathrm{sign}(\delta m^2_{31}) \cos\delta\;,
\end{equation}
but neither $\mathrm{sign}(\delta m^2_{31})$ nor the sign of $\cos\delta$ can be
uniquely determined, even if both neutrino and anti-neutrino beams are used 
\cite{Barger:2001yr}. 

These features are clearly visible in Fig.~\ref{Pmu2e295}, which shows the
probabilities to be probed by the T2K experiment \cite{T2K}.
In phase~2 of T2K, the oscillation event rates for both neutrinos and 
anti-neutrinos are to be measured, the difference from which we can extract $\sin\delta$.
However, this does not provide any information on $\cos\delta$.
It was proposed in Ref.~\cite{AokiHagiwaraOkamura} to measure the event rates
using several beams of different energy, and thereby obtain some information on the
peak position of the oscillation spectrum, but the sign of $\cos\delta$ cannot be
uniquely determined unless the sign of $\delta m^2_{31}$ is known \cite{Hagiwara:2004iq}.
(These difficulties can be best seen visually by utilizing the 
Minakata-Nunokawa plot \cite{MinakataNunokawa}.)

On the other hand, if the oscillation peaks occur in an energy region in which
\begin{equation}
\frac{a}{|\delta m^2_{31}|} \approx \frac{\varepsilon^2}{2\theta_{13}} \;,
\end{equation}
($\alpha\sim -1$) then the measurement of the peak height by itself may not be able to determine either $\mathrm{sign}(\delta m^2_{31})$ or $\sin\delta$, even if $\theta_{13}$
were accurately known.  In particular, there will be a
degeneracy between the two cases in which $\mathrm{sign}(\delta m^2_{31})$ and $\sin\delta$ are both positive, and both negative, as can be clearly seen in Fig.~\ref{Pmu2e1000}.
Measuring peak heights for both the neutrino and anti-neutrino will not help use
here since they both depend on the same linear combination of $a/\delta m^2_{31}$
and $\sin\delta$, though it will help us in determining $\theta_{13}$.
If one also measures the position of the peak,
then the degeneracy in $\mathrm{sign}(\delta m^2_{31})$ and $\cos\delta$ which existed for
the previous case can be lifted, except when $\delta \approx 0$ for the neutrino case, and
$\delta \approx \pi$ for the anti-neutrino case.
Due to these shortcomings in performing a single experiment at either $\alpha\sim -1$
or $\alpha\sim -2$, various scenarios have been suggested which utilize two detectors
set up at different baseline lengths \cite{TwoBaseline,IKMN,HagiwaraOkamuraSenda,SuperNova}.

%%%%%%%%%%%%%%%%%%%%%%%%%%%%%%%%%%%%%%%%%%%%%%%%%%%%%%%%%%%%%%%%%%%%%%%%
%\subsection{$L=2540\,\mathrm{km}$}

For the $L=2540\,\mathrm{km}$ case, an expansion in $\tilde{\Delta}_{21}$ 
is no longer permissible.  The approximation we used above for $\tilde{\theta}_{13}$
also breaks down as $\alpha\rightarrow 0$.  We can nevertheless simplify the
expressions for the oscillation probabilities and understand their behavior analytically.
As an example, consider the oscillation probabiliity $\tilde{P}(\nu_\mu\rightarrow\nu_e)$,
the full expression of which is
\begin{eqnarray}
\tilde{P}(\nu_\mu\rightarrow\nu_e)
& = & 4\,|\tilde{U}_{\mu 2}|^2 |\tilde{U}_{e2}|^2 \sin^2\dfrac{\tilde{\Delta}_{21}}{2}
    + 4\,|\tilde{U}_{\mu 3}|^2 |\tilde{U}_{e3}|^2 \sin^2\dfrac{\tilde{\Delta}_{31}}{2} \cr
& & + 2\,\Re(\tilde{U}^*_{\mu 3}\tilde{U}_{e3}\tilde{U}_{\mu 2}\tilde{U}^*_{e2})
       \left( 4\sin^2\dfrac{\tilde{\Delta}_{21}}{2}\sin^2\dfrac{\tilde{\Delta}_{31}}{2}
            + \sin\tilde{\Delta}_{21} \sin\tilde{\Delta}_{31}
       \right) \cr
& & + 4\,\tilde{J}_{(\mu,e)}
       \left( \sin^2\dfrac{\tilde{\Delta}_{21}}{2} \sin\tilde{\Delta}_{31}
            - \sin^2\dfrac{\tilde{\Delta}_{31}}{2} \sin\tilde{\Delta}_{21}
       \right) \;.
\label{mu2eFull} 
\end{eqnarray}
Using the approximations in Eq.~(\ref{tildetheta4}), except the one for $\tilde{\theta}_{13}$,
we find
\begin{eqnarray}
|\tilde{U}_{\mu 2}|^2 |\tilde{U}_{e2}|^2  
& \approx & s_{23}^2 \tilde{s}_{13}^2 \tilde{c}_{13}^2 - 2\tilde{A}\cos\delta \;,\cr
|\tilde{U}_{\mu 3}|^2 |\tilde{U}_{e3}|^2 
& \approx & s_{23}^2 \tilde{s}_{13}^2 \tilde{c}_{13}^2 \;,\cr
\Re(\tilde{U}^*_{\mu 3}\tilde{U}_{e3}\tilde{U}_{\mu 2}\tilde{U}^*_{e2})
& \approx & -s_{23}^2 \tilde{s}_{13}^2 \tilde{c}_{13}^2 + \tilde{A}\cos\delta \;, \cr
\tilde{J} _{(\mu,e)}
& \approx & \tilde{A}\sin\delta \;,
\end{eqnarray}
where
\begin{equation}
\tilde{A}
\approx \left(\dfrac{\delta m^2_{21}}{2a}\right)
        \sin(2\theta_{12})s_{23}c_{23}\tilde{s}_{13}\tilde{c}_{13}^2 \;.
\end{equation}
Terms of order $(\delta m^2_{21}/a)^2$ and higher have been neglected.
Substituting into Eq.~(\ref{mu2eFull}), we obtain
\begin{eqnarray}
\tilde{P}(\nu_\mu\rightarrow\nu_e)
& \approx & 
  4 s_{23}^2\tilde{s}_{13}^2\tilde{c}_{13}^2 \sin^2\dfrac{\tilde{\Delta}_{32}}{2} 
+ 8\tilde{A}\sin\dfrac{\tilde{\Delta}_{32}}{2}\sin\dfrac{\tilde{\Delta}_{21}}{2}
  \cos\biggl(\dfrac{\tilde{\Delta}_{31}}{2}+\delta\biggr) \cr
& \approx &
\left[ s_{23}\sin(2\tilde{\theta}_{13}) \sin\dfrac{\tilde{\Delta}_{32}}{2}
     +  c_{23}\tilde{c}_{13}
       \left(\dfrac{\delta m^2_{21}}{a}\right)\sin(2\theta_{12})
       \sin\dfrac{\tilde{\Delta}_{21}}{2}
       \cos\biggl(\dfrac{\tilde{\Delta}_{31}}{2}+\delta\biggr) 
\right]^2 \cr
& \approx & \dfrac{1}{2}
\left[ \sin(2\tilde{\theta}_{13})\sin\dfrac{\tilde{\Delta}_{32}}{2}
     + \tilde{c}_{13}\left(\dfrac{\delta m^2_{21}}{a}\right)
       \sin\dfrac{\tilde{\Delta}_{21}}{2}
       \cos\biggl(\dfrac{\tilde{\Delta}_{31}}{2}+\delta\biggr)
\right]^2 \;,
\label{mu2eAPPROX2540}
\end{eqnarray}
where we have used $\sin(2\theta_{12}) = 1 - 2O(\varepsilon)$, 
and $\sin(2\theta_{23}) = 1-\frac{1}{2}O(\varepsilon^2)$.
Now, the $a$-dependence of $\tilde{\theta}_{13}$ is different
depending on the sign of $\delta m^2_{31}$.
If $\delta m^2_{31}>0$ (normal hierarchy), $\tilde{\theta}_{13}$
increases monotonically from $\theta_{13}$ toward $\frac{\pi}{2}$ 
as $a$ increases with energy,
and passes through $\frac{\pi}{4}$ around $a\approx |\delta m^2_{31}|$.
If $\delta m^2_{31}<0$ (inverted hierarchy), $\tilde{\theta}_{13}$
decreases monotonically from $\theta_{13}$ toward $0$ as $a$ increase with energy,
and is about $\theta_{13}/2$ around $a\approx |\delta m^2_{31}|$.
(cf. Table~\ref{aDependenceofAngles})
Therefore, if $\delta m^2_{31}>0$, then the coefficient of the first term 
in the brackets of Eq.~(\ref{mu2eAPPROX2540}) is maximized around
$a\approx |\delta m^2_{31}|$. 
If the oscillation peak where $\tilde{\Delta}_{32}\approx \pi$ matches that energy,
one can expect a maximum oscillation probability as large as $\frac{1}{2}$.
(Note that this maximum probability is determined by $\theta_{23}$ and is
independent of the value of $\theta_{13}$ in vacuum.)
On the other hand, if $\delta m^2_{31}<0$, then the same coefficient
is suppressed to $O(\varepsilon)$, and the oscillation probability
with be suppressed by a factor of $O(\varepsilon^2)$.
This difference is evident in Fig.~\ref{Pmu2e2540}.
(In both cases, the second term in the brackets interferes with the first term 
giving the probability a weak $\delta$-dependence.)
Due to this clear difference, measuring 
the $\nu_e$ appearance probability at the first oscillation peak at a baseline length of
$L=2540\,\mathrm{km}$ has been proposed as an unambiguous method to determine the sign of 
$\delta m^2_{31}$ \cite{Diwan}.

%%%%%%%%%%%%%%%%%%%%
\begin{table}[t]
\begin{tabular}{|c|c|c||c|c|c|c|c|}
\hline
\multicolumn{3}{|c||}{$a/|\delta m^2_{31}|$} 
& $O(\varepsilon^3)$ 
& $O(\varepsilon^2)$ 
& $O(\varepsilon)$
& $O(1)$
& $O(\varepsilon^{-1})$ \\
\hline\hline
\multirow{4}{*}{$\;\nu\;$}
& \multicolumn{2}{|c||}{$\tilde{\theta}_{12}$}
& $\;\approx\theta_{12}\;$ 
& $\nearrow$ 
& \multicolumn{3}{c|}{$\;\approx\dfrac{\pi}{2}\;$} \\
\cline{2-8}
& $\;\tilde{\theta}_{13}\;$ 
& $\;\delta m^2_{31}>0\;$
& \multicolumn{3}{c|}{$\;\approx\theta_{13}\;$} 
& $\nearrow$ 
& $\;\approx\dfrac{\pi}{2}\;$\\
\cline{3-8}
&
& $\;\delta m^2_{31}<0\;$
& \multicolumn{3}{c|}{$\;\approx\theta_{13}\;$} 
& $\searrow$ 
& $\;\approx 0\;$\\
%\cline{2-8}
%& \multicolumn{2}{|c||}{$\tilde{\theta}_{23}$}
%& \multicolumn{5}{c|}{$\;\approx\theta_{23}\;$} \\
\hline\hline
\multirow{4}{*}{$\;\bar{\nu}\;$}
& \multicolumn{2}{|c||}{$\edlit{\theta}_{12}$}
& $\;\approx\theta_{12}\;$ 
& $\searrow$ 
& \multicolumn{3}{c|}{$\;\approx 0\;$} \\
\cline{2-8}
& $\;\edlit{\theta}_{13}\;$ 
& $\;\delta m^2_{31}>0\;$
& \multicolumn{3}{c|}{$\;\approx\theta_{13}\;$} 
& $\searrow$ 
& $\;\approx 0\;$\\
\cline{3-8}
&
& $\;\delta m^2_{31}<0\;$
& \multicolumn{3}{c|}{$\;\approx\theta_{13}\;$} 
& $\nearrow$ 
& $\;\approx \dfrac{\pi}{2}\;$\\
%\cline{2-8}
%& \multicolumn{2}{|c||}{$\edlit{\theta}_{23}$}
%& \multicolumn{5}{c|}{$\;\approx\theta_{23}\;$} \\
\hline
\end{tabular}
\caption{The dependence of the effective mixing angles on $a/|\delta m^2_{31}|$.} 
\label{aDependenceofAngles}
\end{table}
%%%%%%%%%%%%%%%%%%%%

%%%%%%%%%%%%%%%%%%%
\begin{figure}[p]
\begin{center}
\includegraphics[scale=0.8]{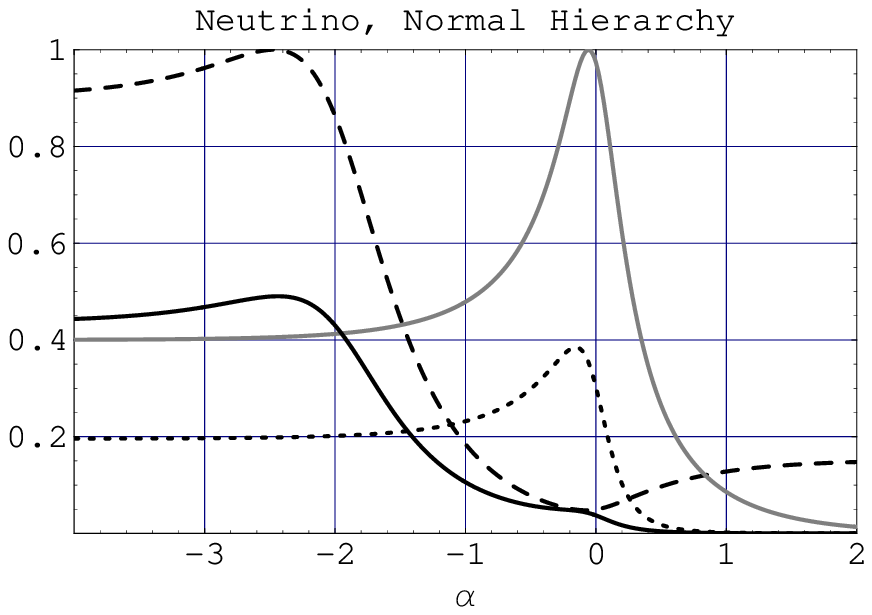}
\includegraphics[scale=0.8]{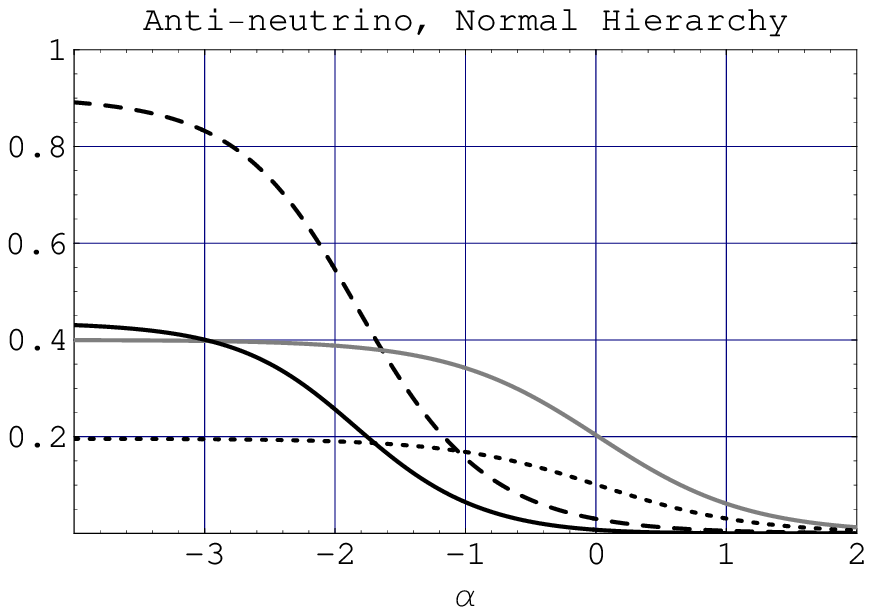}
\includegraphics[scale=0.8]{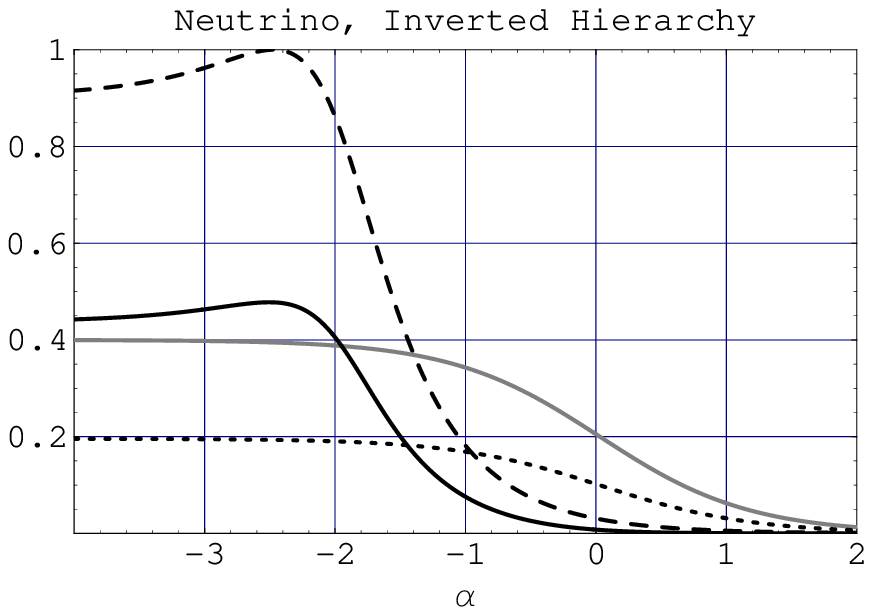}
\includegraphics[scale=0.8]{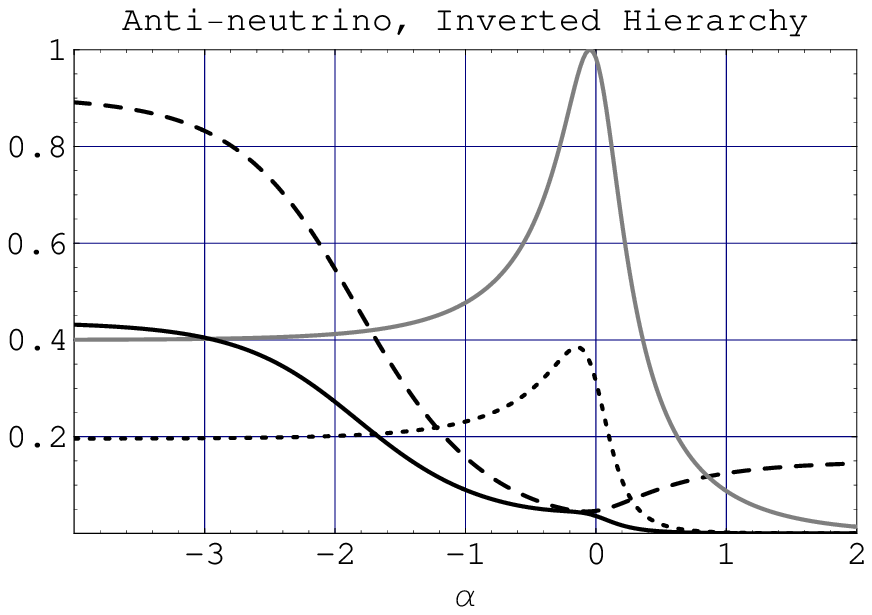}
\caption{The dependence of
$\sin(2\tilde{\theta}_{13})$, $\sin(2\edlit{\theta}_{13})$ (solid gray lines),
$\sin(2\tilde{\theta}_{12})$, $\sin(2\edlit{\theta}_{12})$ (dashed black lines),
$\tilde{s}_{13}(1-\tilde{s}_{13}^2)$, $\edlit{s}_{13}(1-\edlit{s}_{13}^2)$ (dotted black lines),
and $\tilde{A}/A_\mathrm{max}$, $\edlit{A}/A_\mathrm{max}$ (solid black lines)
on $\alpha=\log_{1/\varepsilon}(a/|\delta m^2_{31}|)$.
The input parameters were those listed in Eq.~(\protect{\ref{exampleparameterset}}).
}
\label{AandSines}
\end{center}
\end{figure}
%%%%%%%%%%%%%%%%%%%

%%%%%%%%%%%%%%%%%%%%%%%%%%%%%%%%%%%%%%%%%%%%%%%%%%%
Comparing the $L=295\,\mathrm{km}$,
$L=1000\,\mathrm{km}$, and $L=2540\,\mathrm{km}$
cases considered above, we can discern a generic trend that the
oscillation probability $\tilde{P}(\nu_\mu\rightarrow\nu_e)$ (or
$\edlit{P}(\bar{\nu}_\mu\rightarrow\bar{\nu}_e)$) 
is more sensitive to the CP violating phase $\delta$ at
lower energies (shorter baselines), 
and more sensitive to the mass hierarchy (sign of $\delta m^2_{31}$)
at higher energies (longer baselines).  
This can be understood as follows.
First, the angle $\tilde{\theta}_{13}$, which has a different 
energy dependence depending on the sign of $\delta m^2_{31}$, enters 
$\tilde{P}(\nu_\mu\rightarrow\nu_e)$ dominantly in the combination
\begin{eqnarray}
\sin\tilde{\theta}_{13}\sin\tilde{\theta}_{\mathrm{atm}}
& = & \sin\tilde{\theta}_{13}\cos\tilde{\theta}_{13}\sin\tilde{\theta}_{23} \cr
& = & \frac{1}{2}\sin(2\tilde{\theta}_{13})\sin\tilde{\theta}_{23} \cr
& \approx & \frac{1}{2}\sin(2\tilde{\theta}_{13})\sin\theta_{23}\;.
\end{eqnarray}
Therefore, $\sin(2\tilde{\theta}_{13})$ determines the sensitivity of 
$\tilde{P}(\nu_\mu\rightarrow\nu_e)$ on the behavior of $\tilde{\theta}_{13}$.
On the other hand, the size of CP violation is governed by the Jarskog invariant,
\begin{equation}
\tilde{J}=\tilde{A}\sin\tilde{\delta}\;,
\end{equation}
where
\begin{eqnarray}
\tilde{A} 
& = &\tilde{s}_{12}\tilde{c}_{12}\tilde{s}_{13}\tilde{c}_{13}^2\tilde{s}_{23}\tilde{c}_{23} \cr
& = &\frac{1}{4}\sin(2\tilde{\theta}_{12})\sin(2\tilde{\theta}_{23})
     \tilde{s}_{13}(1-\tilde{s}_{13}^2)  \cr
& \approx & \frac{1}{4}\sin(2\tilde{\theta}_{12})\sin(2{\theta}_{23})
            \tilde{s}_{13}(1-\tilde{s}_{13}^2)\;.
\end{eqnarray}
In our convention where the mixing angles are in the first quadrant, $\tilde{A}$ is bounded by
\begin{equation}
0 \;\le\; \tilde{A} \;\le\; \dfrac{1}{6\sqrt{3}} \equiv A_\mathrm{max}\;.
\end{equation}
In Fig.~\ref{AandSines}, we plot the $\alpha$-dependence of 
$\sin(2\tilde{\theta}_{13})$, $\sin(2\tilde{\theta}_{12})$, $\tilde{s}_{13}(1-\tilde{s}^2_{31})$, 
and $\tilde{A}/A_\mathrm{max}$, and their anti-neutrino counterparts, 
for the input parameters of Eq.~(\ref{exampleparameterset}).
It is clear from the figure that the differences in 
$\sin(2\tilde{\theta}_{13})$ and $\sin(2\edlit{\theta}_{13})$ between the
$\delta m^2_{31}>0$ and $\delta m^2_{31}<0$ cases is most prominent
at $\alpha\approx 0$ where there exists a large peak in one case which is
absent in the other.  This is due to the fact $\tilde{\theta}_{13}$ ($\edlit{\theta}_{13}$)
crosses $\pi/4$ at $a\approx |\delta m^2_{31}|$ when $\delta m^2_{31}>0$ 
($\delta m^2_{31}<0$).  (cf. Table~\ref{aDependenceofAngles}.)
Therefore, experiments that are performed closer to $\alpha = 0$
are more sensitive to $\mathrm{sign}(\delta m^2_{31})$.
On the other hand, the $\alpha$-dependence of $\tilde{A}$ ($\edlit{A}$)
is dominated by that of $\sin(2\tilde{\theta}_{12})$ 
($\sin(2\edlit{\theta}_{12}$) which starts out as a number of 
$O(1)$ at low $\alpha$, but drops off quickly toward zero as 
$\alpha$ is increased from $-2$ to $-1$.  
Therefore, experiments that are performed at $\alpha\alt -2$ are more
sensitive to $\delta$ than those performed at $\alpha\agt -1$.

%%%%%%%%%%%%%%%%%%%%%%%%%%%%%%%%%%%%%%%%%%%%%%%%%%%%%%%%%%%%%%%%%%%%%%%%%%%%%
\section{Summary}

In this paper, we considered the matter effect on neutrino oscillations, and 
derived simple analytical approximations to the effective
mixing angles and effective mass-squared differences in constant density matter.
Our results are summarized in Table~\ref{SummaryTable}.
These expressions can be utilized in calculating, analyzing, and understanding 
the behavior of the oscillation probabilities in LBL neutrino oscillation experiments.

The formalism developed in this paper can be further extended to incorporate
matter effects due to the violation of universality in neutral current interactions,
or additional matter effects due to new interactions.
This will be presented in a subsequent paper \cite{HOT2}.

%%%%%%%%%%%%%%%%%%%%%%%%%%%%%%%%%%%%%%%%%%%%%%%%%%%%%%%%%%%%%%%%%%%%%%%%%%%%%

\begin{turnpage}

\begingroup
\squeezetable
%\begin{center}
\begin{table}
\begin{tabular}{|c|c||c|c|c|c|}
\hline
\multicolumn{2}{|c||}{}
& \multicolumn{2}{c|}{$\delta m^2_{31}<0$ (Inverted Hierarchy)} 
& \multicolumn{2}{c|}{$\delta m^2_{31}>0$ (Normal Hierarchy)} \\
\cline{3-6}
\multicolumn{2}{|c||}{}
& $\dfrac{a}{|\delta m^2_{31}|} \ge O(1)$ 
& $\quad O(\varepsilon) \ge \dfrac{a}{|\delta m^2_{31}|}\quad$  
& $\quad \dfrac{a}{|\delta m^2_{31}|} \le O(\varepsilon)\quad$  
& $O(1) \le \dfrac{a}{|\delta m^2_{31}|}$ \\
\hline\hline
\multirow{7}{*}{%
$\begin{array}{l}
\mbox{Neutrino Case} \\ \\
\tan 2\varphi = \dfrac{a \sin 2\theta_{12}}{\delta m^2_{21} - a\cos 2\theta_{12}} \\ \\
\tan 2\phi    = \dfrac{a \sin 2\theta_{13}}{\delta m^2_{31} - a\cos 2\theta_{13}} 
\end{array}$
} 
%$\;\;\nu\;\;$
& $\;\tilde{\theta}_{13}\;$ 
& \multicolumn{4}{c|}{$\theta'_{13}=\theta_{13}+\phi$} \\    
\cline{2-6}
& $\;\tilde{\theta}_{12}\;$ 
& \multicolumn{3}{c|}{$\theta'_{12}=\theta_{12}+\varphi$} 
& $\dfrac{\pi}{2}-\dfrac{c_{13}}{c'_{13}}\left(\dfrac{\delta m^2_{21}}{2a}\right)\sin(2\theta_{12})$ \\ 
\cline{2-6}
& $\;\tilde{\theta}_{23}\;$ 
& \multicolumn{3}{c|}{$\theta_{23}$} 
& $\theta_{23}+\dfrac{s_\phi}{c'_{13}}\left(\dfrac{\delta m^2_{21}}{2a}\right)\sin(2\theta_{12})\cos\delta$ \\
\cline{2-6}
& $\;\tilde{\delta}\;$      
& \multicolumn{3}{c|}{$\delta$} 
& $\delta-\dfrac{s_\phi}{c'_{13}}\left(\dfrac{\delta m^2_{21}}{a}\right)
          \dfrac{\sin(2\theta_{12})}{\tan(2\theta_{23})}\sin\delta$ \\
\cline{2-6}
& $\;\lambda_1\;$           
& \multicolumn{3}{c|}{$\dfrac{ (\delta m^2_{21}+a)-\sqrt{ (\delta m^2_{21}-a)^2 + 4a\delta m^2_{21}s^2_{12}} }{ 2 }$}
& $\delta m^2_{21} c^2_{12}$ \\
\cline{2-6}
& $\;\lambda_2\;$           
& \multicolumn{3}{c|}{$\dfrac{ (\delta m^2_{21}+a)+\sqrt{ (\delta m^2_{21}-a)^2 + 4a\delta m^2_{21}s^2_{12}} }{ 2 }$} 
& $\;\dfrac{ (\delta m^2_{31}+a)-\sqrt{ (\delta m^2_{31}-a)^2 + 4a\delta m^2_{31}s^2_{13}} }{ 2 }\;$ \\
\cline{2-6}
& $\;\lambda_3\;$           
& \multicolumn{3}{c|}{$\delta m^2_{31}$} 
& $\;\dfrac{ (\delta m^2_{31}+a)+\sqrt{ (\delta m^2_{31}-a)^2 + 4a\delta m^2_{31}s^2_{13}} }{ 2 }\;$ \\
\hline\hline
\multirow{7}{*}{%
$\begin{array}{l}
\mbox{Anti-neutrino Case} \\ \\
\tan 2\bar{\varphi} = -\dfrac{a \sin 2\theta_{12}}{\delta m^2_{21} + a\cos 2\theta_{12}} \\ \\
\tan 2\bar{\phi}    = -\dfrac{a \sin 2\theta_{13}}{\delta m^2_{31} + a\cos 2\theta_{13}} 
\end{array}$
} 
%$\;\;\bar{\nu}\;\;$
& $\;\edlit{\theta}_{13}\;$ 
& \multicolumn{4}{c|}{$\bar{\theta}'_{13}=\theta_{13}+\bar{\phi}$} \\
\cline{2-6}
& $\;\edlit{\theta}_{12}\;$ 
& $\dfrac{c_{13}}{\bar{c}'_{13}}\left(\dfrac{\delta m^2_{21}}{2a}\right)\sin(2\theta_{12})$
& \multicolumn{3}{c|}{$\bar{\theta}'_{12}=\theta_{12}+\bar{\varphi}$} \\ 
\cline{2-6}
& $\;\edlit{\theta}_{23}\;$ 
& $\theta_{23}-\dfrac{\bar{s}_{\phi}}{\bar{c}'_{13}}\left(\dfrac{\delta m^2_{21}}{2a}\right)\sin(2\theta_{12})\cos\delta$
& \multicolumn{3}{c|}{$\theta_{23}$} \\
\cline{2-6}
& $\;\edlit{\delta}\;$     
& $\delta + \dfrac{\bar{s}_\phi}{\bar{c}'_{13}}\left(\dfrac{\delta m^2_{21}}{a}\right)
\dfrac{\sin(2\theta_{12})}{\tan(2\theta_{23})}\sin\delta$ 
& \multicolumn{3}{c|}{$\delta$} \\
\cline{2-6}
& $\;\bar{\lambda}_1\;$     
& $\;\dfrac{ (\delta m^2_{31}-a)+\sqrt{ (\delta m^2_{31}+a)^2 - 4a\delta m^2_{31}s^2_{13}} }{ 2 }\;$ 
& \multicolumn{3}{c|}{$\dfrac{ (\delta m^2_{21}-a)-\sqrt{ (\delta m^2_{21}+a)^2 - 4a\delta m^2_{21}s^2_{12}} }{ 2 }$} \\
\cline{2-6}
& $\;\bar{\lambda}_2\;$     
& $\delta m^2_{21} c^2_{12}$ 
& \multicolumn{3}{c|}{$\dfrac{ (\delta m^2_{21}-a)+\sqrt{ (\delta m^2_{21}+a)^2 - 4a\delta m^2_{21}s^2_{12}} }{ 2 }$} \\
\cline{2-6}
& $\;\bar{\lambda}_3\;$
& $\;\dfrac{ (\delta m^2_{31}-a)-\sqrt{ (\delta m^2_{31}+a)^2 - 4a\delta m^2_{31}s^2_{13}} }{ 2 }\;$ 
& \multicolumn{3}{c|}{$\delta m^2_{31}$} \\
\hline
\end{tabular}
\caption{The approximate formulae for the effective mixing angles and effective mass-squared differences derived in this paper.}
\label{SummaryTable}
\end{table}
%\end{center}
\endgroup
\end{turnpage}

%%%%%%%%%%%%%%%%%%%%%%%%%%%%%%%%%%%%%%%%%%%%%%%%%%%%%%%%%%%%%%%%%%%%%%%%%%%%%
%\newpage

\section*{Acknowledgments}

We would like to thank Masafumi Koike and Masako Saito
for helpful communications.
Takeuchi would like to thank the hospitality of the particle theory group at
Ochanomizu Women's University,
where a major portion of this work was carried out during the summer of 2005.
This research was supported in part by the U.S. Department of Energy, 
grant DE--FG05--92ER40709, Task A (T.T.).

%%%%%%%%%%%%%%%%%%%%%%%%%%%%%%%%%%%%%%%%%%%%%%%%%%%%%%%%%%%%%%%%%%%%%%%%%%%%%

%%%%%%%%%%%%%%%%%%%%%%%%%%%%%%%%%%%%%%%%%%%%%%%%%%%%%%%%%%%%%%%%%%
\end{document}